\documentclass[aps,prx,twocolumn,groupedaddress,floatfix,longbibliography]{revtex4-2}

\usepackage{graphicx}
\usepackage{color}
\usepackage{amsmath,amssymb}
\usepackage{mathtools}
\usepackage{bbm}
\usepackage{natbib}
\usepackage{hyperref}
\hypersetup{
    colorlinks=true,
    citecolor=blue,
    linkcolor=blue,
    urlcolor=blue,
}
\allowdisplaybreaks

\begin{document}

\title{Uncovering the effect of RNA polymerase steric interactions on gene expression noise: analytical distributions of nascent and mature RNA numbers}

\author{Juraj Szavits-Nossan}
\email{Juraj.Szavits.Nossan@ed.ac.uk}
\affiliation{School of Biological Sciences, University of Edinburgh, Edinburgh EH9 3JH, United Kingdom}

\author{Ramon Grima}
\email{Ramon.Grima@ed.ac.uk}
\affiliation{School of Biological Sciences, University of Edinburgh, Edinburgh EH9 3JH, United Kingdom}

\date{\today}

\begin{abstract}
The telegraph model is the standard model of stochastic gene expression, which can be solved exactly to obtain the distribution of mature RNA numbers per cell. A modification of this model also leads to an analytical distribution of nascent RNA numbers. These solutions are routinely used for the analysis of single-cell data, including the inference of transcriptional parameters. However, these models neglect important mechanistic features of transcription elongation, such as the stochastic movement of RNA polymerases and their steric (excluded-volume) interactions. Here we construct a model of gene expression describing promoter switching between inactive and active states, binding of RNA polymerases in the active state, their stochastic movement including steric interactions along the gene, and their unbinding leading to a mature transcript that subsequently decays. We derive the steady-state distributions of the nascent and mature RNA numbers in two important limiting cases: constitutive expression and slow promoter switching. We show that RNA fluctuations are suppressed by steric interactions between RNA polymerases, and that this suppression can in some instances even lead to sub-Poissonian fluctuations; these effects are most pronounced for nascent RNA and less prominent for mature RNA, since the latter is not a direct sensor of transcription. We find a relationship between the parameters of our microscopic mechanistic model and those of the standard models that ensures excellent consistency in their prediction of the first and second RNA number moments over vast regions of parameter space, encompassing slow, intermediate, and rapid promoter switching, provided the RNA number distributions are Poissonian or super-Poissonian. Furthermore, we identify the limitations of inference from mature RNA data, specifically showing that it cannot differentiate between highly distinct RNA polymerase traffic patterns on a gene. 
\end{abstract}

\maketitle

\section{Introduction}

The widespread availability of RNA data in single cells has spurred a large amount of theoretical work on gene expression over the past two decades \cite{paulsson2005models,friedman2006linking,shahrezaei2008analytical,kumar2014exact,Lim2015,cao2020analytical,ham2020extrinsic,Jia2021,gupta2022frequency}. Snapshot measurements of the transcript numbers over a population of cells reveal large cell-to-cell variability in bacteria \cite{so2011general}, yeast\cite{zenklusen2008single} and mammalian cells \cite{suter2011mammalian}. This variability can be more directly appreciated by following transcription dynamics in a single cell, which demonstrates that transcription does not occur continuously, but rather at random times and in bursts \cite{donovan2019live}. A main focus of mathematical models of gene expression has been to understand the origin of this noise, and also how cells tolerate, control and possibly exploit it from the perspective of biological function \cite{rao2002control,singh2009optimal,balazsi2011cellular,kellogg2015noise}.   

Stochastic models of gene expression are predominantly based on the (random) telegraph model \cite{Peccoud1995}, which describes switching between two promoter states (active and inactive), synthesis of a mature transcript from the active state and its subsequent degradation. By assuming that the dynamics are Markovian, one can write a time-evolution equation for the joint probability distribution of the promoter state and the number of RNA molecules, which can be solved exactly in steady-state and also in time \cite{iyer2009stochasticity}. It has become common to fit the steady-state solution of this model to distributions of the number of RNA per cell obtained from single molecule fluorescence in situ hybridization (smFISH) \cite{zenklusen2008single,halpern2015bursty} or single-cell RNA sequencing (scRNA-seq) experiments \cite{kim2013inferring,larsson2019genomic}. Provided the RNA decay rate is estimated experimentally, this fitting leads to estimates of the transcriptional parameters (the rate of switching to the active state, the rate of switching to the off state and the synthesis rate) for any gene of interest. 

A criticism of this fitting procedure is that mature RNA numbers are not a direct sensor of transcription, as they are affected by other processes downstream of transcription, such as splicing and nuclear export. To overcome this criticism, models were developed to predict the distributions of nascent RNA, i.e. RNA that is attached to transcribing RNA polymerases (RNAPs) moving along a gene during transcriptional elongation \cite{Choubey2015,Xu2016}. These models are non-Markovian because nascent RNA does not decay via a first-order reaction (as mature RNA), but rather its removal is assumed to occur after a fixed deterministic time equal to the total time of elongation and termination. Fitting the steady-state solution to single-cell nascent RNA data also leads to estimation of transcriptional parameters, which may differ significantly from those estimated using mature mRNA data \cite{fu2022quantifying}. This makes a strong case for developing more accurate mathematical models of nascent RNA fluctuations. 

A disadvantage of current stochastic models of nascent RNA dynamics with explicit analytical solutions \cite{Choubey2015,Xu2016,Choubey2018,jiang2021neural,braichenko2021distinguishing,Szavits2023} is that they implicitly assume RNAPs move deterministically along the gene, hence they do not interact with each other. This is clearly an over-simplification, since the frequency of such interactions should at least be significant for highly transcribed genes \cite{Klumpp2011}. A significant number of computational studies have been undertaken to model the fine-scale dynamics of transcriptional elongation, such as volume-excluded (steric) interactions between RNAPs, ubiquitous pausing and backtracking, interaction of RNAPs with nucleosomes and the mechanochemical cycle of RNAP movement \cite{Tripathi2008-constitutive,Tripathi2008-telegraph,Voliotis2008,Klumpp2008,Kim_2018,Tripathi2009,Dobrzynski2009,Ribeiro_2010,Rajala_2010,Klumpp2011,Chowdhury2013,Sahoo2013,Heberling2016,Cholewa2019,Ali2020,Szavits2020-roadblocks,turowski2020nascent}. However, none of these studies have analytically derived the steady-state single-cell distributions of nascent and mature RNA numbers.  

In this paper, we take a first step towards achieving this goal by deriving the nascent and mature RNA number distributions for a stochastic model of gene expression that explicitly includes volume-excluded interactions between RNAPs. Specifically, the movement of RNAPs in this model is the same as that of the well known totally asymmetric simple exclusion process (TASEP) with uniform hopping rates \cite{Schutz1993,Derrida1993}. Whilst our model does not have all the detailed fine-scale biological detail of some of the aforementioned simulation-based studies, it provides a minimalist description of RNAP traffic on a gene that makes the transcript number distributions analytically tractable.

The paper is structured as follows. The constitutive model with RNAP volume exclusion is studied in Section \ref{constitutive-model}. The telegraph model with RNAP volume exclusion is studied in Section \ref{telegraph-model}. The results from the two models are summarized and discussed in Section \ref{summary-discussion}.   

\section{The constitutive model with RNAP volume exclusion}
\label{constitutive-model}

\begin{figure}[hbt!]
    \centering
    \includegraphics[width=8.6cm]{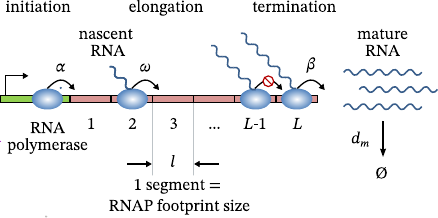}
    \caption{Cartoon illustrating the constitutive model of transcription with RNAP volume exclusion. The gene is divided into $L$ segments, each having a size equal to the footprint of a single RNAP. Transcription initiation is modelled by a single-step reaction that occurs at rate $\alpha$, provided the first segment is empty. RNAPs move along the gene with rate $\omega$, provided the segment in front is empty. As elongation progresses, the nascent RNA tail attached to the RNAP grows. Termination with rate $\beta$ leads to the unbinding of the RNAP-nascent RNA complex (transcription elongation complex) from the gene and its dissociation into the free RNAP and the free RNA (mature RNA) that is subsequently degraded with rate $d_M$.}
    \label{fig1}
\end{figure}

We first consider a model of constitutive gene expression, i.e. a gene that is constantly expressed and is not subject to regulation. The gene body is coarse-grained into $L$ segments of length $\ell\approx 35$ nucleotides, which is the footprint size of RNAP \cite{tongaonkar2005histones,schneider2006rna,claypool2004tor}. Hence, each segment $i=1,\dots,L$ is either empty or is occupied by an RNAP. Transcription starts by the binding of an RNAP to a promoter, which is followed by a sequence of steps including promoter opening, promoter escape, promoter-proximal pausing and pause release after which the elongation of the nascent transcript starts. We lump all these initiation phase processes together into a single-step reaction with rate $\alpha$. We note that the elongation phase can only start, i.e. RNAP can only be released into the first segment, provided this segment is empty. Subsequently, the RNAP moves forward stochastically one segment at a time, provided the neighboring segment in front of the RNAP is empty. We denote by $\omega$ the hopping rate at which the RNAP moves from a segment to the next one. This rate has units of inverse time, and it is related to the measured elongation rate in the absence of volume exclusion (in nucleotides per second) by the formula
\begin{equation}
    \omega=\frac{\text{elongation rate [nt/s]}}{\text{RNAP footprint size [nt]}}.
\end{equation}
Transcription termination occurs at the rate $\beta$ from the last segment, after which the RNAP is removed from the lattice and a mature RNA is produced. The mature RNA degrades stochastically with rate $d_M$. The model is schematically presented in Fig. \ref{fig1}. The full model can be summarized as
\begin{subequations}
\label{TASEP1-reactions}
\begin{align}
    & \text{free RNAP}+\emptyset_{1}\xrightarrow[]{\alpha}\text{RNAP}_1,\label{TASEP1-initiation}\\
    & \text{RNAP}_i+\emptyset_{i+1}\xrightarrow[]{\omega}\emptyset_{i}+\text{RNAP}_{i+1}, \quad i \in [1,L-1] \label{TASEP1-elopngation}\\
    & \text{RNAP}_L\xrightarrow[]{\beta}\emptyset_{L}+\text{free RNAP}+\text{RNA},\label{TASEP1-termination}\\
    & \text{RNA}\xrightarrow[]{d_M}\emptyset\label{TASEP1-degradation},
\end{align}
\end{subequations}
where free RNAP denotes an RNAP that is not actively engaged in transcription, $\text{RNAP}_i$ denotes an RNAP positioned at the segment $i$, $\emptyset_i$ denotes that the segment $i$ is empty, and RNA denotes the mature RNA. Since the number of free RNAPs is large, we can approximate the second-order reaction in Eq. (\ref{TASEP1-initiation}) by a quasi first-order reaction $\emptyset_{1}\xrightarrow[]{}\text{RNAP}_1$, where the rate $\alpha$ is proportional to the number of free RNAPs.

We note that a subset of the model describing the RNAP dynamics [Eqs. (\ref{TASEP1-initiation})-(\ref{TASEP1-termination})] has been studied in various contexts, some of which are non-biological, and is known as the totally asymmetric simple exclusion process (TASEP) \cite{MacDonald1968}. In the TASEP, there is no explicit tracking of mature RNA. Hence, to clearly distinguish our model from the TASEP and from other models that we consider later on, we refer to the model defined by Eqs. (\ref{TASEP1-initiation})-(\ref{TASEP1-degradation}) as the constitutive model with RNAP volume exclusion (vCM for short). We note that since our model describes movement on the coarse length scale of an RNAP, not at the single nucleotide level, it does not have an explicit description of processes occurring over few base pairs such as backtracking or pausing; to some extent, these processes can be captured by a suitable renormalization of the hopping rate (see Section \ref{summary-discussion} for further discussion).

In order to track RNAPs along the gene, we introduce a variable $\tau_i$ such that $\tau_i=0$ if the segment $i$ is empty, and $\tau_i=1$ if the segment $i$ is occupied by an RNAP. The number of RNAPs on the gene, denoted by $n$, is given by
\begin{equation}
    n=\sum_{i=1}^{L}\tau_i.
\end{equation}
We denote the number of mature RNA by $m$. We refer to $C=\{\tau_1,\dots,\tau_L\}$ as a configuration of RNAPs along the gene. The joint probability to find RNAPs in a configuration $C$ and $m$ copies of mature RNA at time $t$ is denoted by $P(C,m,t)$. We are interested in the steady state, in which case we drop the time dependence and consider only the joint probability $P(C,m)$. The marginal distribution of RNAPs along the gene is obtained by summing $P(C,m)$ over all $m$, $P(C)=\sum_{m=0}^{\infty}P(C,m)$. We define a local density of RNAPs at a segment $i$ as
\begin{equation}
    \rho_i=\langle \tau_i\rangle=\sum_{C}P(C)\delta_{\tau_i,1},
\end{equation}
where $\delta_{i,j}$ is the Kronecker delta. The transcription rate at which new mature RNA is produced is defined as the local density $\rho_L$, multiplied by the termination rate $\beta$, $k_{\text{syn}}=\rho_L\beta$. In the steady state, the transcription rate is equal to the current of RNAPs along the gene,
\begin{equation}
    k_{\text{syn}}=J=\alpha(1-\langle\tau_1\rangle)=\omega\langle\tau_i(1-\tau_{i+1})\rangle=\beta\rho_L.
\end{equation}
The probability distributions of the number of RNAPs on the gene and the number of mature RNA are defined as, respectively,
\begin{subequations}
\begin{align}
    \label{TASEP1-nascent-RNA-def}
    & P_N(n)=\sum_{C}P(C)\delta_{\sum_{i}\tau_i,n},\\
    \label{TASEP1-mature-RNA-def}
    & P_M(m)=\sum_{C}P(C,m).
\end{align}    
\end{subequations}
We note that the distribution of the number of RNAPs on the gene is the same as the distribution of the number of nascent RNA since to each gene-bound RNAP, a nascent RNA tail is attached. Analytical results are known for $P(C)$, $J$ and $\rho_{i}$ from the exact solution of the TASEP \cite{Derrida1993}, and are summarized in Subsection \ref{TASEP1-known-results}. Results for the distributions $P_{N}(n)$ and $P_{M}(m)$ are new, and are derived in Subsections \ref{TASEP1-nascent} and \ref{TASEP1-mature}, respectively. 

Before we present the results for $P_{N}(n)$ and $P_{M}(m)$, we consider a much simpler model in which elongation and termination processes are assumed to be deterministic, and in which there are no RNAP volume-exclusion effects. This model is exactly solvable and will serve as a useful benchmark for understanding the effect of RNAP collisions on the distributions of nascent and mature RNA numbers.

\subsection{The delay constitutive model}
\label{delay-constitutive-model}

This model approximates elongation and termination by a series of delay reactions (indicated by double arrows) that take a fixed amount of time to finish. The model is effectively defined by the reaction scheme
\begin{subequations}
\begin{align}
    & \text{free RNAP}\xrightarrow[]{\alpha}\text{RNAP}_1,\\
    & \text{RNAP}_i\xRightarrow[]{1/\omega}\text{RNAP}_{i+1} \ {\rm{for}} \quad i \in [1,L-1],\\
    & \text{RNAP}_L\xRightarrow[]{1/\beta}\text{free RNAP}+\text{RNA},\\
    & \text{RNA}\xrightarrow[]{d_M}\emptyset,
\end{align}
\end{subequations}
Here, $1/\omega$ is the fixed time it takes an RNAP to move across one segment, and $1/\beta$ is the fixed time it takes an RNAP to terminate from the last segment. As before, we assume that the number of free RNAPs is large so that it can be absorbed in the initiation rate $\alpha$. We refer to this model as the delay constitutive model.

The total time of elongation and termination is fixed and equal to,
\begin{equation}
    \label{delay1-elongation-time}
    T_{\text{el}}=\frac{L-1}{\omega}+\frac{1}{\beta}.
\end{equation}
Since elongation and termination are deterministic, the transcription rate $k_{\text{syn}}$ equals the rate of initiation $\alpha$,
\begin{equation}
    \label{delay1-J}
    k_{\text{syn}}=\alpha.
\end{equation}
The local density of RNAPs on the gene can be computed from that fact that the number of initiation events in a given time interval $t$ is a Poisson random variable with parameter $\alpha t$. The time an RNAP stays in the segment $i$ is equal to $1/\omega$ for $i=1,\dots,L-1$ and $1/\beta$ for $i=L$, hence the local density $\rho_i$ is equal to
\begin{equation}
    \label{delay1-rho-i}
    \rho_{i}=\begin{dcases}
    \frac{\alpha}{\omega} & i=1,\dots,L-1\\
    \frac{\alpha}{\beta}, & i=L.
    \end{dcases}
\end{equation}
Since the model ignores excluded volume interactions between RNAPs, the local density $\rho_{i}$ becomes larger than $1$ if $\alpha>\omega$ or $\alpha>\beta$, which is not physical. This model is therefore justified only for $\alpha<\omega$ and $\alpha<\beta$.

The number of RNAPs on the gene $n$ is equal to the number of initiations in the time interval $T_{\text{el}}$, hence
\begin{equation}
    \label{delay1-nascent}
    P_{N}(n)=\frac{(\alpha T_{\text{el}})^{n}}{n!}e^{-\alpha T_{\text{el}}}.
\end{equation}
The mean and the variance of this distribution are 
\begin{align}
    \label{delay1-nascent-mean}
    \mu_N=\alpha T_{\text{el}},\quad \sigma_{N}^{2}=\alpha T_{\text{el}},
\end{align}
and the Fano factor FF$_N$ (the variance divided by the mean) is equal to $1$. Combining Eqs. (\ref{delay1-J}) and (\ref{delay1-nascent-mean}) gives
\begin{equation}
    \mu_N=k_{\text{syn}} T_{\text{el}}.
\end{equation}
In queuing theory, this relationship is known as Little's law \cite{Little1961,Elgart2010-applications}, which states that the long-time average number of customers in a queue is equal to the arrival rate $k_{\text{syn}}$ multiplied by the long-time average of the time spent in the queue, $T_{\text{el}}$.

Due to the deterministic nature of elongation and termination, the mature RNA effectively follows a birth-death process with the birth rate equal to $\alpha$ and the death rate equal to $d_M$,  
\begin{equation}
    \emptyset\xrightarrow[]{\alpha}\text{RNA}\xrightarrow[]{d_M}\emptyset.
\end{equation}
The steady-state probability distribution $P_M(m)$ is the Poisson distribution with the parameter $\alpha/d_M$,
\begin{equation}
    \label{delay1-mature}
    P_{M}(m)=\frac{(\alpha/d_M)^m}{m!}e^{-\alpha/d_M}.
\end{equation}
The mean and the variance of this distribution are 
\begin{equation}
    \mu_M=\frac{\alpha}{d_M},\quad \sigma^{2}_{M}=\frac{\alpha}{d_M},
\end{equation}
and the Fano factor FF$_{M}$ is equal to $1$. 

\subsection{A summary of known results for the TASEP in the steady-state}
\label{TASEP1-known-results}

The steady-state probability $P(C)$ is known exactly and can be written in the following matrix-product form \cite{Derrida1993},
\begin{equation}
    \label{TASEP1-exact-solution}
    P(\tau_i,\dots,\tau_L)=\frac{1}{Z_L}\bigl \langle W\big|\prod_{i=1}^{L}\left[\tau_i D+(1-\tau_i)E\right]\big| V\bigr \rangle,
\end{equation}
where $D$ and $E$ are infinite-dimensional matrices that satisfy
\begin{equation}
    \label{TASEP1-D-E}
    DE=D+E,
\end{equation}
$\langle W\vert$ and $\vert V\rangle$ are infinite-dimensional vectors that satisfy
\begin{equation}
    \label{TASEP1-W-V}
    \langle W\vert E=\frac{\omega}{\alpha}\langle W\vert, \quad D\vert V\rangle=\frac{\omega}{\beta}\vert V\rangle,
\end{equation}
and $Z_L$ is the normalization,
\begin{equation}
    \label{TASEP1-Z-L}
    Z_L=\bigl \langle W\big|(D+E)^{L}\big|V \bigr \rangle.
\end{equation}
We note that the matrices $D$ and $E$ are not uniquely defined \cite{Derrida1993}; however the relations above are sufficient for all practical calculations.

An explicit formula for $Z_L$ in terms of $\alpha$, $\beta$ and $\omega$ is
\begin{equation}
    \label{TASEP1-Z-L-2}
    Z_L=\sum_{p=0}^{L}B_{L,p}\sum_{q=0}^{p}\left(\frac{\omega}{\alpha}\right)^q\left(\frac{\omega}{\beta}\right)^{p-q},
\end{equation}
where 
\begin{equation}
    \label{TASEP1-B-k-p}
    B_{k,p}=\begin{cases}
    \frac{p}{2k-p}\binom{2k-p}{k},& p=1,\dots,k,\\
    0 & \text{otherwise}.
    \end{cases}
\end{equation}
The steady-state RNAP current $J$, which equals the transcription rate $k_{\text{syn}}$, reads
\begin{equation}
    \label{TASEP1-J-exact}
    k_{\text{syn}}=J=\frac{\beta\langle W\vert (D+E)^{L-1}D\vert V\rangle}{Z_L}=\frac{\omega Z_{L-1}}{Z_L},
\end{equation}
and the local RNAP density $\rho_i$ is given by
\begin{align}
    \label{TASEP1-rho-i-exact}
    \rho_{i} &=\frac{1}{Z_L}\langle W\vert (D+E)^{i-1}D(D+E)^{L-i}\vert V\rangle\nonumber\\
    &=\sum_{p=1}^{L-i}B_{p,1}\frac{Z_{L-p}}{Z_L}+\frac{Z_{i-1}}{Z_L}\sum_{p=1}^{L-i}\frac{B_{L-i,p}}{\beta^{p+1}}.
\end{align}
In the limit $L\rightarrow\infty$, $J$ simplifies to
\begin{equation}
    \label{TASEP1-J-infinite}
    J=\begin{cases}
    \alpha\left(1-\frac{\alpha}{\omega}\right), & \alpha<\frac{\omega}{2},\;\beta>\alpha,\\
    \beta\left(1-\frac{\beta}{\omega}\right), & \beta<\frac{\omega}{2},\;\alpha>\beta,\\
    \frac{\omega}{4}, & \alpha,\beta>\frac{\omega}{2},
    \end{cases}
\end{equation}
Away from the boundaries, the local density $\rho_i$ is approximately constant and equal to
\begin{equation}
    \label{TASEP1-rho-infinite}
    \rho=\begin{dcases}
    \frac{\alpha}{\omega}, & \alpha<\frac{\omega}{2},\;\beta>\alpha,\\
    1-\frac{\beta}{\omega}, & \beta<\frac{\omega}{2},\;\alpha>\beta,\\
    \frac{1}{2}, & \alpha,\beta>\frac{\omega}{2}.
    \end{dcases}
\end{equation}
The exception is for $\alpha=\beta<\omega/2$, for which the local density increases linearly from $\alpha/\omega$ at the left boundary to $1-\beta/\omega$ at the right boundary.

Different regimes of the TASEP depending on the initiation and termination rates with respect to the hopping rate $\omega$ are summarized in Table \ref{table1}. In the low-density or initiation-limited (IL) regime ($\alpha<\omega/2$ and $\beta>\alpha$), $\rho=\alpha/\omega$ and the transcription rate is controlled by the initiation rate, $k_{\text{syn}}=\alpha(1-\alpha/\omega)$. In this regime, transcription is rate-limited by initiation, which is usually stated as the most likely scenario under physiological conditions. In the high-density or termination-limited (TL) regime ($\beta<\omega/2$ and $\alpha>\beta$), $\rho=1-\beta/\omega$, and the transcription rate is controlled by the termination rate, $k_{\text{syn}}=\beta(1-\beta/\omega)$. In this regime, there is a long queue of RNAPs spanning from the termination site towards the beginning of the gene. We are unaware of such scenario being observed \textit{in vivo}, though. On the coexistence (IL/TL) line between these two phases ($\alpha=\beta<\omega/2$), the local density increases linearly along the gene and the transcription rate is equal to $k_{\text{syn}}=\alpha(1-\alpha/\omega)$. This is a very specific regime that occurs only when the initiation rate is equal to the termination rate. Finally, in the maximum-current or elongation-limited (EL) regime ($\alpha>\omega/2$ and $\beta>\omega/2$), $\rho=1/2$, and the transcription rate depends only on the hopping rate $\omega$, $k_{\text{syn}}=\omega/4$. In this regime, transcription dynamics is fully controlled by the elongation rate $\omega$. Any further increase in the initiation and termination rates has no effect on the dynamics. 

In comparison to these results, the delay constitutive model predicts $\rho=\alpha/\omega$ and $k_{\text{syn}}=\alpha$, which makes sense only if $\alpha<\omega$.

\begin{table*}[htb]
\caption{\label{table1} Different regimes of the steady-state TASEP.}
\begin{ruledtabular}
\begin{tabular}{lccc}
Regime & Parameter range &  Local RNAP density $\rho_i$ & Transcription rate $J$\\
\hline
initiation-limited regime (IL) & $\alpha<\omega/2$ and $\beta>\alpha$ & $\alpha/\omega$ & $\alpha(1-\alpha/\omega)$\\
termination-limited regime (TL) & $\beta<\omega/2$ and $\alpha>\beta$ & $1-\beta/\omega$ & $\beta(1-\beta/\omega)$\\
coexistence line (IL/TL) & $\alpha=\beta<\omega/2$ & linearly increasing & $\alpha(1-\alpha/\omega)$\\
elongation-limited regime (EL) & $\alpha,\beta>\omega/2$ & $1/2$ & $\omega/4$
\end{tabular}
\end{ruledtabular}
\end{table*}

\subsection{Probability distribution of the number of nascent RNA}
\label{TASEP1-nascent}

Our strategy is to compute the probability generating function $G_{N}$ defined as
\begin{equation}
    G_{N}(z)=\sum_{n=0}^{L}P_{N}(n)z^n.
 \end{equation}
Using Eqs. (\ref{TASEP1-nascent-RNA-def}) and (\ref{TASEP1-exact-solution}), it is straightforward to show that
 \begin{equation}
    \label{TASEP1-nascent-RNA-GN}
    G_{N}(z)=\frac{1}{Z_L}\bigl \langle W\big|(zD+E)^{L}\big|V \bigr \rangle.
\end{equation}
The calculation of $G_N(z)$ is presented in Appendix \ref{appendix_a}, and the final result is 
\begin{equation}
    \label{TASEP1-nascent-RNA-GN-exact}
    G_{N}(z)=\frac{1}{Z_L}\sum_{k=1}^{L}\frac{k}{L}u_{L-k,L}(z)v_k(z),
\end{equation}
where $u_{p,n}(z)$ and $v_k(z)$ are given by
\begin{subequations}
\begin{align}
    & u_{p,n}(z)=\sum_{m=1}^{n}\binom{n}{m}\binom{p+m-1}{p}(z-1)^{n-m},\\
    & v_0(z)=1,\\
    & v_k(z)=\sum_{i=0}^{k}\left(\frac{\alpha z-\alpha+1}{\alpha z}\right)^i\frac{1}{\beta^{k-i}}\nonumber\\
    &\;-\frac{z-1}{z}\sum_{i=0}^{k-1}\left(\frac{\alpha z-\alpha+1}{\alpha z}\right)^i\frac{1}{\beta^{k-1-i}},\; k\geq 1.
\end{align}
\end{subequations}
From here, we get $P_{N}(n)$ by expanding $G_{N}(z)$ in $z$ and collecting the terms containing $z^n$,
\begin{equation}
    \label{TASEP1-nascent-RNA-exact}
    P_{N}(n)=\frac{1}{n!}\left.\frac{d^n}{dz^n}G_{N}(z)\right\vert_{z=0}.
\end{equation}
The mean $\mu_N$ is equal to the spatial average of the local density $\rho_{i}$ given by Eq. (\ref{TASEP1-rho-i-exact}),
\begin{equation}
    \label{TASEP1-nascent-RNA-mean}
    \mu_N=\frac{1}{L}\sum_{i=1}^{L}\rho_{i},
\end{equation}
and the variance $\sigma_{N}^{2}$ can be computed from
\begin{equation}
    \label{TASEP1-nascent-RNA-variance}
    \sigma_{N}^{2}=\left.\frac{d^2 G_{N}(z)}{dz^2}\right\vert_{z=1}+\mu_N(1-\mu_N).
\end{equation}
The Fano factor of the nascent RNA number is given by 
\begin{equation}
    \label{TASEP1-nascent-RNA-FF-exact}
    \text{FF}_N=1-\mu_N+
    \frac{1}{\mu_N}\left.\frac{d^2 G_{N}(z)}{dz^2}\right\vert_{z=1}
\end{equation}
We were not able to find a simple expression for the variance and the Fano factor FF$_N$ for arbitrary $\alpha/\omega$ and $\beta/\omega$. For the special case $\alpha/\omega=\beta/\omega=1$, $\sigma_{N}^{2}=L(L+2)/(8L+4)$ was computed previously in Ref. \cite{Derrida93-correlations}. In the limit in which $L\rightarrow\infty$, this result yields FF$_N=1/4$. We have checked numerically for $L=100$ that FF$_N\approx 1/4$ for other values of $\alpha$ and $\beta$ in the elongation-limited regime.

While the distribution $P_{N}(n)$ does not seem to be related to any known distribution, we know from the exact solution of the TASEP that for $\alpha+\beta=\omega$, $P(\tau_1,\dots,\tau_L)$ simplifies to a product of Bernoulli distributions \cite{Derrida1992},
\begin{equation}
    P(\tau_1,\dots,\tau_L)=\prod_{i=1}^{L}[\tau_i\rho+(1-\tau_i)(1-\rho)],
\end{equation}
where $\rho=\alpha/\omega=1-\beta/\omega$. For this distribution, the local density $\rho_{i}$ is equal to $\rho$ at any segment, i.e. the density of RNAP is uniform along the gene body. By inserting this distribution into Eq. (\ref{TASEP1-nascent-RNA-def}), we get a binomial distribution
\begin{equation}
    \label{TASEP1-nascent-RNA-binomial}
    P_{N}(n)=\binom{L}{n}\rho^{n}(1-\rho)^{L-n}.
\end{equation}
The mean and the variance of this binomial distribution are $L\rho$ and $L\rho(1-\rho)$, respectively, and the Fano factor is given by 
\begin{equation}
    \label{TASEP1-nascent-RNA-FF-simple}
    \text{FF}_N=1-\rho.  
\end{equation}
This Fano factor is always less than $1$, and becomes vanishingly small in the limit $\rho\rightarrow 1$. The skewness of the binomial distribution in Eq. (\ref{TASEP1-nascent-RNA-binomial}) is given by $(1-2\rho)/\sqrt{\rho(1-\rho)L}$, hence the distribution is right-skewed for $\rho<1/2$, symmetrical for $\rho=1/2$ and left-skewed for $\rho>1/2$. In the limit in which $L\rightarrow\infty$ and $\rho\rightarrow 0$ such that $\lambda\equiv L\rho$ is fixed, the binomial distribution becomes the Poisson distribution with the rate parameter $\lambda$. This limit corresponds to the delay constitutive model.

\begin{figure*}[htb]
    \centering
    \includegraphics[width=\textwidth]{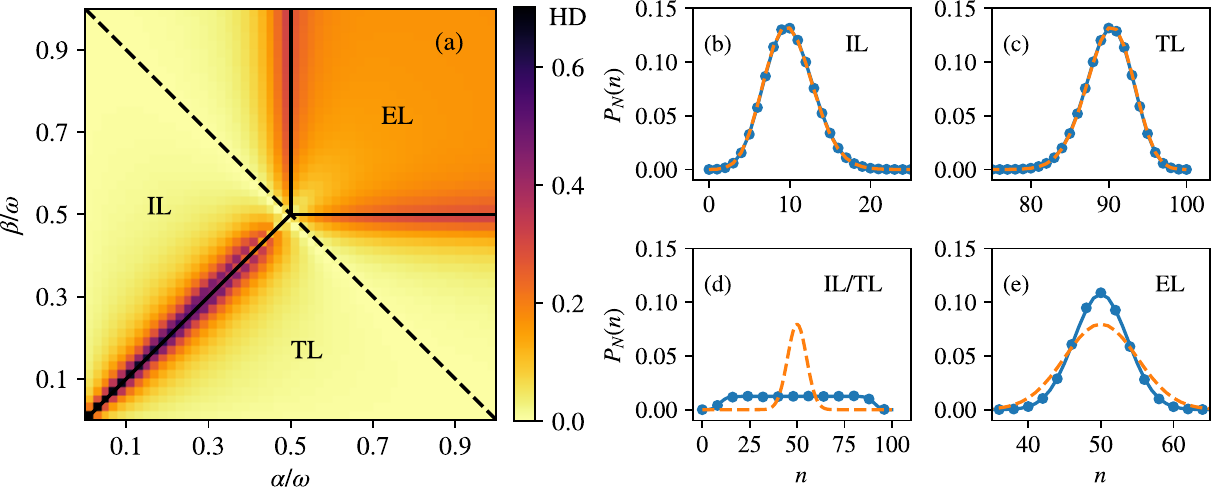}
    \caption{Accuracy of the binomial approximation for the steady-state nascent RNA number distribution in the constitutive model with RNAP volume exclusion. (a) Heat map of the Hellinger distance (HD) between the exact probability distribution $P_N(n)$ in Eq. (\ref{TASEP1-nascent-RNA-exact}) and the binomial distribution in Eq. (\ref{TASEP1-nascent-RNA-binomial}). Solid black lines are regime boundaries and the dashed black line is $\alpha+\beta=\omega$, in which case the Hellinger distance is exactly zero. (b)-(e) compare the nascent RNA number distribution obtained using stochastic simulations (blue points), the exact distribution given by Eq. (\ref{TASEP1-nascent-RNA-exact}) (solid blue lines) and the binomial distribution approximation given by Eq. (\ref{TASEP1-nascent-RNA-binomial}) (dashed orange lines). The number of segments is $L=100$. (b) initiation-limited regime (IL, $\alpha/\omega=0.1$, $\beta/\omega=0.7$). (c) termination-limited regime (TL, $\alpha/\omega=0.7$, $\beta/\omega=0.1$). (d) Coexistence line (IL/TL, $\alpha/\omega=\beta/\omega=0.1$). (e) elongation-limited regime (EL, $\alpha/\omega=0.7$, $\beta/\omega=0.7$).}
    \label{fig2}
\end{figure*}

To check how well the binomial distribution [Eq. (\ref{TASEP1-nascent-RNA-binomial}) with $\rho$ given by Eq. (\ref{TASEP1-rho-infinite})] approximates the exact one [Eq. (\ref{TASEP1-nascent-RNA-exact})] across the parameter space, i.e. without imposing the condition $\alpha+\beta=\omega$, we computed the Hellinger distance (which varies between 0 and 1) between these two distributions for 2500 combinations of $\alpha/\omega$ and $\beta/\omega$, equally spaced between $0$ and $1$ [Fig. \ref{fig2}(a)]. As expected, the Hellinger distance is small in the initiation-limited and termination-limited regimes away from the regime boundaries [Figs. \ref{fig2}(b) and \ref{fig2}(c)], and is exactly zero on the special line $\alpha+\beta=\omega$. The largest difference between the distributions is observed at the coexistence line between the initiation-limited and termination-limited regimes ($\alpha=\beta<\omega/2$) [Fig. \ref{fig2}(d)], where the local density has a linear profile \cite{Derrida1993}. Significant differences are also observed at the boundaries between the initiation-limited and the elongation-limited regimes, between the termination-limited and the elongation-limited regimes, and in the entire elongation-limited regime [Fig. \ref{fig2}(e)]. These discrepancies are due to the algebraic decay of the local density towards its value of $1/2$ in the middle of the gene \cite{Derrida1993}. In the elongation-limited regime, the Fano factor is approximately equal to $1/4$, whereas the binomial distribution predicts the value of $1/2$. Simulations in Fig. \ref{fig2} also confirm the theoretical transition from the left-skewed to the right-skewed nascent RNA distributions as $\rho$ crosses the threshold of $1/2$ [Fig. \ref{fig2}(b) and \ref{fig2}(c)]---similar transitions have been reported in studies of crowding-induced phenomena in other chemical reaction systems \cite{Cianci2016}. The simulations in Fig. \ref{fig2} and in the rest of the paper were performed using the Gillespie algorithm \cite{gillespie1977exact}.

From Eq. (\ref{delay1-nascent-mean}), it follows that the mean number of nascent RNA predicted by the delay constitutive model is $\alpha T_{\text{el}}\approx L\alpha/\omega$. Comparing this result to that from the constitutive model with RNAP exclusion, $\rho L$, we conclude that the two match only in the initiation-limited regime for small $\alpha$, such that $J\approx\alpha$ and $\rho=\alpha/\omega$. It is, however, possible to extend the constitutive delay model such that its mean nascent RNA number matches that of the constitutive model with RNAP beyond this regime. This can be achieved by replacing the initiation rate $\alpha$ in the delay model with an effective initiation rate given by  $\alpha_{\text{eff}}=J$, where $J$ is the transcription rate in the constitutive model with RNAP volume exclusion [Eq. (\ref{TASEP1-J-exact})]. Similarly, we define an effective hopping rate at segment $i$ as $\omega_{\text{eff},i}=J/\rho_i$, where $\rho_i$ is the local RNAP density in the constitutive model with RNAP volume exclusion [Eq. (\ref{TASEP1-rho-i-exact})]. Finally, to satisfy the Little's law, we define the effective elongation and termination time $T_{\text{el,eff}}=\sum_{i=1}^{L}1/\omega_{\text{eff},i}$. We refer to the constitutive delay model with these parameters as the effective delay constitutive model (edCM for short). Assuming $\rho_i\approx\rho$, which is true for large $L$ and outside the coexistence line where $\rho_i$ is linearly increasing, we get $\omega_{\text{eff},i}\approx\omega(1-\rho)$ and $T_{\text{el,eff}}\approx L/\omega_{\text{eff}}$. We note that the effective hopping rate $\omega_{\text{eff}}$ is the ``true" rate of nascent RNA elongation taking into account slowing down of RNAP due to other RNAPs on the gene, whereas $\omega$ is the ``bare" hopping rate in the absence of other RNAPs.

\subsection{Probability distribution of the number of mature RNA}
\label{TASEP1-mature}

We cannot compute $P_{M}(m)$ directly from Eq. (\ref{TASEP1-mature-RNA-def}), because the joint distribution $P(C,m)$ is unknown. Instead, we find an approximate expression for $P_{M}(m)$ by replacing the process of RNA production and degradation with the following queuing process,
\begin{equation}
    \label{TASEP1-queuing-process}
    \emptyset\xrightarrow[]{f_{\text{ter}}(t)}M\xrightarrow[]{d_M}\emptyset.
\end{equation}
Here, $f_{\text{ter}}(t)$ is the probability density function (pdf) of the waiting time between two successive termination events, and $d_M$ is the degradation rate of mature RNA. This process is known as a $G/M/\infty$ queue in Kendall's notation \cite{Kendall1953}, where $G$ stands for general inter-arrival distribution [mature RNAs arrive at time intervals distributed according to $f_{\text{ter}}(t)$], $M$ stands for Markovian service process (service times are exponentially distributed with rate $d_M$), and the number of servers is infinite (the RNA degradation machinery is assumed to be abundant). 

We note that the queuing process described by Eq. (\ref{TASEP1-queuing-process}) is not an exact representation of the original model. In the queuing process, the waiting times between successive termination events are mutually independent, which is not true in the original model. Therefore, we will refer to Eq. (\ref{TASEP1-queuing-process}) as the \emph{renewal approximation} of the original process, because the production of mature RNA in this approximation constitutes a renewal process (a generalization of the Poisson process to an arbitrary inter-arrival time distribution). 

The advantage of this approximation is that the steady-state distribution of the number of customers (the number of mature RNA in our case) in a $G/M/\infty$ queue can be computed analytically for any $f(t)$ whose mean inter-arrival time is finite \cite{Takacs1958}. Here we write the final result for $P_{M}(m)$ and refer the reader to the original paper for the derivation. We denote by $\smash{\tilde{f}}(s)$ the Laplace transform of $f(t)$, and by $\mu$ the mean inter-arrival time,
\begin{equation}
    \tilde{f}(s)=\int_{0}^{\infty}dt\;f(t)e^{-st},\quad \mu=\int_{0}^{\infty}dt\;tf(t)<\infty.
\end{equation}
Next, we define a coefficient $C_i$ as
\begin{equation}
    \label{C-i}
    C_0=1,\quad C_i=\prod_{j=1}^{i}\frac{\tilde{f}(jd_M)}{1-\tilde{f}(j d_M)},\quad i=1,2,3\dots.
\end{equation}
The steady-state distribution of the mature RNA number $m$ is then given by
\begin{subequations}
\label{P-M-queuing}	
\begin{align}
	& P_{M}(0)=1-\sum_{k=1}^{\infty}(-1)^{k-1}\frac{C_{k-1}}{\mu d_M k},\\
	& P_{M}(m)=\sum_{k=m}^{\infty}(-1)^{k-m}\binom{k}{m}\frac{C_{k-1}}{\mu d_M k},\;  m\geq 1.
\end{align}
\end{subequations}
Hence, to compute $P_M(m)$ we need to compute the pdf $f_{\text{ter}}(t)$ of the waiting time between two successive termination events. 

To this end, we denote by $\mathcal{G}$ the set of all configurations $C=\{\tau_1,\dots,\tau_L\}$ in which the last segment is occupied by an RNAP, $\mathcal{G}=\{C\;\vert\;\tau_L(C)=1\}$. For a configuration $C\in \mathcal{G}$, we define the gap size $g(C)$ as the number of consecutive empty segments in front of the last segment. For example, $g(C)=0$ if the closest RNAP trailing behind is at the segment $L-1$, and $g(C)=L-1$ if there are no trailing RNAPs. Next, we denote by $P_{\text{gap}}(k,L)$ the probability that a configuration $C\in \mathcal{G}$ has a gap of size $k$,
\begin{equation}
    P_{\text{gap}}(k;L)=\frac{\sum_{C\in A}P(C)\delta_{g(C),k}}{\sum_{C\in A}P(C)},
\end{equation}
where $\delta_{i,j}$ is the Kronecker delta. Note that the denominator is equal to the local density $\rho_L$.

Let us now assume that an RNAP terminated transcription at time $t=0$, and that the gap size right before that was equal to $k$. We denote by $r_k(t)$ the probability density function (pdf) of the waiting time $t$ until the next termination event. The pdf $f_{\text{ter}}(t)$ is then equal to $P_{\text{gap}}(k;L)$ multiplied by $r_k(t)$ and summed over all $k=0,\dots,L-1$,
\begin{equation}
    \label{TASEP1-wt-def}
    f_{\text{ter}}(t)=\sum_{k=0}^{L-1}P_{\text{gap}}(k;L)r_k(t).
\end{equation}
For $k<L-1$, the time between two successive termination events is equal to the time it takes an RNAP at the segment $L-k-1$ to move $k+1$ segments and terminate from the last segment. Hence, the pdf $r_k(t)$ is a convolution of the Erlang distribution with shape $k+1$ and rate $\omega$, and the exponential distribution with rate $\beta$, which gives
\begin{equation}
    \label{TASEP1-r-k-uniform}
    r_k(t)=\beta e^{-\beta t}\left(\frac{\omega}{\omega-\beta}\right)^{k+1}\frac{\gamma(k+1,(\omega-\beta)t)}{\Gamma(k+1)},
\end{equation}
where $\gamma(n,x)$ is a lower incomplete Gamma function. For $k=L-1$, the time between two successive termination events is equal to the time it takes a free RNAP to initiate transcription, move $L-1$ segments, and terminate from the last segment. Hence, the pdf $r_{L-1}(t)$ is given by a convolution of $r_{L-2}(t)$ and the exponential distribution with the rate parameter $\alpha$, which gives
\begin{align}
    & r_{L-1}(t)=\frac{\alpha\beta}{\alpha-\beta}e^{-\beta t}\left(\frac{\omega}{\omega-\beta}\right)^{L-1}\frac{\gamma(L-1,(\omega-\beta)t)}{\Gamma(L-1)}\nonumber\\
    &\quad-\frac{\alpha\beta}{\alpha-\beta}e^{-\alpha t}\left(\frac{\omega}{\omega-\alpha}\right)^{L-1}\frac{\gamma(L-1,(\omega-\alpha)t)}{\Gamma(L-1)}.
\end{align}
The final expression for $f_{\text{ter}}(t)$ is complicated, so we consider the asymptotic limit of $L\rightarrow\infty$, in which $f_{\text{ter}}(t)$ becomes
\begin{equation}
    \label{TASEP1-wt-infinite-def}
    f_{\text{ter,as}}(t)=\sum_{k=0}^{\infty}P_{\text{gap}}(k)r_k(t),
\end{equation}
where $P_{\text{gap}}(k)=\lim_{L\rightarrow\infty}P_{\text{gap}}(k;L)$. The gap size distribution $P_{\text{gap}}(k,L)$ and the limiting distribution $P_{\text{gap}}(k)$ were previously derived in Ref. \cite{Krbalek2011}. The limiting distribution $P_{\text{gap}}(k)$ reads
\begin{subequations}
\label{TASEP1-Pgap-infinite}
\begin{align}
    P_{\text{gap}}(k)&=\frac{\alpha}{\beta}\left(1-\frac{\alpha}{\omega}\right)^{k+1}\frac{\beta-\alpha}{\omega-2\alpha}+\frac{\omega-\alpha}{\beta}\left(\frac{\alpha}{\omega}\right)^{k+1}\nonumber\\
    &\times\left(1-\frac{\beta-\alpha}{\omega-2\alpha}\right),\; \alpha<\frac{\omega}{2},\;\beta>\alpha,\\
    P_{\text{gap}}(k)&=\left(1-\frac{\beta}{\omega}\right)\left(\frac{\beta}{\omega}\right)^{k},\; \beta<\frac{\omega}{2},\;\alpha\geq \beta,\\
    P_{\text{gap}}(k)&=\frac{\omega}{\beta 2^{k+1}}\left(\frac{1-k}{2}+\frac{\beta k}{\omega}\right),\; \alpha,\beta>\frac{\omega}{2}.
\end{align}
\end{subequations}
This limiting distribution is valid as long as the typical gap size is much less than $L$ and $L$ is large, in which case extending $L$ to infinity does not have much impact on the gap size distribution. If we now insert Eqs. (\ref{TASEP1-r-k-uniform}) and (\ref{TASEP1-Pgap-infinite}) into Eq. (\ref{TASEP1-wt-infinite-def}), we get a remarkably simple result for $f_{\text{ter,as}}(t)$,
\begin{equation}
    \label{TASEP1-wt-infinite}
    f_{\text{ter,as}}(t)=\frac{\omega\nu(1-\nu)}{1-2\nu}\left(e^{-\omega\nu t}-e^{-\omega(1-\nu)t}\right),
\end{equation}
where $\nu$ is given by
\begin{equation}
    \label{TASEP1-nu}
    \nu=\begin{dcases}
    \frac{\alpha}{\omega}, & \alpha<\frac{\omega}{2},\;\beta>\alpha,\\
    1-\frac{\beta}{\omega}, & \beta<\frac{\omega}{2},\;\alpha\geq\beta,\\
    \frac{1}{2}, & \alpha,\beta>\frac{\omega}{2}.
    \end{dcases}
\end{equation}
There is a subtle difference between $\nu$ in Eq. (\ref{TASEP1-nu}) and $\rho$ in Eq. (\ref{TASEP1-rho-infinite}): $\nu=\rho$ everywhere except at the coexistence line $\alpha=\beta<1/2$ for which $\nu=1-\beta/\omega$, whereas $\rho$ is not properly defined on this line (the local density increases linearly from $\alpha/\omega$ at the left boundary to $1-\alpha/\beta$ at the right boundary). The expression similar to the one in Eq. (\ref{TASEP1-wt-infinite}) was previously derived for the discrete-time TASEP with parallel \cite{Ghosh1998,Tripathi2008-telegraph} and random-sequential \cite{Hrabak2020} hopping.

\begin{figure*}[ht]
	\centering
	\includegraphics[width=\textwidth]{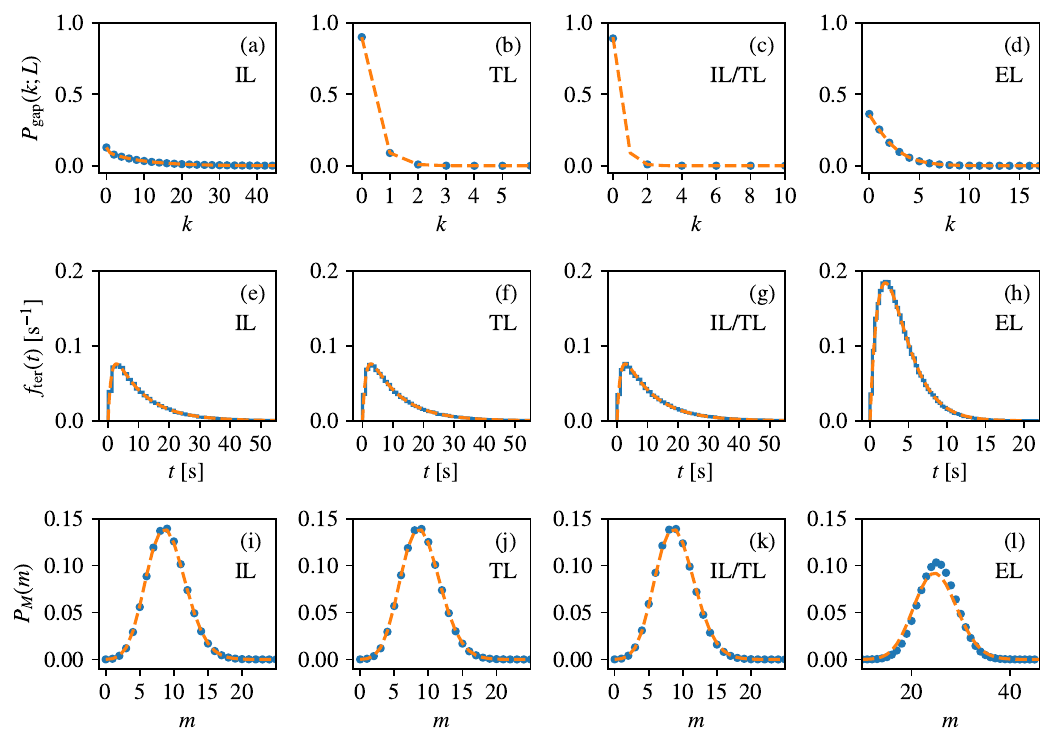}
	\caption{Accuracy of the theoretical results for the gap size distribution, the waiting time distribution between successive termination events and the mature RNA number distribution in the constitutive model with RNAP volume exclusion. (a)-(d) compare the gap size distribution $P_{\text{gap}}(k;L)$ computed using stochastic simulations (blue points) and $P_{\text{gap}}(k)$ computed from the asymptotic theory in Eq. (\ref{TASEP1-Pgap-infinite}) (dashed orange line). (e)-(h) compare the pdf $f_{\text{ter}}(t)$ of the waiting time between two successive termination events computed using stochastic simulations (blue solid line) and $f_{\text{ter,as}}(t)$ computed from the asymptotic theory in Eq. (\ref{TASEP1-wt-infinite}) (dashed orange line). (i)-(l) compare the mature RNA number distribution $P_M(m)$ computed using stochastic simulations (blue points) and the one computed from the renewal approximation in Eq. (\ref{TASEP1-mature-RNA-renewal}) (dashed orange line). Note that the renewal approximation theory is also based on the asymptotic theory for the waiting time between two successive termination events. The parameters are: $\alpha=0.1$ s$^{-1}$, $\beta=0.7$ s$^{-1}$, $\omega=1.0$ s$^{-1}$ and $d_M=0.01$ s$^{-1}$ for the initiation-limited regime (IL, first column), $\alpha=0.7$ s$^{-1}$, $\beta=0.1$ s$^{-1}$, $\omega=1.0$ s$^{-1}$ and $d_M=0.01$ s$^{-1}$ for the termination-limited regime (TL, second column), $\alpha=0.1$ s$^{-1}$, $\beta=0.1$ s$^{-1}$, $\omega=1.0$ s$^{-1}$ and $d_M=0.01$ s$^{-1}$ for the coexistence line (IL/TL, third column) and $\alpha=0.7$ s$^{-1}$, $\beta=0.7$ s$^{-1}$, $\omega=1.0$ s$^{-1}$ and $d_M=0.01$ s$^{-1}$ for the elongation-limited regime (EL, fourth column). The system size is $L=100$.}
	\label{fig3}
\end{figure*}

Eq. (\ref{TASEP1-wt-infinite}) describes a hypoexponential pdf of the sum of two exponentially distributed random variables with rates $\omega\nu$ and $\omega(1-\nu)$. As expected, the mean waiting time $\mu_{\text{ter}}$ is equal to $1/J$, where $J$ is the transcription rate given by Eq. (\ref{TASEP1-J-infinite}). Note that $f_{\text{ter,as}}(t)$ is invariant to the exchange of $\nu\leftrightarrow 1-\nu$. In the elongation-limited regime, $f_{\text{ter,as}}(t)$ becomes the gamma distribution with the shape parameter $2$ and the rate parameter $\omega/2$.

We are now ready to compute the probability distribution of the number of mature RNA in the renewal approximation of our model. The Laplace transform of $f_{\text{ter,as}}(t)$ in Eq. (\ref{TASEP1-wt-infinite}) is given by
\begin{equation}
    \tilde{f}_{\text{ter,as}}(s)=\frac{\omega^2\nu(1-\nu)}{(s+\omega\nu)[s+\omega(1-\nu)]}.
\end{equation}
Inserting this expression into Eq. (\ref{P-M-queuing}) yields the following expression for the mature RNA distribution in the constitutive model,
\begin{equation}
	\label{TASEP1-mature-RNA-renewal}
    P_{M}(m)=\frac{(ab)^m}{m!(a+b)_m}{}_0F_{1}(a+b+m,-ab),
\end{equation}
where $(x)_n=\Gamma(x+n)/\Gamma(x)$ is the Pochhammer symbol, ${}_0 F_{1}$ is the confluent hypergeometric limit function, and $a$ and $b$ are given by
\begin{equation}
    a=\frac{\omega\nu}{d_M},\quad b=\frac{\omega(1-\nu)}{d_M}.
\end{equation}
The waiting time distribution matching procedure that we used to obtain an expression for the approximate mature RNA distribution is similar in principle to the model reduction technique described in \cite{braichenko2021distinguishing}. The probability generating function for $P_{M}(m)$ reads
\begin{align}
    G_{M}(z)&=\sum_{m=0}^{\infty}P_{M}(m)z^m={}_0F_{1}(a+b,ab(z-1)).
\end{align}
The mean and the variance of the RNA number $m$ are given by
\begin{equation}
    \mu_M=\frac{J}{d_M},\quad \sigma_{M}^{2}=\mu_M\left(1-\frac{J}{\omega+d_M}\right).
\end{equation}
The Fano factor of the mature RNA number in the renewal approximation is equal to
\begin{equation}
    \text{FF}_{M}=1-\frac{J}{\omega+d_M}=1-\frac{\omega\rho(1-\rho)}{\omega+d_M}\leq 1.
    \label{TASEP1-mature-RNA-FF}
\end{equation}
The Fano factor is always less than 1, which means that the RNA number distribution is sub-Poissonian. When $\omega$ is much larger than $d_M$, we get a simple expression for FF$_{M}$ that depends only on $\rho$, $\text{FF}_{M}=1-\rho(1-\rho)$. We note that FF$_{M}=1$ in the (effective) delay constitutive model. Hence, the deviation of the mature RNA distribution from the Poisson distribution is directly related to the level of RNAP traffic on the gene, as measured by the RNAP density $\rho$. By comparison of Eqs. (\ref{TASEP1-nascent-RNA-FF-simple}) and (\ref{TASEP1-mature-RNA-FF}), it is also clear that the Fano factor of the nascent RNA number is always less than that of the mature RNA number distribution. This indicates that RNAP volume exclusion effects become less apparent for RNA involved in processes downstream of transcription.

\begin{figure}[hbt]
	\centering
	\includegraphics[width=8.6cm]{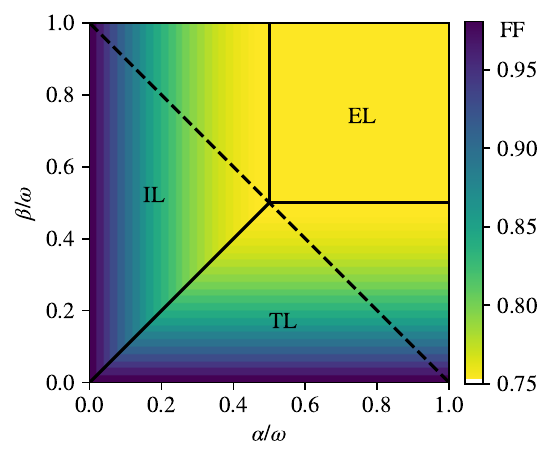}
	\caption{Heat map of the Fano factor (FF) of the mature RNA distribution for the constitutive model with RNAP volume exclusion, computed from the renewal approximation given by Eq. (\ref{TASEP1-mature-RNA-FF}). The Fano factor is bounded between $3/4$ and $1$. The model parameters are: $L=100$, $\omega=1$ s$^{-1}$ and $d_M=0.01$ s$^{-1}$ ($\omega \gg d_M$ is common for many genes).}
	\label{fig4}
\end{figure}

In Fig. \ref{fig3}, we compare the predictions of our asymptotic theory for $P_{\text{gap}}(k)$, $f_{\text{ter,as}}(t)$ and $P_M(m)$ with the results of stochastic simulations. The results were obtained for four sets of parameters $\alpha$, $\beta$ and $\omega$ representing the initiation-limited regime (IL), the termination-limited regime (TL), the coexistence line (IL/TL) and the elongation-limited regime (EL), respectively. We find an excellent agreement between $P_{\text{gap}}(k,L)$ obtained using stochastic simulations and $P_{\text{gap}}(k)$ predicted by Eq. (\ref{TASEP1-Pgap-infinite}) (the top row). Consequently, the pdf $f_{\text{ter,as}}(t)$ computed from Eq. (\ref{TASEP1-wt-infinite}) and the one obtained using stochastic simulations are practically indistinguishable (the middle row). The analytical and simulated results for the mature RNA distribution $P_M(m)$ agree in all the regimes, however a small but noticeable disagreement is observed in the elongation-limited regime (the bottom row). Since $f_{\text{ter}}(t)$ is well approximated by $f_{\text{ter,as}}(t)$ in the elongation-limited regime [Fig. \ref{fig3}(h)], we conclude that this disagreement must have originated from the renewal approximation (the assumption that the waiting times between successive termination events are uncorrelated). 

We emphasize that the good agreement of theory and simulations for $L = 100$ implies that the asymptotic theory for the waiting time distributions between successive termination events and the renewal theory for mature RNA distributions provide accurate results for genes of length larger than approximately $3500$ bp (since each segment is the length of an RNAP footprint). This value is much smaller than the typical gene length in humans (the median value is $26.4$ kb for protein-coding and  $11.2$ kb for non-coding genes, respectively \cite{Piovesan2016}), but larger than the typical gene length in \textit{S. cerevisiae} (the average gene length is $1.4$ kb). For such short genes, the asymptotic result in Eq. (\ref{TASEP1-wt-infinite}) may not be applicable if the mean gap size is of the order of the system size $L$, which occurs if the initiation rate is sufficiently small. However, in that case the RNAP volume exclusion can be ignored, and the results of the delay constitutive model can be used instead.

In Fig. \ref{fig4}, we show the Fano factor FF$_{M}$ computed from Eq. (\ref{TASEP1-mature-RNA-FF}) across the whole phase diagram of the constitutive model. The smallest value of $3/4$ is achieved in the elongation-limited regime in which $\rho=1/2$, whereas the largest value $1$ is achieved in the limit of small transcription rate $J$, which is either when the initiation rate $\alpha$ or the termination rate $\beta$ are much smaller than the hopping rate $\omega$.

\section{The telegraph model with RNAP volume exclusion}
\label{telegraph-model}

\begin{figure}[hbt]
	\centering
	\includegraphics[width=8.6cm]{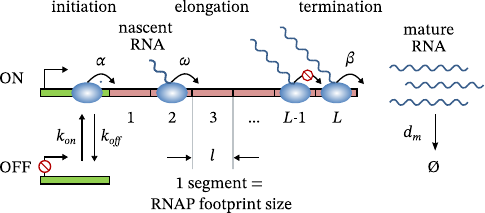}
	\caption{Cartoon illustrating the telegraph model of transcription with RNAP volume exclusion. The gene is divided into $L$ segments, whereby one segment equals in size to one RNAP footprint size ($\approx 35$ bp). The gene switches between two states of activity and inactivity with rates $k_{\text{on}}$ and $k_{\text{off}}$. Transcription initiation occurs from the active state at rate $\alpha$, provided the first segment is empty. RNAPs move along the gene at rate $\omega$ segments per unit time, provided the segment in front is empty. The rates of termination and mature RNA degradation are $\beta$ and $d_M$, respectively.}
	\label{fig5}
\end{figure}

Next, we consider an extension of the constitutive model of gene expression that allows for promoter switching (transitions between two states of activity and inactivity). A cartoon illustrating the new model is shown in Fig. \ref{fig5}. We denote by $k_{\text{on}}$ the rate at which the gene switches to the active state, and by $k_{\text{off}}$ the rate at which the gene switches to the inactive state. Initiation occurs at the rate $\alpha$ if the gene is in the active state, and the first segment is empty. As in the original telegraph model of gene expression \cite{Peccoud1995}, the gene remains in the active state immediately after the initiation. The elongation, termination, and RNA degradation proceed as in the constitutive model. The model can be summarized by the following reactions,
\begin{subequations}
\label{TASEP2-reactions}
\begin{align}
    & G_{\text{off}}\xrightleftharpoons[k_{\text{off}}]{k_{\text{on}}}G_{\text{on}}\\
    & G_{\text{on}}+\text{free RNAP}+\emptyset_1\xrightarrow[]{\alpha}G_{\text{on}}+\text{RNAP}_1\label{TASEP2-reactions-initiation}\\
    & \text{RNAP}_i+\emptyset_{i+1}\xrightarrow[]{\omega}\emptyset_{i}+\text{RNAP}_{i+1},\;i \in [1,L-1]\\
    & \text{RNAP}_L\xrightarrow[]{\beta}\emptyset_{L}+\text{free RNAP}+\text{RNA},\\
    & \text{RNA}\xrightarrow[]{d_M}\emptyset,
\end{align}
\end{subequations}
Similarly as in the constitutive model, we can approximate the
second-order reaction in Eq. (\ref{TASEP2-reactions-initiation}) by a quasi first-order reaction $G_{\text{on}}\rightarrow G_{\text{on}}+\text{RNAP}_1$ where the rate $\alpha$ is proportional
to the number of free RNAPs. Henceforth, we refer to this model as the telegraph model with RNAP volume exclusion.

We introduce a random variable $\sigma$ describing promoter activity, whereby $\sigma=0$ when the promoter is inactive and $\sigma=1$ when the promoter is active. As before, $C$ denotes the configuration of RNAPs on the gene, $n$ the number of RNAPs that are actively engaged in transcription, and $m$ the number of mature RNAs. The probability to find the gene in state $\sigma$ and configuration $C$, along with $m$ copies of mature RNA, is denoted by $P(\sigma,C,m)$. We are interested in computing the distributions $P_{N}(n)$ and $P_M(m)$ defined by
\begin{subequations}
\begin{align}
    & P_{N}(n)=\sum_{C}\sum_{\sigma=0,1}P(\sigma,C)\delta_{\sum_{i}\tau_i,n},\\
    & P_{M}(m)=\sum_{C}\sum_{\sigma=0,1}P(\sigma,C,m).
\end{align}    
\end{subequations}

Before we present results for the telegraph model with RNAP volume exclusion, we consider a simpler benchmark model in which RNAPs move deterministically on the gene, which can be solved in full for any choice of model parameters.

\subsection{The delay telegraph model}
\label{delay-telegraph-model}

In this model, elongation and termination take a fixed amount of time to finish, and the excluded-volume interactions between RNAPs are ignored. The reactions for this model are 
\begin{subequations}
\begin{align}
    & G_{\text{off}}\xrightleftharpoons[k_{\text{off}}]{k_{\text{on}}}G_{\text{on}}\\
    & G_{\text{on}}+\text{free RNAP}\xrightarrow[]{\alpha}G_{\text{on}}+\text{RNAP}_1\\
    & \text{RNAP}_i\xRightarrow[]{1/\omega}\text{RNAP}_{i+1},\; i \in [1,L-1],\\
    & \text{RNAP}_L\xRightarrow[]{1/\beta}\text{free RNAP}+\text{RNA},\\
    & \text{RNA}\xrightarrow[]{d_M}\emptyset.
\end{align}
\end{subequations}
As before, double arrows denote a delay reaction that takes a fixed amount of time to finish. We refer to this model as the delay telegraph model.

This model can be solved using renewal theory, which generalizes the Poisson process to an arbitrary distribution of inter-arrival times \cite{Cox1967}. To this end, we denote by $f_{\text{in}}(t)$ the probability density function of the waiting time between two successive initiation events, which has been computed in Ref. \cite{Dobrzynski2009}. The mean waiting time between two successive initiation events is given by
\begin{equation}
    \label{mu}
    \mu_{\text{in}}=\frac{k_{\text{off}}+k_{\text{on}}}{\alpha k_{\text{on}}},
\end{equation}
and the transcription rate $k_{\text{syn}}$ is equal to $1/\mu_{\text{in}}$,
\begin{equation}
    k_{\text{syn}}=\frac{1}{\mu_{\text{in}}}=\frac{\alpha k_{\text{on}}}{k_{\text{on}}+k_{\text{off}}}.
\end{equation}

The number of RNAPs that reside at segment $i$ at time $t$ is equal to the number of initiation events that occurred between $t-i/\omega$ and $t-(i-1)/\omega$. According to renewal theory, the mean number of initiation events in an interval $\Delta t$ in the steady state is equal to $\Delta t/\mu_{\text{in}}$ \cite{Cox1967}, from which it follows that the local density $\rho_{i}=1/(\mu_{\text{in}}\omega)$ for $i=1,\dots,L-1$ and $\rho_{L}=1/(\mu_{\text{in}}\beta)$. Using the expression for $\mu_{\text{in}}$ in Eq. (\ref{mu}), we get that
\begin{equation}
    \rho_{i}=\begin{dcases}
    \frac{\alpha}{\omega}\frac{k_{\text{on}}}{k_{\text{on}}+k_{\text{off}}} & i=1,\dots,L-1\\
    \frac{\alpha}{\beta}\frac{k_{\text{on}}}{k_{\text{on}}+k_{\text{off}}}, & i=L.
    \end{dcases}
\end{equation}

The probability distribution of the total number of nascent RNA can be computed by noting that the elongation and termination steps can be grouped into one delayed reaction of duration $T_{\text{el}}$ given by Eq. (\ref{delay1-elongation-time}). The distribution $P_{N}(n)$ for this process has been derived in Ref. \cite{Xu2016, Szavits2023}. Alternatively, $P_{N}(n)$ can be computed from the Taylor expansion of the probability generating function of the nascent RNA number given by
\begin{align}
    G_N(u)&=\frac{e^{-\kappa(u)T}}{2(k_{\text{on}}+k_{\text{off}})\delta(u)}\Bigl\{(k_{\text{on}}+k_{\text{off}})\delta(u)\Bigl[e^{\delta(u)T_{\text{el}}}\nonumber\\
    &+1\Bigr]+\Bigl[(k_{\text{on}}+k_{\text{off}})^2+\alpha u(k_{\text{on}}-k_{\text{off}})\Bigr]\Bigl[e^{\delta(u)T_{\text{el}}}\nonumber\\
    &-1\Bigr]\Bigr\},
    \label{G-telegraph}
\end{align}    
where $u=z-1$, $\kappa(u)=[k_{\text{on}}+k_{\text{off}}-k_{\text{syn}}u+\delta(u)]/2$ and $\delta(u)=\sqrt{(k_{\text{on}}+k_{\text{off}}-\alpha u)^2+4k_{\text{on}}\alpha u}$. The mean and the variance of the number of RNAPs are equal to
\begin{subequations}
   \begin{align}
    & \mu_N=\frac{T_{\text{el}}}{\mu_{\text{in}}}=k_{\text{syn}}T_{\text{el}}=\frac{\alpha k_{\text{on}}T_{\text{el}}}{k_{\text{on}}+k_{\text{off}}},\\
    \label{delay2-nascent-variance}
    &\sigma_{N}^{2}=\mu_N\Bigg\{1+\frac{2\alpha k_{\text{off}}}{T_{\text{el}}(k_{\text{on}}+k_{\text{off}})^3}\Bigg[e^{-(k_{\text{on}}+k_{\text{off}})T_{\text{el}}}\nonumber\\
    &\qquad -1+(k_{\text{on}}+k_{\text{off}})T_{\text{el}}\Bigg]\Bigg\}.
\end{align} 
\end{subequations}
The Fano factor of the nascent RNA number is
\begin{align}
    \label{delay2-nascent-FF}
    \text{FF}_{N} =& 1+\frac{2\alpha k_{\text{off}}}{T_{\text{el}}(k_{\text{on}}+k_{\text{off}})^3}\Bigg[e^{-(k_{\text{on}}+k_{\text{off}})T_{\text{el}}}\nonumber\\
    &-1+(k_{\text{on}}+k_{\text{off}})T_{\text{el}}\Bigg].
\end{align}
Note that because $\text{exp}(x)\geq 1+x$ for any real $x$, it follows that FF$_N\geq 1$.

Since elongation and termination are deterministic, the waiting time between two mature RNA production events is the same as the waiting time between two RNAP binding events, hence the probability distribution of the number of mature RNA is that of the telegraph model \cite{Peccoud1995,filatova2021statistics}, 
\begin{align}
    \label{delay2-mature}
    & P_{M}(m)=\frac{(\alpha/d_M)^{m}e^{-\alpha/d_M}}{m!}\frac{(k_{\text{on}}/d_M)_{m}}{(k_{\text{on}}/d_M+k_{\text{off}}/d_M)_{m}}\nonumber\\
    &\quad\times M(k_{\text{off}}/d_M,k_{\text{on}}/d_M+k_{\text{off}}/d_M+m,\alpha/d_M),
\end{align}
where $(x)_n=\Gamma(x+n)/\Gamma(x)$ is the Pochhammer symbol and $M$ is Kummer's (confluent hypergeometric) function. In particular, the mean and the variance of this distribution are
\begin{subequations}
\begin{align}
    \label{delay2-mature-RNA-mean}
    &\mu_M=\frac{k_{\text{on}}}{k_{\text{on}}+k_{\text{off}}}\frac{\alpha}{d_M},\\
    &\sigma_{M}^{2}=\mu_M\Bigg[1+\frac{\alpha k_{\text{off}}}{(k_{\text{on}}+k_{\text{off}})(k_{\text{on}}+k_{\text{off}}+d_M)}\Bigg],
\end{align}    
\end{subequations}
and the Fano factor FF$_{M}$ is given by
\begin{equation}
    \label{delay2-mature-RNA-FF}
    \text{FF}_{M}=1+\frac{\alpha k_{\text{off}}}{(k_{\text{on}}+k_{\text{off}})(k_{\text{on}}+k_{\text{off}}+d_M)}\geq 1.
\end{equation}

\subsection{The telegraph model with RNAP volume exclusion in the slow-switching regime}

The steady-state probability distribution $P(C)$ is known for the constitutive model with RNAP volume exclusion, but not for the telegraph model with RNAP volume exclusion. To make progress, we therefore need to apply some approximation method. Inspired by the success of timescale separation methods to simplify stochastic models of gene regulatory networks (where RNAP dynamics is not explicitly taken into account) \cite{thomas2014phenotypic,jia2020small}, here we focus on the regime of slow switching. 

In the slow-switching regime, we require that the nascent RNA in each gene state reaches the steady state of the TASEP corresponding to that gene state in the absence of switching, \textit{before} the gene switches its state again. We discuss these conditions and the constraints they impose on the parameters of the telegraph model with RNAP volume exclusion in detail in Appendix \ref{appendix_b}. In general, we find that transcription is in the slow-switching regime if the mean time the gene spends in the on and off states is much larger than the total time of elongation and termination of a single RNAP.

To put these conditions in the context of eukaryotic transcription, we analyzed a large dataset of the on and off rates that were inferred from single-cell RNA sequencing data in mouse fibroblasts \cite{larsson2019genomic}. We performed the analysis for three values of the elongation rate ($0.6$, $1.8$ and $2.4$ kb/min), which were reported in Ref. \cite{Jonkers2014}. Out of 3236 genes from the dataset, the slow-switching conditions were met in $40$\% of the genes at $0.6$ kb/min, $62$\% of the genes at $1.8$ kb/min and $66$\% of the genes at $2.4$ kb/min. Details of this analysis are presented in Appendix \ref{appendix_b}. 

The advantage of the slow-switching regime is that the calculation of the RNAP distribution $P(C)$ is simplified, since the conditional probability $P(C\vert \sigma)$ can be approximated by
\begin{subequations}
 \begin{align}
    \label{TASEP2-conditional-0}
    & P(C\vert \sigma=0)\approx\begin{cases}
    1, & \tau_1=\dots=\tau_L=0\\
    0, & \text{otherwise},
    \end{cases}\\
    \label{TASEP2-conditional-1}
    & P(C\vert \sigma=1)\approx P(C)\big\vert_{\text{vCM}}.
\end{align}   
\end{subequations}
where $P(C)\vert_{\text{vCM}}$ is the RNAP distribution in the constitutive model with RNAP exclusion [Eq. (\ref{TASEP1-exact-solution})]. In simple terms, Eqs. (\ref{TASEP2-conditional-0}) and (\ref{TASEP2-conditional-1}) are the steady states of the TASEP in which the initiation rates has been set to zero and $\alpha$, respectively. The probability distribution $P(C)$ can be now computed from
\begin{equation}
    \label{TASEP2-P-C}
    P(C)=\sum_{\sigma=0,1}P(\sigma)P(C\vert \sigma),
\end{equation}
where $P(\sigma)$ is the probability to find the gene in the state $\sigma=0,1$,
\begin{equation}
    P(\sigma)=\frac{k_{\text{on}}\sigma+k_{\text{off}}(1-\sigma)}{k_{\text{on}}+k_{\text{off}}}.
\end{equation}

\begin{figure*}[htb]
    \centering
    \includegraphics[width=\textwidth]{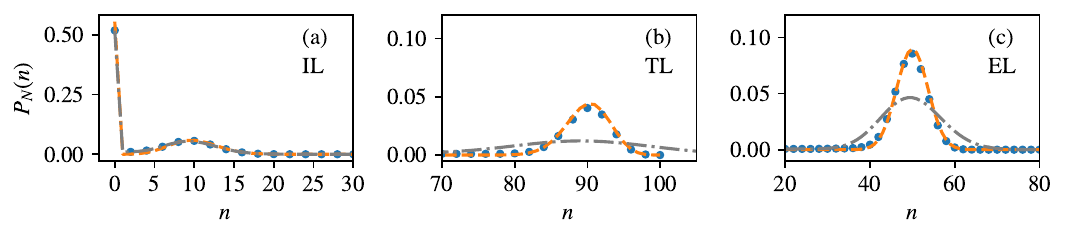}
    \caption{Accuracy of the theoretical prediction of the nascent RNA number distribution $P_N(n)$ in the telegraph model with RNAP volume exclusion in the slow-switching regime. The blue points are from stochastic simulations of the reaction scheme in Eq. (\ref{TASEP2-reactions}), the dashed orange lines are the slow-switching prediction computed from Eq. (\ref{TASEP2-nascent-RNA}) and the dot-dashed gray lines are the prediction from the effective delay telegraph model. Note that the effective delay telegraph model has the same mean nascent RNA number as the telegraph model with RNAP volume exclusion (see main text for details). The slow-switching theory provides accurate results for all levels of RNAP traffic, whereas the effective delay telegraph model is only accurate in the initiation-limited regime. The parameters are: (a) $\alpha=0.1$ s$^{-1}$, $\beta=0.7$ s$^{-1}$, $\omega=1.0$ s$^{-1}$, $k_{\text{on}}=8\cdot 10^{-4}$ s$^{-1}$, $k_{\text{off}}=0.001$ s$^{-1}$, and $d_M=0.01$ s$^{-1}$ (the initiation-limited regime, IL), (b) $\alpha=0.7$ s$^{-1}$, $\beta=0.1$ s$^{-1}$, $\omega=1.0$ s$^{-1}$, $k_{\text{on}}=10^{-4}$ s$^{-1}$, $k_{\text{off}}=2\cdot 10^{-4}$ s$^{-1}$ and $d_M=0.01$ s$^{-1}$ (the termination-limited regime, TL), and (c) $\alpha=0.7$ s$^{-1}$, $\beta=0.7$ s$^{-1}$, $\omega=1.0$ s$^{-1}$, $k_{\text{on}}=5\cdot 10^{-4}$ s$^{-1}$, $k_{\text{off}}=10^{-4}$ s$^{-1}$ and $d_M=0.01$ s$^{-1}$ (the elongation-limited regime, EL). The system size is $L=100$ for all plots.}
    \label{fig6}
\end{figure*}

\subsubsection{Local RNAP density, transcription rate and the nascent RNA number distribution}

Using Eq. (\ref{TASEP2-P-C}) yields the following expressions for the local density $\rho_{i}$, the transcription rate $k_{\text{syn}}$ and the nascent RNA number distribution $P_{N}(n)$, respectively, 
\begin{subequations}
   \begin{align}
    & \rho_{i}=\frac{k_{\text{on}}}{k_{\text{on}}+k_{\text{off}}}\rho_{i}\big\vert_{\text{vCM}},\quad i=1,\dots,L\label{TASEP2-rho-i}\\
    & k_{\text{syn}}=\frac{k_{\text{on}}}{k_{\text{on}}+k_{\text{off}}}J,\label{TASEP2-J}\\
    & P_{N}(n)=\frac{k_{\text{on}}}{k_{\text{on}}+k_{\text{off}}}P_{N}(n)\big\vert_{\text{vCM}}+\frac{k_{\text{off}}}{k_{\text{on}}+k_{\text{off}}}\delta_{n,0},\label{TASEP2-nascent-RNA}
\end{align} 
\end{subequations}
where $\rho_{i}\vert_{\text{vCM}}$ is the local RNAP density [Eq. (\ref{TASEP1-rho-i-exact})], $J$ is the transcription rate [Eq. (\ref{TASEP1-J-exact})] and  $P_{N}(n)\vert_{\text{vCM}}$ is the nascent RNA number distribution [Eq. (\ref{TASEP1-nascent-RNA-exact})] predicted by the constitutive model with RNAP volume exclusion. The expressions for the mean and the variance of $n$ are
\begin{subequations}
\begin{align}
    & \mu_N=\frac{k_{\text{on}}}{k_{\text{on}}+k_{\text{off}}}\mu_N\big\vert_{\text{vCM}},\\
    & \sigma_{N}^{2}=\frac{k_{\text{on}}}{k_{\text{on}}+k_{\text{off}}}\sigma_{N}^{2}\big\vert_{\text{vCM}}+\frac{k_{\text{on}}k_{\text{off}}}{(k_{\text{on}}+k_{\text{off}})^2}\mu_{N}\big\vert_{\text{vCM}},
\end{align}    
\end{subequations}
where $\mu_N\vert_{\text{vCM}}$ and $\sigma^{2}_{M}\vert_{\text{vCM}}$ are given by Eqs. (\ref{TASEP1-nascent-RNA-mean}) and (\ref{TASEP1-nascent-RNA-variance}), respectively. The Fano factor $\text{FF}_{N}$ is thus given by
\begin{equation}
    \label{TASEP2-nascent-RNA-FF}
    \text{FF}_{N}=\text{FF}_{N}\big\vert_{\text{vCM}}+\frac{k_{\text{off}}}{k_{\text{on}}+k_{\text{off}}}\mu_{N}\big\vert_{\text{vCM}}, 
\end{equation}
where $\text{FF}_{N}\vert_{\text{vCM}}$ is given by Eq. (\ref{TASEP1-nascent-RNA-FF-exact}). Note that since $\text{FF}_{N}\vert_{\text{vCM}}<1$, the nascent RNA number distribution can be (depending on the model parameters) either sub-Poissonian (FF$_{N}<1$) or super-Poissonian (FF$_{N}>1$) distribution. The transition from the sub-Poissonian to the super-Poissonian behavior occurs at the ratio $k_{\text{off}}/k_{\text{on}}\simeq 1/L$ in the initiation-limited and termination-limited regimes, and at the ratio $k_{\text{off}}/k_{\text{on}}\simeq 3/(2L)$ in the elongation-limited regime.

It is instructive to compare these results to those of the delay telegraph model analyzed in Section \ref{delay-telegraph-model}. Using the effective parameters discussed in Section \ref{TASEP1-nascent}, we find that the two models agree on the local RNAP density, transcription rate and mean nascent RNA number, but differ on the Fano factor of the nascent RNA number. To see this, we compute the Fano factor for the delay constitutive model in the slow-switching regime by expanding the right-hand side of Eq. (\ref{delay2-nascent-FF}) in $(k_{\text{on}}+k_{\text{off}})T_{\text{el,eff}}$, which gives up to the leading order
\begin{equation}
    \label{delay-telegraph-nascent-RNA-FF-slow-switching}
    \text{FF}_N\big\vert_{\text{edTM}}=1+\frac{k_{\text{off}}}{k_{\text{on}}+k_{\text{off}}}\mu_{N}\big\vert_{\text{vCM}}
\end{equation}
where ``edTM" stands for the effective delay telegraph model. Comparing Eqs. (\ref{TASEP2-nascent-RNA-FF}) and (\ref{delay-telegraph-nascent-RNA-FF-slow-switching}), we conclude that the Fano factor in the telegraph model with RNAP volume exclusion is always smaller than the Fano factor in the effective delay telegraph model. 

In Fig. \ref{fig6}, we compare $P_N(n)$ obtained using stochastic simulations with the predictions of our slow-switching theory and the effective delay telegraph model. The simulations of the reaction scheme in Eq. (\ref{TASEP2-reactions}) were performed in the slow-switching regime for three sets of parameters corresponding to the initiation-limited [IL, Fig. \ref{fig6}(a)], termination-limited [TL, Fig. \ref{fig6}(b)] and elongation-limited regimes [Fig. \ref{fig6}(c)]. Our slow-switching theory agrees well with the simulations for all three levels of RNAP traffic. In contrast, the effective delay telegraph model correctly predicts the nascent RNA number distribution in the initiation-limited regime, but fails to do so in the termination-limited and elongation-limited regimes, which can be explained as follows. In the slow-switching regime, the nascent RNA number distribution predicted by the effective delay telegraph model is a mixture distribution of the delta function at $n=0$ (corresponding to the off state), and the Poisson distribution with the rate parameter $\rho L$ (corresponding to the on state). However, the nascent RNA number distribution in the on state, $P_{N}(n)\vert_{\text{vCM}}$, is close to the Poisson distribution only in the initiation-limited regime, and it is markedly sub-Poissonian in the termination-limited and elongation-limited regimes.

\begin{figure*}[htb]
    \centering
    \includegraphics[width=\textwidth]{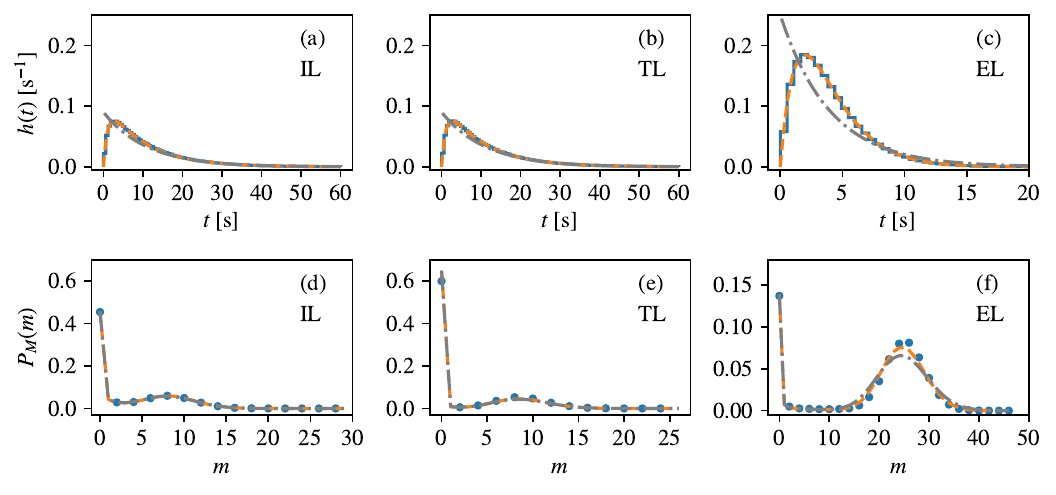}
    \caption{Accuracy of the theoretical predictions of the statistics of the time between termination events and of mature RNA numbers in the slow-switching regime of the telegraph model with RNAP volume exclusion. (a)-(c) compare the pdf $h(t)$ of the waiting time distribution between two successive termination events computed using stochastic simulations (solid blue lines), the slow-switching prediction obtained from $\smash{\tilde{h}}(s)$ in Eq. (\ref{TASEP2-wt-infinite-LT}) (dashed orange lines) and the prediction of the effective delay telegraph model (dash-dotted gray lines). (d)-(f) compare the mature RNA number distribution $P_{M}(m)$ computed using stochastic simulations (blue points), the slow-switching prediction computed from Eq. (\ref{TASEP2-mature-RNA-renewal}) (dashed orange line), and the prediction of the effective delay telegraph model (dash-dotted gray line). Note that the effective delay telegraph model has the same mean nascent and mature RNA numbers as the telegraph model with RNAP volume exclusion (see main text for details). The parameters are: (a) and (d) $\alpha=0.1$ s$^{-1}$, $\beta=0.7$ s$^{-1}$, $\omega=1.0$ s$^{-1}$, $k_{\text{on}}=8\cdot 10^{-4}$ s$^{-1}$, $k_{\text{off}}=0.001$ s$^{-1}$, and $d_M=0.01$ s$^{-1}$ (the initiation-limited regime, IL), (b) and (e) $\alpha=0.7$ s$^{-1}$, $\beta=0.1$ s$^{-1}$, $\omega=1.0$ s$^{-1}$, $k_{\text{on}}=10^{-4}$ s$^{-1}$, $k_{\text{off}}=2\cdot 10^{-4}$ s$^{-1}$ and $d_M=0.01$ s$^{-1}$ (the termination-limited regime, TL), and (c) and (f) $\alpha=0.7$ s$^{-1}$, $\beta=0.7$ s$^{-1}$, $\omega=1.0$ s$^{-1}$, $k_{\text{on}}=5\cdot 10^{-4}$ s$^{-1}$, $k_{\text{off}}=10^{-4}$ s$^{-1}$ and $d_M=0.01$ s$^{-1}$ (the elongation-limited regime, EL). The system size is $L=100$ for all plots.}
    \label{fig7}
\end{figure*}

\subsubsection{Mature RNA number distribution}

To compute the probability distribution of mature RNA number, we replace the TASEP with the following stochastic process for the turnover of mature RNA,
\begin{equation}
    \label{TASEP2-renewal-process}
    G_{\text{off}}\xrightleftharpoons[k_{\text{off}}]{k_{\text{on}}} G_{\text{on}}\xrightarrow[]{f_{\text{ter,as}}(t)}G_{\text{on}}+M,\quad M\xrightarrow[]{d_M}\emptyset,
\end{equation}
where $f_{\text{ter,as}}(t)$ is given by Eq. (\ref{TASEP1-wt-infinite}). The assumption of slow-switching is crucial here, otherwise the pdf of the waiting time between two successive termination events conditioned on not leaving the on state is not necessarily described by $f_{\text{ter,as}}(t)$, which was derived for the TASEP in the steady state. Since we are interested in the probability distribution of the mature RNA number $m$ irrespective of the gene state, we can rewrite this process as the following queuing process,
\begin{equation}
    \label{TASEP2-queuing-process}
    \emptyset\xrightarrow[]{h(t)}M\xrightarrow[]{d_M}\emptyset,
\end{equation}
where $h(t)$ is the waiting time distribution between two successive mature RNA production events in the process defined by Eq. (\ref{TASEP2-renewal-process}). The steady-state distribution of the mature RNA number $m$ for the queuing process (\ref{TASEP2-queuing-process}) is given by Eq. (\ref{P-M-queuing}) in which $\smash{\tilde{f}(s)}$ is replaced by $\smash{\tilde{h}}(s)$, the Laplace transform of $h(t)$. The calculation of $\smash{\tilde{h}}(s)$ is presented in Appendix \ref{appendix_c}, and the final result is given by Eq. (\ref{TASEP2-wt-infinite-LT}). Inserting Eq. (\ref{TASEP2-wt-infinite-LT}) into Eq. (\ref{P-M-queuing}) yields
\begin{subequations}
\label{TASEP2-mature-RNA-renewal}
\begin{align}
    & P_{M}(0)=1-\frac{b_1 b_2}{\mu_h K a_1 a_2}\Bigg[1 \nonumber\\
    & \qquad -{}_2 F_{2}\left(\frac{a_1}{d_M},\frac{a_2}{d_M};\frac{b_1}{d_M},\frac{b_2}{d_M};-\frac{K}{d_M}\right)\Bigg],\\
    & P_{M}(m)=\frac{b_1 b_2\left(\frac{K}{d_M}\right)^{m}\left(\frac{a_1}{d_M}\right)_m \left(\frac{a_2}{d_M}\right)_m}{\mu_h K a_1 a_2 (m!)\left(\frac{b_1}{d_M}\right)_m \left(\frac{b_2}{d_M}\right)_m}\nonumber\\
    & \qquad \times{}_2F_{2}\left(\frac{a_1}{d_M}+m,\frac{a_2}{d_M}+m;\right.\nonumber\\
    &\qquad \left.\frac{b_1}{d_M}+m,\frac{b_2}{d_M}+m;-\frac{K}{d_M}\right),\quad m\geq 1.
\end{align}
\end{subequations}
In the expression above, $\mu_h$ is the first moment of $h(t)$,
\begin{equation}
    \mu_h=\frac{(k_{\text{off}}+k_{\text{on}})(k_{\text{off}}+\omega)}{k_{\text{on}}J\omega}.
\end{equation}
Note that ${}_2 F_2$ is a generalized hypergeometric function, and the constants $K$, $a_1$, $a_2$, $b_1$ and $b_2$ are defined through the factorization of the polynomials in $\tilde{h}/[1-\tilde{h}(s)]$ such that
\begin{equation}
    \frac{\tilde{h}(s)}{1-\tilde{h}(s)}=\frac{K(s+a_1)(s+a_2)}{s(s+b_1)(s+b_2)}.
\end{equation}
The mean and the variance of the mature RNA number $m$ are given by
\begin{subequations}
\begin{align}
    & \mu_M=\frac{1}{\mu_h d_M}=\frac{k_{\text{on}}}{(k_{\text{on}}+k_{\text{off}})}\frac{J\omega}{d_M(\omega+k_{\text{off}})},\label{TASEP2-mature-RNA-mean-slow-switching}\\
    & \sigma_{M}^{2}=\frac{1}{\mu_{h} d_M}\left[1+\frac{\tilde{h}(d_M)}{1-\tilde{h}(d_M)}-\frac{1}{\mu_{h} d_M}\right],
\end{align}    
\end{subequations}
from which we get the following expression for the Fano factor FF$_M$
\begin{equation}
    \label{TASEP2-mature-RNA-FF-slow-switching}
    \text{FF}_M=1+\frac{\tilde{h}(d_M)}{1-\tilde{h}(d_M)}-\frac{1}{\mu_{h} d_M}.
\end{equation}

Interestingly, the mean mature RNA number in Eq. (\ref{TASEP2-mature-RNA-mean-slow-switching}) differs from the one predicted by the effective delay telegraph model [Eq. (\ref{delay2-mature-RNA-mean})]---the former is smaller by a factor of $\omega/(\omega+k_{\text{off}})$, which is close to $1$ only when $k_{\text{off}}\ll\omega$. Given that the elongation rate is typically in the range of $10-100$ nt/s (corresponding to the hopping rate $\omega$ in the range of $0.3-3.0$ segments/s), this condition requires that the average time the gene spends in the on state is much larger than a few seconds, which is in line with values reported in the literature that measure in minutes. Following this argument, we set $\omega=k_{\text{off}}/x$ and expand the Fano factor of the mature RNA number given by Eq. (\ref{TASEP2-mature-RNA-FF-slow-switching}) in $x$, keeping the first two terms. The calculation, not shown here, reveals that the leading term is precisely the Fano factor predicted by the delay telegraph model [Eq. (\ref{delay2-mature-RNA-FF})], whereas the leading correction is always negative. Hence, in the biologically relevant regime of $k_{\text{off}}\ll\omega$, the Fano factor of the mature RNA number predicted by the telegraph model with RNAP volume exclusion is smaller than the one predicted by the delay telegraph model.

\begin{figure*}[htb]
    \centering
    \includegraphics[width=\textwidth]{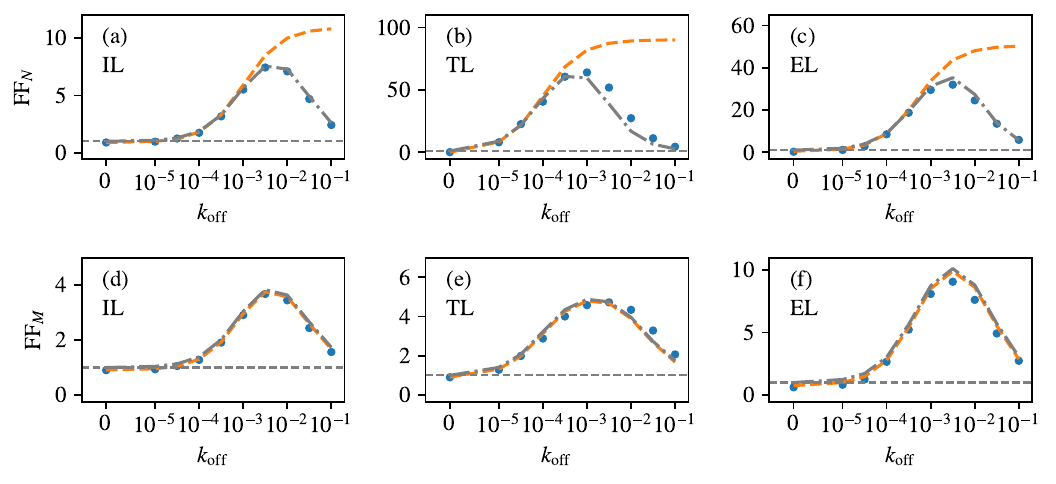}
    \caption{Accuracy of the theoretical predictions for the Fano factor of nascent and mature RNA numbers in the telegraph model with RNA volume exclusion as a function of the switching off rate. (a)-(c) compare the Fano factor FF$_N$ of the nascent RNA number distribution computed using stochastic simulations (blue points), the slow-switching prediction obtained from Eq. (\ref{TASEP2-nascent-RNA-FF}) (dashed orange lines) and the prediction of the effective delay telegraph model (dash-dotted gray lines). (d)-(f) compare the Fano factor FF$_M$ of the mature RNA number distribution computed using stochastic simulations (blue points), the slow-switching prediction computed from Eq. (\ref{TASEP2-mature-RNA-FF-slow-switching}) (dashed orange line), and the prediction of the effective delay telegraph model (dash-dotted gray line). The dashed gray line is at the value of $1$, marking a transition from the sub-Poissonian to the super-Poissonian behavior. The parameters are: (a) and (d) $\alpha=0.1$ s$^{-1}$, $\beta=0.7$ s$^{-1}$, $\omega=1.0$ s$^{-1}$, $k_{\text{on}}=10^{-3}$ s$^{-1}$ and $d_M=0.02$ s$^{-1}$ (the initiation-limited regime, IL), (b) and (e) $\alpha=0.3$ s$^{-1}$, $\beta=0.1$ s$^{-1}$, $\omega=1.0$ s$^{-1}$, $k_{\text{on}}=10^{-4}$ s$^{-1}$ and $d_M=0.02$ s$^{-1}$ (the termination-limited regime, TL), and (c) and (f) $\alpha=0.7$ s$^{-1}$, $\beta=0.7$ s$^{-1}$, $\omega=1.0$ s$^{-1}$, $k_{\text{on}}=5\cdot 10^{-4}$ s$^{-1}$ and $d_M=0.02$ s$^{-1}$ (the elongation-limited regime, EL). The system size is $L=100$ for all plots.}
    \label{fig8}
\end{figure*}

In Fig. \ref{fig7}, we compare $h(t)$ and $P_M(m)$ obtained using stochastic simulations with the predictions of our slow-switching theory and the effective delay telegraph model. The simulations of the reaction scheme in Eq. (\ref{TASEP2-reactions}) were performed in the slow-switching regime for the same three sets of parameters as in Fig. \ref{fig6} corresponding to the initiation-limited, termination-limited and elongation-limited regimes. Overall, our slow-switching theory agrees well with the simulations for all three levels of RNAP traffic. [A small discrepancy between the slow-switching theory and the simulations is found in the elongation-limited regime, which is due to the renewal approximation that ignores correlations between successive termination events.]

The mature RNA number distribution predicted by the effective delay telegraph model matches closely that of the telegraph model with RNAP volume exclusion, especially in the initiation-limited and termination-limited regimes [Fig. \ref{fig7}(d) and \ref{fig7}(e)], whereas a small but visible discrepancy between the distributions is found in the elongation-limited regime [Fig. \ref{fig7}(f)]. While this agreement may seem surprising, we note that the results were obtained in the biologically-relevant regime of $k_{\text{off}}\ll\omega$ in which the mean and the variance of the mature RNA number, as well as the first three moments of the waiting time distribution, agree between the two models (provided the delay telegraph model is used with the effective parameters discussed in Section \ref{TASEP1-nascent}). 

We note that besides the effective delay telegraph model here considered, there are other ways of finding a reduced model for the mature RNA distributions which might lead to equally good fits to the telegraph model with RNAP volume exclusion. For example, following the approach in \cite{Szavits2022}, one could find parameters of the conventional telegraph model (promoter switching and synthesis rate) such that its predictions for the fraction of time that the gene is active, the mean mature RNA number and the variance of the mature RNA fluctuations agree with those of the reaction scheme in Eq. (\ref{TASEP2-reactions}).

\subsection{The telegraph model with RNAP volume exclusion outside the slow-switching regime}
\label{outside-slow-switching}

We consider here what happens when we relax the slow-switching condition in Eq. (\ref{TASEP2-slow-switching-koff}), but keep the condition in Eq. (\ref{TASEP2-slow-switching-kon}) satisfied. Relaxing the condition in Eq. (\ref{TASEP2-slow-switching-koff}) means that the RNAP distribution in the on state is no longer given by the steady-state distribution of the TASEP [Eq. (\ref{TASEP1-exact-solution})], whereas keeping the condition in Eq. (\ref{TASEP2-slow-switching-kon}) satisfied means that the gene in the off state has no active RNAPs. The motivation to study this regime comes from mammalian genes which stay inactive for hours, but are active only for few minutes, which may not be enough for the RNAP dynamics to reach the steady state.

In Fig. \ref{fig8}, we show the Fano factor of the nascent and mature RNA numbers obtained using stochastic simulations for various values of the off rate $k_{\text{off}}$, including values outside the slow-switching regime. These results are compared to the predictions of the slow-switching theory and the predictions of the effective delay telegraph model. As can be seen from Figs. \ref{fig8}(a)-(c) depicting the Fano factor FF$_{N}$ dependence on $k_{\text{off}}$, the slow-switching theory eventually fails as $k_{\text{off}}$ is increased, whereas the effective delay telegraph model prediction provides a good fit for all values of $k_{\text{off}}$, except for a small but visible disagreement in the termination-limited regime [Fig. \ref{fig8}(b)]. In contrast, the Fano factor of the mature RNA number is well accounted for by both the slow-switching theory and the effective delay telegraph model. 

We emphasize that an implicit reason for the agreement of the effective delay telegraph model and the telegraph model with RNAP volume exclusion in Fig. \ref{fig8} is that for all the values of $k_{\text{off}}$ considered (except the value of $k_{\text{off}}=0$), the RNA fluctuations in the latter model are super-Poissonian. If the parameters were such that the RNA fluctuations are sub-Poissonian (by choosing $k_{\text{off}}$ to be much less than $k_{\text{on}}$) than this model matching would be impossible. 

The agreement between the models can be understood by looking at the time evolution of the TASEP, starting from an empty lattice (the initial state at the time of switching to the on state). We note that this problem was previously studied in the context of mRNA translation \cite{Nagar2011} and stochastic resetting \cite{Karthika2020}. We begin with the initiation-limited regime in which $\alpha<\omega/2$ and $\beta>\alpha$ [Figs. \ref{fig8}(a) and \ref{fig8}(d)]. In this case, the RNAP density is a shock wave of density $\alpha/\omega$ travelling at the speed of $\omega(1-\alpha/\omega)$. The nascent RNA production rate is given by $\alpha(1-\alpha/\omega)$, and the total time it takes the RNAP density wave to travel across the gene is equal to $L/[\omega(1-\alpha/\omega)]$. We note that the RNAP density of $\alpha/\omega$ is compatible with the termination rate $\beta>\alpha$. Hence, once the RNAP density wave reaches the end of the gene, the RNAP density does not change, and the rate of mature RNA production is given by $\alpha(1-\alpha/\omega)$. This explains why the effective delay telegraph model provides a good fit for all values of $k_{\text{off}}$. 

We next consider the termination-limited regime in which $\beta<\omega/2$ and $\alpha>\beta$ [Figs. \ref{fig8}(b) and \ref{fig8}(e)]. For $\alpha<\omega/2$, the RNAP density wave is a shock wave of density $\alpha/\omega$ as in the initiation-limited regime. Once the RNAP density wave reaches the end of the gene, the RNAPs begin to accumulate, creating a second shock wave of density $1-\beta/\omega$ that travels backwards towards the promoter. This happens because the termination rate $\beta<\alpha$ is not compatible with the density of $\alpha/\omega$. Because the RNAP density at the end of the gene is now equal to $1-\beta/\omega$, the rate of mature RNA production is given by $\beta(1-\beta/\omega)$. Once the second shock wave reaches the start of the gene, provided the gene stays in the on state long enough, the rate of nascent RNA production changes from $\alpha(1-\alpha/\omega)$ to $\beta(1-\beta/\omega)$. Otherwise, the RNAP density becomes non-uniform along the gene, which explains why the Fano factor of the nascent RNA number predicted by the effective delay telegraph model does not quite match the simulations for large values of $k_{\text{off}}$ [Fig. \ref{fig8}(b)]. 

Finally, we consider the elongation-limited regime in which $\alpha>\omega/2$ and $\beta>\omega/2$ [Figs. \ref{fig8}(b) and \ref{fig8}(e)]. In this case, the RNAP density is not a shock wave, i.e. the initiation rate of $\alpha$ has no influence on the RNAP density. Instead, the RNAP density is a rarefaction wave that decays linearly along the gene, and the rate of nascent RNA production is given by $\omega/4$. Once the tip of the RNAP density wave reaches the end of the gene, which happens after time $L/\omega$, the RNAP density begins to increase towards its steady state value of $1/2$. If the gene stays long enough for this to happen, the total time of elongation and termination is equal to $2L/\omega$, and the rate of mature RNA production is given by $\omega/4$. We would therefore expect the effective delay telegraph model not to provide a good fit for large values of $k_{\text{off}}$. However, that is not what we see in Figs. \ref{fig8}(b) and \ref{fig8}(e), i.e. the effective delay telegraph model fits the data obtained using simulations even for large values of $k_{\text{off}}$ (up to $10$\% value of the elongation rate $\omega$). Presently, we do not have a good explanation for this agreement.

\section{Summary and Discussion}
\label{summary-discussion}

The main question we asked in this study is how the excluded volume interactions between RNAPs affect the distributions of nascent and mature RNA numbers. To answer this question, we developed a stochastic model of gene expression that accounts for: (i) transcription initiation, (ii) transcription elongation by RNAPs moving along the gene, (iii) transcription termination and nascent RNA processing into a mature RNA, and (iv) degradation of mature RNA. The movement of RNAPs was modelled by the totally asymmetric simple exclusion process (TASEP) in which each lattice site corresponded to $\approx 35$ bp (the footprint of RNAP) and RNAPs could only move forward with non-zero probability if the next segment is free. We considered two initiation mechanisms, one in which the gene is always active (the constitutive model with RNAP volume exclusion, see Section \ref{constitutive-model}), and the other in which the gene switches between two states of activity and inactivity (the telegraph model with RNAP volume exclusion, Section \ref{telegraph-model}). In order to determine the importance of RNAP volume exclusion, we compared these models with another set of models in which RNAPs move deterministically and do not interact with each other (the delay constitutive and telegraph models). 

\textit{The constitutive model with RNAP volume exclusion.} For the constitutive model with RNAP volume exclusion, we obtained an exact expression for the distribution of nascent RNA number in the steady state. The shape of this distribution is strongly determined by the RNAP density $\rho$, which equals the average probability of a gene segment being occupied by an RNAP. The distribution is very close to a binomial distribution in the initiation-limited ($\rho<1/2$) and termination-limited ($\rho>1/2$) regimes, but differs substantially from the binomial distribution at the coexistence line between the initiation-limited and termination-limited regimes, and in the elongation-limited regime ($\rho=1/2$). A direct consequence of the excluded volume interactions between RNAPs is that the nascent RNA number distribution is sub-Poissonian. Consequently, the Fano factor of the nascent RNA number ranges between $0$ when $\rho\rightarrow 1$ and $1$ when $\rho\rightarrow 0$. In contrast, the delay constitutive model predicts a Poisson distribution of the nascent RNA number whose Fano factor equals exactly $1$. We note that the Poisson distribution is a special limit of the binomial distribution when the RNAP density is small and the gene length is large. Hence, the two models are expected to agree when the RNAP density is low. A similar result that the nascent RNA number distribution becomes sub-Poissonian due to excluded-volume interaction has been obtained before using stochastic simulations of the TASEP in which RNAPs occupy more than one lattice site \cite{Ali2020}.

The distribution of the mature RNA number was computed in the renewal approximation, in which the correlations between waiting times between successive termination events are ignored. This approximation allowed us to map the production and degradation of mature RNA to a queuing process. The distribution of mature RNA number was computed analytically using known results from queuing theory. This distribution is also sub-Poissonian, whereby its Fano factor ranges from $3/4$ in the elongation-limited regime to $1$ when $\rho\rightarrow 0$ or $\rho\rightarrow 1$. In contrast, the delay constitutive model predicts a Poisson distribution whose Fano factor equals exactly $1$. Our results also show that the Fano factor of the nascent RNA number is less than that of the mature RNA number, implying that a signature of RNAP volume exclusion effects (the sub-Poissonian nature of fluctuations) is gradually erased as RNA progresses through its life cycle. 

It is important here to note that experimentally observed sub-Poissonian fluctuations in nascent and nuclear transcript numbers are generally not solely due to RNAP volume exclusion effects; they can also be caused by multiple rate-limiting steps in initiation \cite{lloyd2016dissecting,braichenko2021distinguishing,weidemann2023minimal}, which we have not considered in this paper. In particular, the same waiting time distribution between two mature RNA production events that was obtained in the constitutive model with RNAP volume exclusion [the hypoexponential distribution in Eq. (\ref{TASEP1-wt-infinite})], can also be obtained in a constitutive model with no RNAP volume exclusion, but with a two-step initiation mechanism of the type: $G_0 \rightarrow G_1 \rightarrow G_0 + \text{RNA}, \text{RNA}\rightarrow\emptyset$ ($G_i$ are the promoter states).

Furthermore, we note that the mature RNA distribution is invariant to $\rho\leftrightarrow 1-\rho$, which originates from the waiting time distribution between two successive termination events possessing the same symmetry. This implies a fundamental limitation to the accurate estimation of the RNAP density from experimental measurements of mature RNA number alone, i.e. without measuring the nascent RNA number.

\textit{The telegraph model with RNAP volume exclusion.} Similarly, for the telegraph model with RNAP volume exclusion, we obtained analytical expressions for the nascent and mature RNA distributions in the slow-switching regime, in which case the TASEP has enough time to reach the steady state it would relax to in the absence of switching, before the next switching event occurs. In this regime, the nascent RNA number distribution simplified to a mixture distribution of the delta function at zero and the nascent RNA number distribution derived for the constitutive model. To compare this distribution to the one predicted by the delay telegraph model, we adjusted the initiation rate and the elongation time of the delay telegraph model to match those predicted by the TASEP, which we called the effective delay telegraph model. This made sure that the two nascent RNA number distributions had the same mean. Nevertheless, significant deviations between the two distributions were observed in the termination-limited and elongation-limited regimes, suggesting that caution is needed when inferring parameters by fitting nascent RNA data to the delay telegraph model \cite{senecal2014transcription,fritzsch2018estrogen,fu2022quantifying}. In contrast, the two models agreed on the mature RNA number distribution both in the initiation-limited and termination-limited regimes, whereas a small discrepancy between the distribution was found in the elongation-limited regime. We also found that the slow promoter switching causes the Fano factor of the nascent RNA number to be larger than in the constitutive model, potentially changing the nature of the fluctuations from sub-Poissonian to super-Poissonian. This transition occurs when $k_{\text{off}}/k_{\text{on}}\simeq 1/L$, i.e. when the gene spends most of the time in the on state. As $k_{\text{off}}$ increases and becomes much larger than $k_{\text{on}}$, a condition that is often associated with bursty gene expression \cite{suter2011mammalian,cao2020analytical}, the effects of RNAP volume exclusion on the nascent RNA fluctuations are expected to diminish. 

Outside the slow-switching regime, limited results were obtained, as the telegraph model with RNAP volume exclusion becomes difficult to solve analytically. Motivated by mammalian genes which have short periods of activity followed by long periods of inactivity, we focused on the regime in which the nascent RNA reaches the steady state in the off state (an empty lattice), but not necessarily in the on state. We computed the Fano factor of the nascent and mature RNA numbers using stochastic simulations, and compared it to the predictions of the effective delay telegraph model. We showed that the effective delay telegraph model provides a good match for the telegraph model with RNAP volume exclusion well beyond the slow-switching regime. 

\textit{Model-based inference of transcription kinetics from single-cell mature RNA data.} Although nascent RNA is a direct reflection of the transcription process, most of the gene expression data comes from measuring the mature RNA. The question is then, what can the mature RNA data tell us about transcription kinetics and in particular about the RNAP density (assuming mature RNA fluctuations are not dominated by post-transcriptional noise)? 

For the constitutive expression and for the slow promoter switching, the mature RNA number distribution is determined by the waiting time distribution between two successive termination events, and the degradation rate $d_m$. The waiting time distribution is parametrized by $\omega\rho$ and $\omega(1-\rho)$, and is invariant under the exchange of $\rho\leftrightarrow 1-\rho$, where $\rho$ is the RNAP density in the on state. Hence, it is not possible to distinguish from the mature RNA data alone whether the transcription is in the initiation-limited or termination-limited regime. 

The same is true outside the slow-switching regime. We remind the reader that in this case, the effective delay telegraph model provides a reasonably good fit to the mature RNA number distribution. For this model, the effective transcription rate in the on state is equal to $J$, the RNAP current of the TASEP [Eq. (\ref{TASEP1-J-infinite})]. Since $J$ is invariant under the exchange of $\alpha\leftrightarrow \beta$, inferring the value of $J$ from the mature RNA data is not enough to determine whether the transcription is in the initiation-limited or termination-limited regime. There is a fundamental loss of information due to the downstream processing of nascent RNA, which makes it difficult to infer transcriptional dynamics from the mature RNA data.

If one is interested in the estimation of transcriptional parameters, then one can use the nascent RNA number distributions derived in this paper to construct the likelihood function. The maximization of this function, given the measured single-cell nascent RNA data, leads to the desired estimates. This approach is presently limited to constitutive expression and the slow-switching regime, for which we have derived analytical expressions for the nascent RNA number distributions. For genes with a substantial RNAP traffic, this approach may produce more accurate estimates than models of nascent RNA dynamics that ignore steric interactions. Examples of such genes are the ribosomal genes. In \textit{E. coli}, one RNA polymerase was observed every 85 bp on rRNA operons \cite{French1989}, which is equivalent to the RNAP density of $41\%$. In yeast, between 30 and 70 RNA polymerases were observed on the 35S rRNA gene \cite{turowski2020nascent}, which given the gene length of 6858 bp amounts to the RNAP density in the range of $15-36$\%. Other genes which might have a large amount of RNAP traffic might be those whose transcription rate scales with the cell volume, particularly when the volume is large \cite{Padovan2015}. 

\textit{Possible extensions to include more details of transcription initiation and elongation.} In general, there might be more than two promoter states \cite{Zhou2012}, for example, to describe off duration times that are non-exponential, which has been measured in some mammalian cells \cite{suter2011mammalian}. As well, extra promoter states could capture multiple rate-limiting steps in initiation \cite{Szavits2023,weidemann2023minimal}. The promoter-proximal pausing of RNAP II in metazoans, which occurs within $100$ bp from the transcription start site, can also be considered as one of the promoter states since the paused RNAP blocks another RNAP from being recruited to the promoter \cite{Shao2017,braichenko2021distinguishing}. The methods developed in this paper can be extended in these directions, provided the RNAP distribution is allowed to reach the steady state in each of the promoter states much sooner than the next change of state occurs.

Since our model describes the movement of RNAPs at the resolution of $\approx 35$ bp (the footprint length of the RNAP), it does not have an explicit description of the processes occurring at the single nucleotide level, such as ubiquitous pausing and backtracking. These processes can be included implicitly in our model by setting $\omega$ to $1/(\ell\tau)$, where $\tau$ is the mean residence time the RNAP spends at each nucleotide and $\ell$ is the footprint length of the RNAP. For example, in a model with ubiquitous pausing in which an RNAP enters the paused state at rate $k_{p}$, returns to the active state at rate $k_{a}$, and moves to the next nucleotide in the active state at rate $\epsilon$ \cite{Klumpp2011}, the mean residence time $\tau$ is equal to $(k_p+k_a)/(k_a\epsilon)$. Backtracking is usually modelled as a biased random walk, in which case $\tau$ is equal to the mean first-passage time it takes the RNAP to return to the original position it backtracked from \cite{Voliotis2008}. We note that setting $\omega$ to $1/(\ell\tau)$ in these examples will work only if the RNAP pausing does not cause a substantial queuing of the RNAPs. Otherwise, the mean residence time $\tau$ is not only determined by the pausing process, but also by the time the RNAP spends waiting in a queue. Finally, we note that our model does not account for long-range cooperation between RNAPs during transcription elongation \cite{Epshtein2003,Kim2019}. This cooperation results in a reduced elongation time when multiple RNAPs transcribe the gene at the same time. Several theoretical studies have been put forward to explain this cooperation via DNA supercoiling \cite{Heberling2016,Chatterjee2021,Tripathi2021,Klindziuk2021}. However, the RNAP density at which these effects become evident is poorly understood. For example, a several-fold  increase in RNAP density on \textit{lacZ} operon showed no effect on the elongation rate \cite{Kim2019}. Future work will focus on resolving how the aforementioned mechanistic details affect the nascent and mature RNA number distributions.

In conclusion, we have developed a theory of stochastic gene expression that elucidates the link between fluctuations in RNA numbers and the strength of mechanical interactions between RNAPs. The specialty of the model is its analytic tractability, which enables us to make several predictions of physical and biological relevance. The theory (i) uncovers a fundamental limitation in the deduction of RNA polymerase traffic patterns on a gene from the steady-state distribution of mature RNA numbers; (ii) proves that steric interactions result in a suppression of fluctuations in the RNA numbers, potentially even leading to sub-Poissonian fluctuations; (iii) provides a novel interpretation of standard models of gene expression by expressing their parameters in terms of the kinetic parameters controlling the microscopic processes of transcription. Our work shows that there is a wealth of information about transcriptional processes hidden in the nascent RNA number fluctuations. This information can be extracted using statistical-mechanical techniques previously developed for the TASEP, giving us in turn an unprecedented view of stochastic gene expression.   

\begin{acknowledgments}
We thank Martin Evans for suggesting Eqs. (\ref{DE-trick}) and (\ref{Lagrange-inversion-formula}). This work was supported by a Leverhulme Trust research award (RPG-2020-327).
\end{acknowledgments}

\appendix
\section{The probability generating function \texorpdfstring{$G_N(n)$}{G\_N(n)} in the constitutive model with RNAP volume exclusion}
\label{appendix_a}

Starting from Eq. (\ref{TASEP1-nascent-RNA-GN}), we introduce
\begin{equation}
    Y_L(z)=\langle W\vert (zD+E)^L\vert V\rangle,
\end{equation}
and denote by $H(w,z)$ the generating function for $Y_L(z)$,
\begin{align}
    \label{H-definition}
    H(w,z)&=\sum_{L=0}^{\infty}Y_{L}(z)w^L\nonumber\\
    &=\bigl \langle W\big|[1-w(zD+E)]^{-1}\big|V \bigr \rangle.
\end{align}
Next, we note that since $DE=D+E$, we can write
\begin{equation}
    \label{DE-trick}
    1-w(zD+E)=(1-aD)(1-bE),
\end{equation}
where $a$ and $b$ satisfy
\begin{equation}
    \label{a-b-equations}
    wz=a(1-b),\quad w=b(1-a).
\end{equation}
Hence, the inverse of $1-w(zD+E)$ is given by
\begin{equation}
    [1-w(zD+E)]^{-1}=(1-bE)^{-1}(1-aD)^{-1}.
\end{equation}
Using Eq. (\ref{a-b-equations}), we express $b$ in terms of $a$ and $z$, which gives $b=a/(z-za+a)$. Altogether, 
\begin{equation}
    \label{H}
    H(a,z)=\frac{1-\left(\frac{z-1}{z}\right)a}{\left[1-\big(\frac{\alpha z-\alpha+1}{\alpha z}\big)a\right]\left(1-\frac{1}{\beta}a\right)},
\end{equation}
where $a$ is a function of $w$ and $z$ defined implicitly by the equation
\begin{equation}
    \label{w-equation}
    w=\frac{a(1-a-z)}{z(1-a)}.
\end{equation}
To find $Y_L(z)$, we look for the coefficient of $w^{L}$ in the Taylor expansion of $H(w,z)$ around $w=0$,
\begin{equation}
    Y_L(z)=\left[w^{L}\right]H(w,z)=\left.\frac{1}{L!}\frac{\partial^L H}{\partial w^L}\right\vert_{w=0},
\end{equation}
where $[w^L]$ is an operator which extracts the coefficient of $w^L$ in the Taylor series of a function of $w$. According to the Lagrange inversion formula, if we can find a function $\phi(a,z)$ such that $w=a/\phi(a,z)$ whereby $\phi(0,z)\neq 0$, then
\begin{equation}
    \label{Lagrange-inversion-formula}
    \left[w^{L}\right]H(w,z)=\frac{1}{L}\left[a^{L-1}\right]\left(\frac{\partial H(a,z)}{\partial a}[\phi(a,z)]^{L}\right).
\end{equation}
The advantage of this formula is that the right-hand side is much easier to calculate than the left-hand side. According to Eq. (\ref{w-equation}), 
\begin{equation}
    w=\frac{a}{\phi(a,z)}\;\text{and}\;\phi(a,z)=z-1+\frac{1}{1-a}.
\end{equation}
By expanding the right-hand side of Eq. (\ref{Lagrange-inversion-formula}) and collecting the terms of the same order in $a$ yields $G_N(z)$ given by Eqs. (\ref{TASEP1-nascent-RNA-GN-exact}).

\section{Slow-switching conditions and their prevalence in mammalian genes}
\label{appendix_b}

We consider the TASEP in the absence of switching and with the initiation rate set to $\alpha$. We assume that the TASEP is initially in the steady state, i.e. the RNAP distribution is given by Eq. (\ref{TASEP1-exact-solution}). At time $t=0$, we set the initiation rate to zero, and observe the relaxation of the TASEP to the new steady state in which the lattice is empty. We denote by $T_{\text{off}}$ the mean time it takes the TASEP to reach that state. For the approximation in Eq. (\ref{TASEP2-conditional-0}) to hold, we require that 
\begin{equation}
    \label{TASEP2-slow-switching-kon}
    T_{\text{off}}\ll \frac{1}{k_{\text{on}}},
\end{equation}
where $1/k_{\text{on}}$ is the mean time that the gene spends in the off state. In order to compute $T_{\text{off}}$, we denote by $k$ the position of the leftmost RNAP on the lattice at $t=0$, and by $P_1(k)$ the probability distribution of $k$. The mean time $T_{\text{off}}$ is then given by 
\begin{equation}
    T_{\text{off}}=\sum_{k=1}^{L}P_1(k)\sum_{i=k}^{L}\frac{\rho_i}{J},
\end{equation}
where $\sum_{i=k}^{L}\rho_i/J$ is the mean time it takes the leftmost RNAP at the segment $k$ to leave the gene. The probability distribution $P_1(k)$ can be written as
\begin{equation}
    P_1(k)=\frac{\langle W\vert E^{k-1}DC^{L-k}\vert V\rangle}{Z_L},\quad k=1,\dots,L.
\end{equation}
For $k=L$, we get 
\begin{equation}
    P_1(L)=\frac{1}{Z_{L}}\left(\frac{\omega}{\alpha}\right)^{L-1}\frac{\omega}{\beta},
\end{equation}
where $Z_L$ is given by Eq. (\ref{TASEP1-Z-L-2}). For $k=1,\dots,L-1$, we use the following identity \cite{Derrida1993},
\begin{align}
    DC^{L-k}&=\sum_{i=0}^{L-k-1}B_{i+1,1}C^{L-k-i}\nonumber\\
    &+\sum_{i=2}^{L-k+1}B_{L-k,i-1}D^i,
\end{align}
where $B_{k,p}$ is defined in Eq. (\ref{TASEP1-B-k-p}). This gives
\begin{align}
    P_1(k)&=\left(\frac{\omega}{\alpha}\right)^{k-1}\sum_{i=0}^{L-k-1}B_{i+1,1}\frac{Z_{L-k-i}}{Z_L}\nonumber\\
    &+\left(\frac{\omega}{\alpha}\right)^{k-1}\sum_{i=2}^{L-k+1}B_{L-k,i-1}\frac{1}{Z_L}\left(\frac{\omega}{\beta}\right)^{i}.
\end{align}
If we assume that $\rho_i\approx\rho$, then $P_1(k)=(1-\rho)^{k-1}\rho$, in which case the expression for $T_{\text{off}}$ simplifies to  
\begin{equation}
    T_{\text{off}}=\frac{L}{\omega(1-\rho)}-\frac{1}{\omega\rho}\left[1-(1-\rho)^L\right].
\end{equation}

We now consider the TASEP in the absence of switching and with the initiation rate set to zero. We assume that the TASEP is initially in the steady state, i.e. the lattice is empty. At time $t=0$, we set the initiation rate to $\alpha$, and observe the relaxation of the TASEP to the new steady state in which the RNAP probability distribution is given by Eq. (\ref{TASEP1-exact-solution}). We denote by $T_{\text{on}}$ the mean time it takes the TASEP to reach that state. For the approximation in Eq. (\ref{TASEP2-conditional-1}) to hold, we require that
\begin{equation}
    \label{TASEP2-slow-switching-koff}
    T_{\text{on}}\ll \frac{1}{k_{\text{off}}}.
\end{equation}
The time-evolution of the TASEP from an empty lattice has been studied in the context of mRNA translation \cite{Nagar2011,Szavits2020-dynamics}, and more recently in the context of stochastic resetting \cite{Karthika2020}. In the initiation-limited regime, $T_{\text{on}}=T_{\text{el}}$, 
where $T_{\text{el}}\approx L/[\omega(1-\rho)]$ is the mean time of elongation and termination of a single RNAP. For the expression for $T_{\text{on}}$ in the termination-limited regime, we refer to Ref. \cite{Karthika2020}. Unfortunately, no analytical expression for $T_{\text{on}}$ is available on the coexistence line and in the elongation-limited regime. However, we know that the large-time relaxation of the TASEP is determined by the eigenvectors of the transition matrix with the largest real parts of the corresponding eigenvalues (note that the real parts of all eigenvalues are non-positive) \cite{deGier2005}. On the coexistence line, the spectral gap $\Delta$ (the difference between the zero eigenvalue corresponding to the steady state and the next largest eigenvalue) approaches zero as $L^{-2}$ for large $L$, signalling a diffusive relaxation time $\Delta^{-1}\propto L^2$, whereas in the elongation-limited regime the spectral gap approaches zero as $L^{-3/2}$ for large $L$ \cite{deGier2005}. Hence, we expect $T_{\text{on}}$ to scale with $L^2$ on the coexistence line, and with $L^{3/2}$ in the elongation-limited regime.

The conditions (\ref{TASEP2-slow-switching-kon}) and (\ref{TASEP2-slow-switching-koff}) were tested on a large dataset of the on and off rates inferred from single-cell RNA sequencing data in mouse fibroblasts \cite{larsson2019genomic}. Gene lengths for each gene in the dataset were obtained from the GRCm39 genome assembly for \textit{Mus musculus} published by Ensembl (release 106, Nov 2022) using \texttt{gffutils} package in Python. The data was analyzed for three values of the elongation rate ($0.6$, $1.8$ and $2.4$ kb/min) reported for this type of cells in Ref. \cite{Jonkers2014}. These elongation rates were measured by observing the depletion of RNAPs after stopping new transcription initiation, therefore we assumed that these values represent the effective elongation rates $\omega_{\text{eff}}=\omega(1-\rho)$, rather than the bare ones ($\omega$). The conditions (\ref{TASEP2-slow-switching-kon}) and (\ref{TASEP2-slow-switching-koff}) were tested assuming $T_{\text{on}}\simeq T_{\text{el}}$ and $T_{\text{off}}\simeq T_{\text{el}}$, where $T_{\text{el}}\approx L/[\omega(1-\rho)]$. A gene was considered to be in the slow-switching regime if both $k_{\text{on}}/T_{\text{el}}<0.2$ and $k_{\text{off}}/T_{\text{el}}<0.2$ conditions were satisfied.

Out of 3236 genes, $78$\% of the genes at $0.6$ kb/min, $96$\% of the genes at $1.8$ kb/min and $98$\% of the genes at $2.4$ kb/min satisfied the condition (\ref{TASEP2-slow-switching-kon}), meaning that for most genes the off state should last long enough for all RNAPs to leave the gene before the gene switches back to the on state. This also means that most of the genes will have only one transcriptional burst of RNAPs active on the gene. In contrast, the conditions (\ref{TASEP2-slow-switching-kon}) and (\ref{TASEP2-slow-switching-koff}) were simultaneously satisfied in $40$\% of the genes at $0.6$ kb/min, $62$\% of the genes at $1.8$ kb/min and $66$\% of the genes at $2.4$ kb/min.

\section{The waiting-time distribution \texorpdfstring{$h(t)$}{h(t)} between two successive termination events in the telegraph model with RNAP volume exclusion}
\label{appendix_c}

For the process described in Eq. (\ref{TASEP2-renewal-process}), we define the probability density functions $f_{0\rightarrow 1}(t)$, $f_{1\rightarrow 0}(t)$ and $f_{1\rightarrow 2}(t)$ as follows: $f_{0\rightarrow 1}(t)$ is the probability density function of the waiting time in the off state, after which the gene switches to the on state, $f_{1\rightarrow 0}(t)$ is the probability density function of the waiting time in the on state after which the gene switches to the off state, and $f_{1\rightarrow 2}(t)$ is the probability density function of the waiting time in the on state, after which the gene produces a nascent RNA (and remains in the on state). We denote their respective Laplace transforms by $\tilde{f}_{0\rightarrow 1}(s)$, $\tilde{f}_{1\rightarrow 0}(s)$ and $\tilde{f}_{1\rightarrow 2}(s)$. 

Using these definitions, the Laplace transform of $h(t)$, denoted by $\tilde{h}(s)$, can be written as
\begin{align}
    \tilde{h}(s)&=\sum_{n=0}^{\infty}[\tilde{f}_{1\rightarrow 0}(s)\tilde{f}_{0\rightarrow 1}(s)]^n \tilde{f}_{1\rightarrow 2}(s)\nonumber\\
    &=\frac{\tilde{f}_{1\rightarrow 2}(s)}{1-\tilde{f}_{1\rightarrow 0}(s)\tilde{f}_{0\rightarrow 1}(s)}.
\end{align}
The expression in the square brackets accounts for the progression through the cycle $G_{\text{on}}\rightarrow G_{\text{off}}\rightarrow G_{\text{on}}$ during which no nascent RNA is produced. The integer $n$ denotes the number of such cycles before a nascent RNA is produced, and the summation goes over all $n\geq 0$. 

The remaining problem is to compute $\tilde{f}_{0\rightarrow 1}(s)$, $\tilde{f}_{1\rightarrow 0}(s)$ and $\tilde{f}_{1\rightarrow 2}(s)$. Since the waiting time in the off state is exponentially distributed, and there is only one reaction that leads from the off state, $\tilde{f}_{0 \rightarrow 1}(s)$ is given by
\begin{equation}
    \tilde{f}_{0 \rightarrow 1}(s)=\int_{0}^{\infty}k_{\text{on}}e^{-(s+k_{\text{on}})t}=\frac{k_{\text{on}}}{s+k_{\text{on}}}.
\end{equation}
From the definitions of $\tilde{f}_{1\rightarrow 0}(s)$ and $\tilde{f}_{1\rightarrow 2}(s)$, it follows that
\begin{align}
    & f_{1\rightarrow 0}(t)=P(1\rightarrow 0)p(t),\\
    & f_{1\rightarrow 2}(t)=P(1\rightarrow 2)p(t),
\end{align}
where $p(t)$ is the probability density function of the waiting time in the on state until the gene either switches to the off state or produces a nascent RNA, $P(1\rightarrow 0)$ is the probability that the gene switches to off state, and $P(1\rightarrow 2)$ is the probability that the gene produces a nascent RNA in the on state. We note that
\begin{equation}
    \int_{0}^{\infty}p(t)dt=1, \quad P(1\rightarrow 0)+P(1\rightarrow 2)=1.
\end{equation}
In order to find $p(t)$, we denote by $t_1$ a random variable whose pdf is given by $f_1(t_1)$ and by $t_2$ a random variable whose pdf is given by $f_2(t)$, where $f_1(t_1)$ and $f_2(t_2)$ are given by
\begin{equation}
    f_1(t_1)=k_{\text{off}}e^{-k_{\text{off}}t},\quad f_2(t_2)=f_{\text{ter,as}}(t_2),
\end{equation}
and $f_{\text{ter,as}}(t)$ is given by Eq. (\ref{TASEP1-wt-infinite}). The waiting time $t$ in the on state before either switching to the off state or producing a nascent RNA is the minimum of $t_1$ and $t_2$, $t=\text{min}\{t_1,t_2\}$. Hence,
\begin{equation}
    P(t>x)=e^{-k_{\text{off}}} P_2(t_2>x).
\end{equation}
From here, we get $p(x)$ by taking the derivative of $P(t\leq x)=1-P(t>x)$,
\begin{equation}
    p(x)=e^{-k_{\text{off}}x}\left[k_{\text{off}}P_2(t_2>x)+f_2(x)\right].
\end{equation}
The probability $P(1\rightarrow 0)$ can be found from the condition that $t_1<t_2$,
\begin{align}
    P(1\rightarrow 0)&=P(t_1<t_2)=\int_{0}^{\infty}dt_1 f_1(t_1)\int_{t_1}^{\infty}dt_2 f_2(t_2)\nonumber\\
    &=k_{\text{off}}\int_{0}^{\infty}dt_1 e^{-k_{\text{off}}t_1}P_2(t_2>t_1).
\end{align}
Similarly, $P(1\rightarrow 2)$ follows from the condition that $t_2<t_1$,
\begin{align}
    P(1\rightarrow 2)&=P(t_2<t_1)=\int_{0}^{\infty}dt_2 f_2(t_2)\int_{t_2}^{\infty}dt_1 f_1(t_1)\nonumber\\
    &=\int_{0}^{\infty}dt_2 e^{-k_{\text{off}}t_2}f_2(t_2).
\end{align}
It is convenient to introduce $\lambda_1=\omega\nu$ and $\lambda_2=\omega(1-\nu)$, in which case $f_{\text{ter,as}}(t)$ can be written as
\begin{equation}
    f_{\text{ter,as}}(t)=\frac{\lambda_1\lambda_2}{\lambda_1-\lambda_2}\left(e^{-\lambda_2 t}-e^{-\lambda_1 t}\right).
\end{equation}
Inserting this expression into the expressions for $p(t)$, $P(1\rightarrow 0)$ and $P(1\rightarrow 2)$ yields,
\begin{align}
    & p(t)=\frac{\lambda_2 (k_{\text{off}}+\lambda_1)}{\lambda_2-\lambda_1}e^{-(k_{\text{off}}+\lambda_1)t}\nonumber\\
    &\qquad-\frac{\lambda_1 (k_{\text{off}}+\lambda_2)}{\lambda_2-\lambda_1}e^{-(k_{\text{off}}+\lambda_2)t},\label{p-t}\\
    & P(1\rightarrow 0)=\frac{k_{\text{off}}(k_{\text{off}}+\lambda_1+\lambda_2)}{(k_{\text{off}}+\lambda_1)(k_{\text{off}}+\lambda_2)},\\
    & P(1\rightarrow 2)=\frac{\lambda_1\lambda_2}{(k_{\text{off}}+\lambda_1)(k_{\text{off}}+\lambda_2)}.
\end{align}
From Eq. (\ref{p-t}), it follows that the Laplace transform of $p(t)$ is given by
\begin{equation}
    \tilde{p}(s)=\frac{k_{\text{off}}s+(k_{\text{off}}+\lambda_1)(k_{\text{off}}+\lambda_2)}{(k_{\text{off}}+\lambda_1+s)(k_{\text{off}}+\lambda_2+s)}.
\end{equation}
Inserting $\tilde{p}(s)$ into $\tilde{f}_{1\rightarrow 0}(s)=P(1\rightarrow 0)\tilde{p}(s)$ and $\tilde{f}_{1\rightarrow 2}(s)=P(1\rightarrow 2)\tilde{p}(s)$, and then in the expression for $\tilde{h}(s)$ yields
\begin{equation}
    \label{TASEP2-wt-infinite-LT}
    \tilde{h}(s)=\frac{A(s)}{B(s)},
\end{equation}
where $A(s)$ and $B(s)$ are given by
\begin{widetext}
\begin{subequations}
    \begin{align}
    & A(s)=\lambda_1\lambda_2(s+k_{\text{on}})[k_{\text{off}} s+(k_{\text{off}}+\lambda_1)(k_{\text{off}}+\lambda_2)]\\
    & B(s)=(k_{\text{off}}+\lambda_1)(k_{\text{off}}+\lambda_2)(s+k_{\text{off}}+\lambda_1)(s+k_{\text{off}}+\lambda_2)(s+k_{\text{on}})\nonumber\\
    &\qquad-k_{\text{on}}k_{\text{off}}(k_{\text{off}}+\lambda_1+\lambda_2)[k_{\text{off}} s+(k_{\text{off}}+\lambda_1)(k_{\text{off}}+\lambda_2)].
\end{align} 
\end{subequations}
\end{widetext}
This expression can be inverted to get $h(t)$ by finding the roots of $B(s)$ and performing the partial fraction decomposition.


\begin{thebibliography}{87}%
	\makeatletter
	\providecommand \@ifxundefined [1]{%
		\@ifx{#1\undefined}
	}%
	\providecommand \@ifnum [1]{%
		\ifnum #1\expandafter \@firstoftwo
		\else \expandafter \@secondoftwo
		\fi
	}%
	\providecommand \@ifx [1]{%
		\ifx #1\expandafter \@firstoftwo
		\else \expandafter \@secondoftwo
		\fi
	}%
	\providecommand \natexlab [1]{#1}%
	\providecommand \enquote  [1]{``#1''}%
	\providecommand \bibnamefont  [1]{#1}%
	\providecommand \bibfnamefont [1]{#1}%
	\providecommand \citenamefont [1]{#1}%
	\providecommand \href@noop [0]{\@secondoftwo}%
	\providecommand \href [0]{\begingroup \@sanitize@url \@href}%
	\providecommand \@href[1]{\@@startlink{#1}\@@href}%
	\providecommand \@@href[1]{\endgroup#1\@@endlink}%
	\providecommand \@sanitize@url [0]{\catcode `\\12\catcode `\$12\catcode
		`\&12\catcode `\#12\catcode `\^12\catcode `\_12\catcode `\%12\relax}%
	\providecommand \@@startlink[1]{}%
	\providecommand \@@endlink[0]{}%
	\providecommand \url  [0]{\begingroup\@sanitize@url \@url }%
	\providecommand \@url [1]{\endgroup\@href {#1}{\urlprefix }}%
	\providecommand \urlprefix  [0]{URL }%
	\providecommand \Eprint [0]{\href }%
	\providecommand \doibase [0]{https://doi.org/}%
	\providecommand \selectlanguage [0]{\@gobble}%
	\providecommand \bibinfo  [0]{\@secondoftwo}%
	\providecommand \bibfield  [0]{\@secondoftwo}%
	\providecommand \translation [1]{[#1]}%
	\providecommand \BibitemOpen [0]{}%
	\providecommand \bibitemStop [0]{}%
	\providecommand \bibitemNoStop [0]{.\EOS\space}%
	\providecommand \EOS [0]{\spacefactor3000\relax}%
	\providecommand \BibitemShut  [1]{\csname bibitem#1\endcsname}%
	\let\auto@bib@innerbib\@empty
	\bibitem [{\citenamefont {Paulsson}(2005)}]{paulsson2005models}%
	\BibitemOpen
	\bibfield  {author} {\bibinfo {author} {\bibfnamefont {J.}~\bibnamefont
			{Paulsson}},\ }\bibfield  {title} {\bibinfo {title} {Models of stochastic
			gene expression},\ }\href
	{https://doi.org/https://doi.org/10.1016/j.plrev.2005.03.003} {\bibfield
		{journal} {\bibinfo  {journal} {Physics of life reviews}\ }\textbf {\bibinfo
			{volume} {2}},\ \bibinfo {pages} {157} (\bibinfo {year} {2005})}\BibitemShut
	{NoStop}%
	\bibitem [{\citenamefont {Friedman}\ \emph {et~al.}(2006)\citenamefont
		{Friedman}, \citenamefont {Cai},\ and\ \citenamefont
		{Xie}}]{friedman2006linking}%
	\BibitemOpen
	\bibfield  {author} {\bibinfo {author} {\bibfnamefont {N.}~\bibnamefont
			{Friedman}}, \bibinfo {author} {\bibfnamefont {L.}~\bibnamefont {Cai}},\ and\
		\bibinfo {author} {\bibfnamefont {X.~S.}\ \bibnamefont {Xie}},\ }\bibfield
	{title} {\bibinfo {title} {Linking stochastic dynamics to population
			distribution: an analytical framework of gene expression},\ }\href
	{https://doi.org/10.1103/PhysRevLett.97.168302} {\bibfield  {journal}
		{\bibinfo  {journal} {Physical review letters}\ }\textbf {\bibinfo {volume}
			{97}},\ \bibinfo {pages} {168302} (\bibinfo {year} {2006})}\BibitemShut
	{NoStop}%
	\bibitem [{\citenamefont {Shahrezaei}\ and\ \citenamefont
		{Swain}(2008)}]{shahrezaei2008analytical}%
	\BibitemOpen
	\bibfield  {author} {\bibinfo {author} {\bibfnamefont {V.}~\bibnamefont
			{Shahrezaei}}\ and\ \bibinfo {author} {\bibfnamefont {P.~S.}\ \bibnamefont
			{Swain}},\ }\bibfield  {title} {\bibinfo {title} {Analytical distributions
			for stochastic gene expression},\ }\href
	{https://doi.org/10.1073/pnas.0803850105} {\bibfield  {journal} {\bibinfo
			{journal} {Proceedings of the National Academy of Sciences}\ }\textbf
		{\bibinfo {volume} {105}},\ \bibinfo {pages} {17256} (\bibinfo {year}
		{2008})}\BibitemShut {NoStop}%
	\bibitem [{\citenamefont {Kumar}\ \emph {et~al.}(2014)\citenamefont {Kumar},
		\citenamefont {Platini},\ and\ \citenamefont {Kulkarni}}]{kumar2014exact}%
	\BibitemOpen
	\bibfield  {author} {\bibinfo {author} {\bibfnamefont {N.}~\bibnamefont
			{Kumar}}, \bibinfo {author} {\bibfnamefont {T.}~\bibnamefont {Platini}},\
		and\ \bibinfo {author} {\bibfnamefont {R.~V.}\ \bibnamefont {Kulkarni}},\
	}\bibfield  {title} {\bibinfo {title} {Exact distributions for stochastic
			gene expression models with bursting and feedback},\ }\href
	{https://doi.org/10.1103/PhysRevLett.113.268105} {\bibfield  {journal}
		{\bibinfo  {journal} {Physical review letters}\ }\textbf {\bibinfo {volume}
			{113}},\ \bibinfo {pages} {268105} (\bibinfo {year} {2014})}\BibitemShut
	{NoStop}%
	\bibitem [{\citenamefont {Lim}\ \emph {et~al.}(2015)\citenamefont {Lim},
		\citenamefont {Kim}, \citenamefont {Park}, \citenamefont {Yang},
		\citenamefont {Song}, \citenamefont {Chang}, \citenamefont {Lee},\ and\
		\citenamefont {Sung}}]{Lim2015}%
	\BibitemOpen
	\bibfield  {author} {\bibinfo {author} {\bibfnamefont {Y.~R.}\ \bibnamefont
			{Lim}}, \bibinfo {author} {\bibfnamefont {J.-H.}\ \bibnamefont {Kim}},
		\bibinfo {author} {\bibfnamefont {S.~J.}\ \bibnamefont {Park}}, \bibinfo
		{author} {\bibfnamefont {G.-S.}\ \bibnamefont {Yang}}, \bibinfo {author}
		{\bibfnamefont {S.}~\bibnamefont {Song}}, \bibinfo {author} {\bibfnamefont
			{S.-K.}\ \bibnamefont {Chang}}, \bibinfo {author} {\bibfnamefont {N.~K.}\
			\bibnamefont {Lee}},\ and\ \bibinfo {author} {\bibfnamefont {J.}~\bibnamefont
			{Sung}},\ }\bibfield  {title} {\bibinfo {title} {Quantitative understanding
			of probabilistic behavior of living cells operated by vibrant intracellular
			networks},\ }\href {https://doi.org/10.1103/PhysRevX.5.031014} {\bibfield
		{journal} {\bibinfo  {journal} {Phys. Rev. X}\ }\textbf {\bibinfo {volume}
			{5}},\ \bibinfo {pages} {031014} (\bibinfo {year} {2015})}\BibitemShut
	{NoStop}%
	\bibitem [{\citenamefont {Cao}\ and\ \citenamefont
		{Grima}(2020)}]{cao2020analytical}%
	\BibitemOpen
	\bibfield  {author} {\bibinfo {author} {\bibfnamefont {Z.}~\bibnamefont
			{Cao}}\ and\ \bibinfo {author} {\bibfnamefont {R.}~\bibnamefont {Grima}},\
	}\bibfield  {title} {\bibinfo {title} {Analytical distributions for detailed
			models of stochastic gene expression in eukaryotic cells},\ }\href
	{https://doi.org/10.1073/pnas.1910888117} {\bibfield  {journal} {\bibinfo
			{journal} {Proceedings of the National Academy of Sciences}\ }\textbf
		{\bibinfo {volume} {117}},\ \bibinfo {pages} {4682} (\bibinfo {year}
		{2020})}\BibitemShut {NoStop}%
	\bibitem [{\citenamefont {Ham}\ \emph {et~al.}(2020)\citenamefont {Ham},
		\citenamefont {Brackston},\ and\ \citenamefont {Stumpf}}]{ham2020extrinsic}%
	\BibitemOpen
	\bibfield  {author} {\bibinfo {author} {\bibfnamefont {L.}~\bibnamefont
			{Ham}}, \bibinfo {author} {\bibfnamefont {R.~D.}\ \bibnamefont {Brackston}},\
		and\ \bibinfo {author} {\bibfnamefont {M.~P.}\ \bibnamefont {Stumpf}},\
	}\bibfield  {title} {\bibinfo {title} {Extrinsic noise and heavy-tailed laws
			in gene expression},\ }\href {https://doi.org/10.1103/PhysRevLett.124.108101}
	{\bibfield  {journal} {\bibinfo  {journal} {Physical review letters}\
		}\textbf {\bibinfo {volume} {124}},\ \bibinfo {pages} {108101} (\bibinfo
		{year} {2020})}\BibitemShut {NoStop}%
	\bibitem [{\citenamefont {Jia}\ and\ \citenamefont {Grima}(2021)}]{Jia2021}%
	\BibitemOpen
	\bibfield  {author} {\bibinfo {author} {\bibfnamefont {C.}~\bibnamefont
			{Jia}}\ and\ \bibinfo {author} {\bibfnamefont {R.}~\bibnamefont {Grima}},\
	}\bibfield  {title} {\bibinfo {title} {Frequency domain analysis of
			fluctuations of {mRNA} and protein copy numbers within a cell lineage: Theory
			and experimental validation},\ }\href
	{https://doi.org/10.1103/PhysRevX.11.021032} {\bibfield  {journal} {\bibinfo
			{journal} {Phys. Rev. X}\ }\textbf {\bibinfo {volume} {11}},\ \bibinfo
		{pages} {021032} (\bibinfo {year} {2021})}\BibitemShut {NoStop}%
	\bibitem [{\citenamefont {Gupta}\ and\ \citenamefont
		{Khammash}(2022)}]{gupta2022frequency}%
	\BibitemOpen
	\bibfield  {author} {\bibinfo {author} {\bibfnamefont {A.}~\bibnamefont
			{Gupta}}\ and\ \bibinfo {author} {\bibfnamefont {M.}~\bibnamefont
			{Khammash}},\ }\bibfield  {title} {\bibinfo {title} {Frequency spectra and
			the color of cellular noise},\ }\href
	{https://doi.org/10.1038/s41467-022-31263-x} {\bibfield  {journal} {\bibinfo
			{journal} {Nature communications}\ }\textbf {\bibinfo {volume} {13}},\
		\bibinfo {pages} {4305} (\bibinfo {year} {2022})}\BibitemShut {NoStop}%
	\bibitem [{\citenamefont {So}\ \emph {et~al.}(2011)\citenamefont {So},
		\citenamefont {Ghosh}, \citenamefont {Zong}, \citenamefont {Sep{\'u}lveda},
		\citenamefont {Segev},\ and\ \citenamefont {Golding}}]{so2011general}%
	\BibitemOpen
	\bibfield  {author} {\bibinfo {author} {\bibfnamefont {L.-H.}\ \bibnamefont
			{So}}, \bibinfo {author} {\bibfnamefont {A.}~\bibnamefont {Ghosh}}, \bibinfo
		{author} {\bibfnamefont {C.}~\bibnamefont {Zong}}, \bibinfo {author}
		{\bibfnamefont {L.~A.}\ \bibnamefont {Sep{\'u}lveda}}, \bibinfo {author}
		{\bibfnamefont {R.}~\bibnamefont {Segev}},\ and\ \bibinfo {author}
		{\bibfnamefont {I.}~\bibnamefont {Golding}},\ }\bibfield  {title} {\bibinfo
		{title} {General properties of transcriptional time series in escherichia
			coli},\ }\href {https://doi.org/10.1038/ng.821} {\bibfield  {journal}
		{\bibinfo  {journal} {Nature genetics}\ }\textbf {\bibinfo {volume} {43}},\
		\bibinfo {pages} {554} (\bibinfo {year} {2011})}\BibitemShut {NoStop}%
	\bibitem [{\citenamefont {Zenklusen}\ \emph {et~al.}(2008)\citenamefont
		{Zenklusen}, \citenamefont {Larson},\ and\ \citenamefont
		{Singer}}]{zenklusen2008single}%
	\BibitemOpen
	\bibfield  {author} {\bibinfo {author} {\bibfnamefont {D.}~\bibnamefont
			{Zenklusen}}, \bibinfo {author} {\bibfnamefont {D.~R.}\ \bibnamefont
			{Larson}},\ and\ \bibinfo {author} {\bibfnamefont {R.~H.}\ \bibnamefont
			{Singer}},\ }\bibfield  {title} {\bibinfo {title} {Single-{RNA} counting
			reveals alternative modes of gene expression in yeast},\ }\href
	{https://doi.org/10.1038/nsmb.1514} {\bibfield  {journal} {\bibinfo
			{journal} {Nature structural \& molecular biology}\ }\textbf {\bibinfo
			{volume} {15}},\ \bibinfo {pages} {1263} (\bibinfo {year}
		{2008})}\BibitemShut {NoStop}%
	\bibitem [{\citenamefont {Suter}\ \emph {et~al.}(2011)\citenamefont {Suter},
		\citenamefont {Molina}, \citenamefont {Gatfield}, \citenamefont {Schneider},
		\citenamefont {Schibler},\ and\ \citenamefont {Naef}}]{suter2011mammalian}%
	\BibitemOpen
	\bibfield  {author} {\bibinfo {author} {\bibfnamefont {D.~M.}\ \bibnamefont
			{Suter}}, \bibinfo {author} {\bibfnamefont {N.}~\bibnamefont {Molina}},
		\bibinfo {author} {\bibfnamefont {D.}~\bibnamefont {Gatfield}}, \bibinfo
		{author} {\bibfnamefont {K.}~\bibnamefont {Schneider}}, \bibinfo {author}
		{\bibfnamefont {U.}~\bibnamefont {Schibler}},\ and\ \bibinfo {author}
		{\bibfnamefont {F.}~\bibnamefont {Naef}},\ }\bibfield  {title} {\bibinfo
		{title} {Mammalian genes are transcribed with widely different bursting
			kinetics},\ }\href {https://doi.org/10.1126/science.1198817} {\bibfield
		{journal} {\bibinfo  {journal} {science}\ }\textbf {\bibinfo {volume}
			{332}},\ \bibinfo {pages} {472} (\bibinfo {year} {2011})}\BibitemShut
	{NoStop}%
	\bibitem [{\citenamefont {Donovan}\ \emph {et~al.}(2019)\citenamefont
		{Donovan}, \citenamefont {Huynh}, \citenamefont {Ball}, \citenamefont
		{Patel}, \citenamefont {Poirier}, \citenamefont {Larson}, \citenamefont
		{Ferguson},\ and\ \citenamefont {Lenstra}}]{donovan2019live}%
	\BibitemOpen
	\bibfield  {author} {\bibinfo {author} {\bibfnamefont {B.~T.}\ \bibnamefont
			{Donovan}}, \bibinfo {author} {\bibfnamefont {A.}~\bibnamefont {Huynh}},
		\bibinfo {author} {\bibfnamefont {D.~A.}\ \bibnamefont {Ball}}, \bibinfo
		{author} {\bibfnamefont {H.~P.}\ \bibnamefont {Patel}}, \bibinfo {author}
		{\bibfnamefont {M.~G.}\ \bibnamefont {Poirier}}, \bibinfo {author}
		{\bibfnamefont {D.~R.}\ \bibnamefont {Larson}}, \bibinfo {author}
		{\bibfnamefont {M.~L.}\ \bibnamefont {Ferguson}},\ and\ \bibinfo {author}
		{\bibfnamefont {T.~L.}\ \bibnamefont {Lenstra}},\ }\bibfield  {title}
	{\bibinfo {title} {Live-cell imaging reveals the interplay between
			transcription factors, nucleosomes, and bursting},\ }\href
	{https://doi.org/https://doi.org/10.15252/embj.2018100809} {\bibfield
		{journal} {\bibinfo  {journal} {The EMBO journal}\ }\textbf {\bibinfo
			{volume} {38}},\ \bibinfo {pages} {e100809} (\bibinfo {year}
		{2019})}\BibitemShut {NoStop}%
	\bibitem [{\citenamefont {Rao}\ \emph {et~al.}(2002)\citenamefont {Rao},
		\citenamefont {Wolf},\ and\ \citenamefont {Arkin}}]{rao2002control}%
	\BibitemOpen
	\bibfield  {author} {\bibinfo {author} {\bibfnamefont {C.~V.}\ \bibnamefont
			{Rao}}, \bibinfo {author} {\bibfnamefont {D.~M.}\ \bibnamefont {Wolf}},\ and\
		\bibinfo {author} {\bibfnamefont {A.~P.}\ \bibnamefont {Arkin}},\ }\bibfield
	{title} {\bibinfo {title} {Control, exploitation and tolerance of
			intracellular noise},\ }\href {https://doi.org/10.1038/nature01258}
	{\bibfield  {journal} {\bibinfo  {journal} {Nature}\ }\textbf {\bibinfo
			{volume} {420}},\ \bibinfo {pages} {231} (\bibinfo {year}
		{2002})}\BibitemShut {NoStop}%
	\bibitem [{\citenamefont {Singh}\ and\ \citenamefont
		{Hespanha}(2009)}]{singh2009optimal}%
	\BibitemOpen
	\bibfield  {author} {\bibinfo {author} {\bibfnamefont {A.}~\bibnamefont
			{Singh}}\ and\ \bibinfo {author} {\bibfnamefont {J.~P.}\ \bibnamefont
			{Hespanha}},\ }\bibfield  {title} {\bibinfo {title} {Optimal feedback
			strength for noise suppression in autoregulatory gene networks},\ }\href
	{https://doi.org/https://doi.org/10.1016/j.bpj.2009.02.064} {\bibfield
		{journal} {\bibinfo  {journal} {Biophysical journal}\ }\textbf {\bibinfo
			{volume} {96}},\ \bibinfo {pages} {4013} (\bibinfo {year}
		{2009})}\BibitemShut {NoStop}%
	\bibitem [{\citenamefont {Bal{\'a}zsi}\ \emph {et~al.}(2011)\citenamefont
		{Bal{\'a}zsi}, \citenamefont {Van~Oudenaarden},\ and\ \citenamefont
		{Collins}}]{balazsi2011cellular}%
	\BibitemOpen
	\bibfield  {author} {\bibinfo {author} {\bibfnamefont {G.}~\bibnamefont
			{Bal{\'a}zsi}}, \bibinfo {author} {\bibfnamefont {A.}~\bibnamefont
			{Van~Oudenaarden}},\ and\ \bibinfo {author} {\bibfnamefont {J.~J.}\
			\bibnamefont {Collins}},\ }\bibfield  {title} {\bibinfo {title} {Cellular
			decision making and biological noise: from microbes to mammals},\ }\href
	{https://doi.org/https://doi.org/10.1016/j.cell.2011.01.030} {\bibfield
		{journal} {\bibinfo  {journal} {Cell}\ }\textbf {\bibinfo {volume} {144}},\
		\bibinfo {pages} {910} (\bibinfo {year} {2011})}\BibitemShut {NoStop}%
	\bibitem [{\citenamefont {Kellogg}\ and\ \citenamefont
		{Tay}(2015)}]{kellogg2015noise}%
	\BibitemOpen
	\bibfield  {author} {\bibinfo {author} {\bibfnamefont {R.~A.}\ \bibnamefont
			{Kellogg}}\ and\ \bibinfo {author} {\bibfnamefont {S.}~\bibnamefont {Tay}},\
	}\bibfield  {title} {\bibinfo {title} {Noise facilitates transcriptional
			control under dynamic inputs},\ }\href
	{https://doi.org/https://doi.org/10.1016/j.cell.2015.01.013} {\bibfield
		{journal} {\bibinfo  {journal} {Cell}\ }\textbf {\bibinfo {volume} {160}},\
		\bibinfo {pages} {381} (\bibinfo {year} {2015})}\BibitemShut {NoStop}%
	\bibitem [{\citenamefont {Peccoud}\ and\ \citenamefont
		{Ycart}(1995)}]{Peccoud1995}%
	\BibitemOpen
	\bibfield  {author} {\bibinfo {author} {\bibfnamefont {J.}~\bibnamefont
			{Peccoud}}\ and\ \bibinfo {author} {\bibfnamefont {B.}~\bibnamefont
			{Ycart}},\ }\bibfield  {title} {\bibinfo {title} {Markovian modeling of
			gene-product synthesis},\ }\href
	{https://doi.org/https://doi.org/10.1006/tpbi.1995.1027} {\bibfield
		{journal} {\bibinfo  {journal} {Theoretical Population Biology}\ }\textbf
		{\bibinfo {volume} {48}},\ \bibinfo {pages} {222} (\bibinfo {year}
		{1995})}\BibitemShut {NoStop}%
	\bibitem [{\citenamefont {Iyer-Biswas}\ \emph {et~al.}(2009)\citenamefont
		{Iyer-Biswas}, \citenamefont {Hayot},\ and\ \citenamefont
		{Jayaprakash}}]{iyer2009stochasticity}%
	\BibitemOpen
	\bibfield  {author} {\bibinfo {author} {\bibfnamefont {S.}~\bibnamefont
			{Iyer-Biswas}}, \bibinfo {author} {\bibfnamefont {F.}~\bibnamefont {Hayot}},\
		and\ \bibinfo {author} {\bibfnamefont {C.}~\bibnamefont {Jayaprakash}},\
	}\bibfield  {title} {\bibinfo {title} {Stochasticity of gene products from
			transcriptional pulsing},\ }\href
	{https://doi.org/10.1103/PhysRevE.79.031911} {\bibfield  {journal} {\bibinfo
			{journal} {Physical Review E}\ }\textbf {\bibinfo {volume} {79}},\ \bibinfo
		{pages} {031911} (\bibinfo {year} {2009})}\BibitemShut {NoStop}%
	\bibitem [{\citenamefont {Halpern}\ \emph {et~al.}(2015)\citenamefont
		{Halpern}, \citenamefont {Tanami}, \citenamefont {Landen}, \citenamefont
		{Chapal}, \citenamefont {Szlak}, \citenamefont {Hutzler}, \citenamefont
		{Nizhberg},\ and\ \citenamefont {Itzkovitz}}]{halpern2015bursty}%
	\BibitemOpen
	\bibfield  {author} {\bibinfo {author} {\bibfnamefont {K.~B.}\ \bibnamefont
			{Halpern}}, \bibinfo {author} {\bibfnamefont {S.}~\bibnamefont {Tanami}},
		\bibinfo {author} {\bibfnamefont {S.}~\bibnamefont {Landen}}, \bibinfo
		{author} {\bibfnamefont {M.}~\bibnamefont {Chapal}}, \bibinfo {author}
		{\bibfnamefont {L.}~\bibnamefont {Szlak}}, \bibinfo {author} {\bibfnamefont
			{A.}~\bibnamefont {Hutzler}}, \bibinfo {author} {\bibfnamefont
			{A.}~\bibnamefont {Nizhberg}},\ and\ \bibinfo {author} {\bibfnamefont
			{S.}~\bibnamefont {Itzkovitz}},\ }\bibfield  {title} {\bibinfo {title}
		{Bursty gene expression in the intact mammalian liver},\ }\href
	{https://doi.org/https://doi.org/10.1016/j.molcel.2015.01.027} {\bibfield
		{journal} {\bibinfo  {journal} {Molecular cell}\ }\textbf {\bibinfo {volume}
			{58}},\ \bibinfo {pages} {147} (\bibinfo {year} {2015})}\BibitemShut
	{NoStop}%
	\bibitem [{\citenamefont {Kim}\ and\ \citenamefont
		{Marioni}(2013)}]{kim2013inferring}%
	\BibitemOpen
	\bibfield  {author} {\bibinfo {author} {\bibfnamefont {J.~K.}\ \bibnamefont
			{Kim}}\ and\ \bibinfo {author} {\bibfnamefont {J.~C.}\ \bibnamefont
			{Marioni}},\ }\bibfield  {title} {\bibinfo {title} {Inferring the kinetics of
			stochastic gene expression from single-cell {RNA}-sequencing data},\ }\href
	{https://doi.org/10.1186/gb-2013-14-1-r7} {\bibfield  {journal} {\bibinfo
			{journal} {Genome biology}\ }\textbf {\bibinfo {volume} {14}},\ \bibinfo
		{pages} {1} (\bibinfo {year} {2013})}\BibitemShut {NoStop}%
	\bibitem [{\citenamefont {Larsson}\ \emph {et~al.}(2019)\citenamefont
		{Larsson}, \citenamefont {Johnsson}, \citenamefont {Hagemann-Jensen},
		\citenamefont {Hartmanis}, \citenamefont {Faridani}, \citenamefont {Reinius},
		\citenamefont {Segerstolpe}, \citenamefont {Rivera}, \citenamefont {Ren},\
		and\ \citenamefont {Sandberg}}]{larsson2019genomic}%
	\BibitemOpen
	\bibfield  {author} {\bibinfo {author} {\bibfnamefont {A.~J.}\ \bibnamefont
			{Larsson}}, \bibinfo {author} {\bibfnamefont {P.}~\bibnamefont {Johnsson}},
		\bibinfo {author} {\bibfnamefont {M.}~\bibnamefont {Hagemann-Jensen}},
		\bibinfo {author} {\bibfnamefont {L.}~\bibnamefont {Hartmanis}}, \bibinfo
		{author} {\bibfnamefont {O.~R.}\ \bibnamefont {Faridani}}, \bibinfo {author}
		{\bibfnamefont {B.}~\bibnamefont {Reinius}}, \bibinfo {author} {\bibfnamefont
			{{\AA}.}~\bibnamefont {Segerstolpe}}, \bibinfo {author} {\bibfnamefont
			{C.~M.}\ \bibnamefont {Rivera}}, \bibinfo {author} {\bibfnamefont
			{B.}~\bibnamefont {Ren}},\ and\ \bibinfo {author} {\bibfnamefont
			{R.}~\bibnamefont {Sandberg}},\ }\bibfield  {title} {\bibinfo {title}
		{Genomic encoding of transcriptional burst kinetics},\ }\href
	{https://doi.org/10.1038/s41586-018-0836-1} {\bibfield  {journal} {\bibinfo
			{journal} {Nature}\ }\textbf {\bibinfo {volume} {565}},\ \bibinfo {pages}
		{251} (\bibinfo {year} {2019})}\BibitemShut {NoStop}%
	\bibitem [{\citenamefont {Choubey}\ \emph {et~al.}(2015)\citenamefont
		{Choubey}, \citenamefont {Kondev},\ and\ \citenamefont
		{Sanchez}}]{Choubey2015}%
	\BibitemOpen
	\bibfield  {author} {\bibinfo {author} {\bibfnamefont {S.}~\bibnamefont
			{Choubey}}, \bibinfo {author} {\bibfnamefont {J.}~\bibnamefont {Kondev}},\
		and\ \bibinfo {author} {\bibfnamefont {A.}~\bibnamefont {Sanchez}},\
	}\bibfield  {title} {\bibinfo {title} {Deciphering transcriptional dynamics
			in vivo by counting nascent {RNA} molecules},\ }\href
	{https://doi.org/10.1371/journal.pcbi.1004345} {\bibfield  {journal}
		{\bibinfo  {journal} {PLOS Computational Biology}\ }\textbf {\bibinfo
			{volume} {11}},\ \bibinfo {pages} {1} (\bibinfo {year} {2015})}\BibitemShut
	{NoStop}%
	\bibitem [{\citenamefont {Xu}\ \emph {et~al.}(2016)\citenamefont {Xu},
		\citenamefont {Skinner}, \citenamefont {Sokac},\ and\ \citenamefont
		{Golding}}]{Xu2016}%
	\BibitemOpen
	\bibfield  {author} {\bibinfo {author} {\bibfnamefont {H.}~\bibnamefont
			{Xu}}, \bibinfo {author} {\bibfnamefont {S.~O.}\ \bibnamefont {Skinner}},
		\bibinfo {author} {\bibfnamefont {A.~M.}\ \bibnamefont {Sokac}},\ and\
		\bibinfo {author} {\bibfnamefont {I.}~\bibnamefont {Golding}},\ }\bibfield
	{title} {\bibinfo {title} {Stochastic kinetics of nascent {RNA}},\ }\href
	{https://doi.org/10.1103/PhysRevLett.117.128101} {\bibfield  {journal}
		{\bibinfo  {journal} {Phys. Rev. Lett.}\ }\textbf {\bibinfo {volume} {117}},\
		\bibinfo {pages} {128101} (\bibinfo {year} {2016})}\BibitemShut {NoStop}%
	\bibitem [{\citenamefont {Fu}\ \emph {et~al.}(2022)\citenamefont {Fu},
		\citenamefont {Patel}, \citenamefont {Coppola}, \citenamefont {Xu},
		\citenamefont {Cao}, \citenamefont {Lenstra},\ and\ \citenamefont
		{Grima}}]{fu2022quantifying}%
	\BibitemOpen
	\bibfield  {author} {\bibinfo {author} {\bibfnamefont {X.}~\bibnamefont
			{Fu}}, \bibinfo {author} {\bibfnamefont {H.~P.}\ \bibnamefont {Patel}},
		\bibinfo {author} {\bibfnamefont {S.}~\bibnamefont {Coppola}}, \bibinfo
		{author} {\bibfnamefont {L.}~\bibnamefont {Xu}}, \bibinfo {author}
		{\bibfnamefont {Z.}~\bibnamefont {Cao}}, \bibinfo {author} {\bibfnamefont
			{T.~L.}\ \bibnamefont {Lenstra}},\ and\ \bibinfo {author} {\bibfnamefont
			{R.}~\bibnamefont {Grima}},\ }\bibfield  {title} {\bibinfo {title}
		{Quantifying how post-transcriptional noise and gene copy number variation
			bias transcriptional parameter inference from {mRNA} distributions},\ }\href
	{https://doi.org/10.7554/eLife.82493} {\bibfield  {journal} {\bibinfo
			{journal} {Elife}\ }\textbf {\bibinfo {volume} {11}},\ \bibinfo {pages}
		{e82493} (\bibinfo {year} {2022})}\BibitemShut {NoStop}%
	\bibitem [{\citenamefont {Choubey}(2018)}]{Choubey2018}%
	\BibitemOpen
	\bibfield  {author} {\bibinfo {author} {\bibfnamefont {S.}~\bibnamefont
			{Choubey}},\ }\bibfield  {title} {\bibinfo {title} {Nascent {RNA} kinetics:
			Transient and steady state behavior of models of transcription},\ }\href
	{https://doi.org/10.1103/PhysRevE.97.022402} {\bibfield  {journal} {\bibinfo
			{journal} {Phys. Rev. E}\ }\textbf {\bibinfo {volume} {97}},\ \bibinfo
		{pages} {022402} (\bibinfo {year} {2018})}\BibitemShut {NoStop}%
	\bibitem [{\citenamefont {Jiang}\ \emph {et~al.}(2021)\citenamefont {Jiang},
		\citenamefont {Fu}, \citenamefont {Yan}, \citenamefont {Li}, \citenamefont
		{Du}, \citenamefont {Cao}, \citenamefont {Qian},\ and\ \citenamefont
		{Grima}}]{jiang2021neural}%
	\BibitemOpen
	\bibfield  {author} {\bibinfo {author} {\bibfnamefont {Q.}~\bibnamefont
			{Jiang}}, \bibinfo {author} {\bibfnamefont {X.}~\bibnamefont {Fu}}, \bibinfo
		{author} {\bibfnamefont {S.}~\bibnamefont {Yan}}, \bibinfo {author}
		{\bibfnamefont {R.}~\bibnamefont {Li}}, \bibinfo {author} {\bibfnamefont
			{W.}~\bibnamefont {Du}}, \bibinfo {author} {\bibfnamefont {Z.}~\bibnamefont
			{Cao}}, \bibinfo {author} {\bibfnamefont {F.}~\bibnamefont {Qian}},\ and\
		\bibinfo {author} {\bibfnamefont {R.}~\bibnamefont {Grima}},\ }\bibfield
	{title} {\bibinfo {title} {Neural network aided approximation and parameter
			inference of non-markovian models of gene expression},\ }\href
	{https://doi.org/10.1038/s41467-021-22919-1} {\bibfield  {journal} {\bibinfo
			{journal} {Nature communications}\ }\textbf {\bibinfo {volume} {12}},\
		\bibinfo {pages} {2618} (\bibinfo {year} {2021})}\BibitemShut {NoStop}%
	\bibitem [{\citenamefont {Braichenko}\ \emph {et~al.}(2021)\citenamefont
		{Braichenko}, \citenamefont {Holehouse},\ and\ \citenamefont
		{Grima}}]{braichenko2021distinguishing}%
	\BibitemOpen
	\bibfield  {author} {\bibinfo {author} {\bibfnamefont {S.}~\bibnamefont
			{Braichenko}}, \bibinfo {author} {\bibfnamefont {J.}~\bibnamefont
			{Holehouse}},\ and\ \bibinfo {author} {\bibfnamefont {R.}~\bibnamefont
			{Grima}},\ }\bibfield  {title} {\bibinfo {title} {Distinguishing between
			models of mammalian gene expression: telegraph-like models versus mechanistic
			models},\ }\href {https://doi.org/10.1098/rsif.2021.0510} {\bibfield
		{journal} {\bibinfo  {journal} {Journal of the Royal Society Interface}\
		}\textbf {\bibinfo {volume} {18}},\ \bibinfo {pages} {20210510} (\bibinfo
		{year} {2021})}\BibitemShut {NoStop}%
	\bibitem [{\citenamefont {Szavits-Nossan}\ and\ \citenamefont
		{Grima}(2023)}]{Szavits2023}%
	\BibitemOpen
	\bibfield  {author} {\bibinfo {author} {\bibfnamefont {J.}~\bibnamefont
			{Szavits-Nossan}}\ and\ \bibinfo {author} {\bibfnamefont {R.}~\bibnamefont
			{Grima}},\ }\bibfield  {title} {\bibinfo {title} {Steady-state distributions
			of nascent {RNA} for general initiation mechanisms},\ }\href
	{https://doi.org/10.1103/PhysRevResearch.5.013064} {\bibfield  {journal}
		{\bibinfo  {journal} {Phys. Rev. Res.}\ }\textbf {\bibinfo {volume} {5}},\
		\bibinfo {pages} {013064} (\bibinfo {year} {2023})}\BibitemShut {NoStop}%
	\bibitem [{\citenamefont {Klumpp}(2011)}]{Klumpp2011}%
	\BibitemOpen
	\bibfield  {author} {\bibinfo {author} {\bibfnamefont {S.}~\bibnamefont
			{Klumpp}},\ }\bibfield  {title} {\bibinfo {title} {Pausing and backtracking
			in transcription under dense traffic conditions},\ }\href
	{https://doi.org/10.1007/s10955-011-0120-3} {\bibfield  {journal} {\bibinfo
			{journal} {Journal of Statistical Physics}\ }\textbf {\bibinfo {volume}
			{142}},\ \bibinfo {pages} {1252} (\bibinfo {year} {2011})}\BibitemShut
	{NoStop}%
	\bibitem [{\citenamefont {Tripathi}\ and\ \citenamefont
		{Chowdhury}(2008)}]{Tripathi2008-constitutive}%
	\BibitemOpen
	\bibfield  {author} {\bibinfo {author} {\bibfnamefont {T.}~\bibnamefont
			{Tripathi}}\ and\ \bibinfo {author} {\bibfnamefont {D.}~\bibnamefont
			{Chowdhury}},\ }\bibfield  {title} {\bibinfo {title} {Interacting {RNA}
			polymerase motors on a dna track: Effects of traffic congestion and intrinsic
			noise on {RNA} synthesis},\ }\href
	{https://doi.org/10.1103/PhysRevE.77.011921} {\bibfield  {journal} {\bibinfo
			{journal} {Phys. Rev. E}\ }\textbf {\bibinfo {volume} {77}},\ \bibinfo
		{pages} {011921} (\bibinfo {year} {2008})}\BibitemShut {NoStop}%
	\bibitem [{\citenamefont {Tr\textbf{}ipathi}\ and\ \citenamefont
		{Chowdhury}(2008)}]{Tripathi2008-telegraph}%
	\BibitemOpen
	\bibfield  {author} {\bibinfo {author} {\bibfnamefont {T.}~\bibnamefont
			{Tr\textbf{}ipathi}}\ and\ \bibinfo {author} {\bibfnamefont {D.}~\bibnamefont
			{Chowdhury}},\ }\bibfield  {title} {\bibinfo {title} {Transcriptional bursts:
			A unified model of machines and mechanisms},\ }\href
	{https://doi.org/10.1209/0295-5075/84/68004} {\bibfield  {journal} {\bibinfo
			{journal} {Europhysics Letters}\ }\textbf {\bibinfo {volume} {84}},\ \bibinfo
		{pages} {68004} (\bibinfo {year} {2008})}\BibitemShut {NoStop}%
	\bibitem [{\citenamefont {Voliotis}\ \emph {et~al.}(2008)\citenamefont
		{Voliotis}, \citenamefont {Cohen}, \citenamefont {Molina-París},\ and\
		\citenamefont {Liverpool}}]{Voliotis2008}%
	\BibitemOpen
	\bibfield  {author} {\bibinfo {author} {\bibfnamefont {M.}~\bibnamefont
			{Voliotis}}, \bibinfo {author} {\bibfnamefont {N.}~\bibnamefont {Cohen}},
		\bibinfo {author} {\bibfnamefont {C.}~\bibnamefont {Molina-París}},\ and\
		\bibinfo {author} {\bibfnamefont {T.~B.}\ \bibnamefont {Liverpool}},\
	}\bibfield  {title} {\bibinfo {title} {Fluctuations, pauses, and backtracking
			in dna transcription},\ }\href
	{https://doi.org/https://doi.org/10.1529/biophysj.107.105767} {\bibfield
		{journal} {\bibinfo  {journal} {Biophysical Journal}\ }\textbf {\bibinfo
			{volume} {94}},\ \bibinfo {pages} {334} (\bibinfo {year} {2008})}\BibitemShut
	{NoStop}%
	\bibitem [{\citenamefont {Klumpp}\ and\ \citenamefont
		{Hwa}(2008)}]{Klumpp2008}%
	\BibitemOpen
	\bibfield  {author} {\bibinfo {author} {\bibfnamefont {S.}~\bibnamefont
			{Klumpp}}\ and\ \bibinfo {author} {\bibfnamefont {T.}~\bibnamefont {Hwa}},\
	}\bibfield  {title} {\bibinfo {title} {Stochasticity and traffic jams in the
			transcription of ribosomal {RNA}: Intriguing role of termination and
			antitermination},\ }\href {https://doi.org/10.1073/pnas.0806084105}
	{\bibfield  {journal} {\bibinfo  {journal} {Proceedings of the National
				Academy of Sciences}\ }\textbf {\bibinfo {volume} {105}},\ \bibinfo {pages}
		{18159} (\bibinfo {year} {2008})}\BibitemShut {NoStop}%
	\bibitem [{\citenamefont {Kim}\ and\ \citenamefont
		{Jacobs-Wagner}(2018)}]{Kim_2018}%
	\BibitemOpen
	\bibfield  {author} {\bibinfo {author} {\bibfnamefont {S.}~\bibnamefont
			{Kim}}\ and\ \bibinfo {author} {\bibfnamefont {C.}~\bibnamefont
			{Jacobs-Wagner}},\ }\bibfield  {title} {\bibinfo {title} {Effects of {mRNA}
			degradation and site-specific transcriptional pausing on protein expression
			noise},\ }\href {https://doi.org/https://doi.org/10.1016/j.bpj.2018.02.010}
	{\bibfield  {journal} {\bibinfo  {journal} {Biophysical Journal}\ }\textbf
		{\bibinfo {volume} {114}},\ \bibinfo {pages} {1718} (\bibinfo {year}
		{2018})}\BibitemShut {NoStop}%
	\bibitem [{\citenamefont {Tripathi}\ \emph {et~al.}(2009)\citenamefont
		{Tripathi}, \citenamefont {Schütz},\ and\ \citenamefont
		{Chowdhury}}]{Tripathi2009}%
	\BibitemOpen
	\bibfield  {author} {\bibinfo {author} {\bibfnamefont {T.}~\bibnamefont
			{Tripathi}}, \bibinfo {author} {\bibfnamefont {G.~M.}\ \bibnamefont
			{Schütz}},\ and\ \bibinfo {author} {\bibfnamefont {D.}~\bibnamefont
			{Chowdhury}},\ }\bibfield  {title} {\bibinfo {title} {{RNA} polymerase
			motors: dwell time distribution, velocity and dynamical phases},\ }\href
	{https://doi.org/10.1088/1742-5468/2009/08/P08018} {\bibfield  {journal}
		{\bibinfo  {journal} {Journal of Statistical Mechanics: Theory and
				Experiment}\ }\textbf {\bibinfo {volume} {2009}},\ \bibinfo {pages} {P08018}
		(\bibinfo {year} {2009})}\BibitemShut {NoStop}%
	\bibitem [{\citenamefont {Dobrzy\'{n}ski}\ and\ \citenamefont
		{Bruggeman}(2009)}]{Dobrzynski2009}%
	\BibitemOpen
	\bibfield  {author} {\bibinfo {author} {\bibfnamefont {M.}~\bibnamefont
			{Dobrzy\'{n}ski}}\ and\ \bibinfo {author} {\bibfnamefont {F.~J.}\
			\bibnamefont {Bruggeman}},\ }\bibfield  {title} {\bibinfo {title} {Elongation
			dynamics shape bursty transcription and translation},\ }\href
	{https://doi.org/10.1073/pnas.0803507106} {\bibfield  {journal} {\bibinfo
			{journal} {Proceedings of the National Academy of Sciences}\ }\textbf
		{\bibinfo {volume} {106}},\ \bibinfo {pages} {2583} (\bibinfo {year}
		{2009})}\BibitemShut {NoStop}%
	\bibitem [{\citenamefont {Ribeiro}\ \emph {et~al.}(2010)\citenamefont
		{Ribeiro}, \citenamefont {Häkkinen}, \citenamefont {Healy},\ and\
		\citenamefont {Yli-Harja}}]{Ribeiro_2010}%
	\BibitemOpen
	\bibfield  {author} {\bibinfo {author} {\bibfnamefont {A.~S.}\ \bibnamefont
			{Ribeiro}}, \bibinfo {author} {\bibfnamefont {A.}~\bibnamefont {Häkkinen}},
		\bibinfo {author} {\bibfnamefont {S.}~\bibnamefont {Healy}},\ and\ \bibinfo
		{author} {\bibfnamefont {O.}~\bibnamefont {Yli-Harja}},\ }\bibfield  {title}
	{\bibinfo {title} {Dynamical effects of transcriptional pause-prone sites},\
	}\href {https://doi.org/https://doi.org/10.1016/j.compbiolchem.2010.04.003}
	{\bibfield  {journal} {\bibinfo  {journal} {Computational Biology and
				Chemistry}\ }\textbf {\bibinfo {volume} {34}},\ \bibinfo {pages} {143}
		(\bibinfo {year} {2010})}\BibitemShut {NoStop}%
	\bibitem [{\citenamefont {Rajala}\ \emph {et~al.}(2010)\citenamefont {Rajala},
		\citenamefont {Häkkinen}, \citenamefont {Healy}, \citenamefont {Yli-Harja},\
		and\ \citenamefont {Ribeiro}}]{Rajala_2010}%
	\BibitemOpen
	\bibfield  {author} {\bibinfo {author} {\bibfnamefont {T.}~\bibnamefont
			{Rajala}}, \bibinfo {author} {\bibfnamefont {A.}~\bibnamefont {Häkkinen}},
		\bibinfo {author} {\bibfnamefont {S.}~\bibnamefont {Healy}}, \bibinfo
		{author} {\bibfnamefont {O.}~\bibnamefont {Yli-Harja}},\ and\ \bibinfo
		{author} {\bibfnamefont {A.~S.}\ \bibnamefont {Ribeiro}},\ }\bibfield
	{title} {\bibinfo {title} {Effects of transcriptional pausing on gene
			expression dynamics},\ }\href {https://doi.org/10.1371/journal.pcbi.1000704}
	{\bibfield  {journal} {\bibinfo  {journal} {PLOS Computational Biology}\
		}\textbf {\bibinfo {volume} {6}},\ \bibinfo {pages} {1} (\bibinfo {year}
		{2010})}\BibitemShut {NoStop}%
	\bibitem [{\citenamefont {Chowdhury}(2013)}]{Chowdhury2013}%
	\BibitemOpen
	\bibfield  {author} {\bibinfo {author} {\bibfnamefont {D.}~\bibnamefont
			{Chowdhury}},\ }\bibfield  {title} {\bibinfo {title} {Stochastic
			mechano-chemical kinetics of molecular motors: A multidisciplinary enterprise
			from a physicist’s perspective},\ }\href
	{https://doi.org/https://doi.org/10.1016/j.physrep.2013.03.005} {\bibfield
		{journal} {\bibinfo  {journal} {Physics Reports}\ }\textbf {\bibinfo {volume}
			{529}},\ \bibinfo {pages} {1} (\bibinfo {year} {2013})}\BibitemShut {NoStop}%
	\bibitem [{\citenamefont {Sahoo}\ and\ \citenamefont
		{Klumpp}(2013)}]{Sahoo2013}%
	\BibitemOpen
	\bibfield  {author} {\bibinfo {author} {\bibfnamefont {M.}~\bibnamefont
			{Sahoo}}\ and\ \bibinfo {author} {\bibfnamefont {S.}~\bibnamefont {Klumpp}},\
	}\bibfield  {title} {\bibinfo {title} {Backtracking dynamics of {RNA}
			polymerase: pausing and error correction},\ }\href
	{https://doi.org/10.1088/0953-8984/25/37/374104} {\bibfield  {journal}
		{\bibinfo  {journal} {Journal of Physics: Condensed Matter}\ }\textbf
		{\bibinfo {volume} {25}},\ \bibinfo {pages} {374104} (\bibinfo {year}
		{2013})}\BibitemShut {NoStop}%
	\bibitem [{\citenamefont {Heberling}\ \emph {et~al.}(2016)\citenamefont
		{Heberling}, \citenamefont {Davis}, \citenamefont {Gedeon}, \citenamefont
		{Morgan},\ and\ \citenamefont {Gedeon}}]{Heberling2016}%
	\BibitemOpen
	\bibfield  {author} {\bibinfo {author} {\bibfnamefont {T.}~\bibnamefont
			{Heberling}}, \bibinfo {author} {\bibfnamefont {L.}~\bibnamefont {Davis}},
		\bibinfo {author} {\bibfnamefont {J.}~\bibnamefont {Gedeon}}, \bibinfo
		{author} {\bibfnamefont {C.}~\bibnamefont {Morgan}},\ and\ \bibinfo {author}
		{\bibfnamefont {T.}~\bibnamefont {Gedeon}},\ }\bibfield  {title} {\bibinfo
		{title} {A mechanistic model for cooperative behavior of co-transcribing
			{RNA} polymerases},\ }\href {https://doi.org/10.1371/journal.pcbi.1005069}
	{\bibfield  {journal} {\bibinfo  {journal} {PLOS Computational Biology}\
		}\textbf {\bibinfo {volume} {12}},\ \bibinfo {pages} {1} (\bibinfo {year}
		{2016})}\BibitemShut {NoStop}%
	\bibitem [{\citenamefont {Cholewa-Waclaw}\ \emph {et~al.}(2019)\citenamefont
		{Cholewa-Waclaw}, \citenamefont {Shah}, \citenamefont {Webb}, \citenamefont
		{Chhatbar}, \citenamefont {Ramsahoye}, \citenamefont {Pusch}, \citenamefont
		{Yu}, \citenamefont {Greulich}, \citenamefont {Waclaw},\ and\ \citenamefont
		{Bird}}]{Cholewa2019}%
	\BibitemOpen
	\bibfield  {author} {\bibinfo {author} {\bibfnamefont {J.}~\bibnamefont
			{Cholewa-Waclaw}}, \bibinfo {author} {\bibfnamefont {R.}~\bibnamefont
			{Shah}}, \bibinfo {author} {\bibfnamefont {S.}~\bibnamefont {Webb}}, \bibinfo
		{author} {\bibfnamefont {K.}~\bibnamefont {Chhatbar}}, \bibinfo {author}
		{\bibfnamefont {B.}~\bibnamefont {Ramsahoye}}, \bibinfo {author}
		{\bibfnamefont {O.}~\bibnamefont {Pusch}}, \bibinfo {author} {\bibfnamefont
			{M.}~\bibnamefont {Yu}}, \bibinfo {author} {\bibfnamefont {P.}~\bibnamefont
			{Greulich}}, \bibinfo {author} {\bibfnamefont {B.}~\bibnamefont {Waclaw}},\
		and\ \bibinfo {author} {\bibfnamefont {A.~P.}\ \bibnamefont {Bird}},\
	}\bibfield  {title} {\bibinfo {title} {Quantitative modelling predicts the
			impact of dna methylation on {RNA} polymerase ii traffic},\ }\href
	{https://doi.org/10.1073/pnas.1903549116} {\bibfield  {journal} {\bibinfo
			{journal} {Proceedings of the National Academy of Sciences}\ }\textbf
		{\bibinfo {volume} {116}},\ \bibinfo {pages} {14995} (\bibinfo {year}
		{2019})}\BibitemShut {NoStop}%
	\bibitem [{\citenamefont {Ali}\ \emph {et~al.}(2020)\citenamefont {Ali},
		\citenamefont {Choubey}, \citenamefont {Das},\ and\ \citenamefont
		{Brewster}}]{Ali2020}%
	\BibitemOpen
	\bibfield  {author} {\bibinfo {author} {\bibfnamefont {M.~Z.}\ \bibnamefont
			{Ali}}, \bibinfo {author} {\bibfnamefont {S.}~\bibnamefont {Choubey}},
		\bibinfo {author} {\bibfnamefont {D.}~\bibnamefont {Das}},\ and\ \bibinfo
		{author} {\bibfnamefont {R.~C.}\ \bibnamefont {Brewster}},\ }\bibfield
	{title} {\bibinfo {title} {Probing mechanisms of transcription elongation
			through cell-to-cell variability of {RNA} polymerase},\ }\href
	{https://doi.org/https://doi.org/10.1016/j.bpj.2020.02.002} {\bibfield
		{journal} {\bibinfo  {journal} {Biophysical Journal}\ }\textbf {\bibinfo
			{volume} {118}},\ \bibinfo {pages} {1769} (\bibinfo {year}
		{2020})}\BibitemShut {NoStop}%
	\bibitem [{\citenamefont {Szavits-Nossan}\ and\ \citenamefont
		{Waclaw}(2020)}]{Szavits2020-roadblocks}%
	\BibitemOpen
	\bibfield  {author} {\bibinfo {author} {\bibfnamefont {J.}~\bibnamefont
			{Szavits-Nossan}}\ and\ \bibinfo {author} {\bibfnamefont {B.}~\bibnamefont
			{Waclaw}},\ }\bibfield  {title} {\bibinfo {title} {Current-density relation
			in the exclusion process with dynamic obstacles},\ }\href
	{https://doi.org/10.1103/PhysRevE.102.042117} {\bibfield  {journal} {\bibinfo
			{journal} {Phys. Rev. E}\ }\textbf {\bibinfo {volume} {102}},\ \bibinfo
		{pages} {042117} (\bibinfo {year} {2020})}\BibitemShut {NoStop}%
	\bibitem [{\citenamefont {Turowski}\ \emph {et~al.}(2020)\citenamefont
		{Turowski}, \citenamefont {Petfalski}, \citenamefont {Goddard}, \citenamefont
		{French}, \citenamefont {Helwak},\ and\ \citenamefont
		{Tollervey}}]{turowski2020nascent}%
	\BibitemOpen
	\bibfield  {author} {\bibinfo {author} {\bibfnamefont {T.~W.}\ \bibnamefont
			{Turowski}}, \bibinfo {author} {\bibfnamefont {E.}~\bibnamefont {Petfalski}},
		\bibinfo {author} {\bibfnamefont {B.~D.}\ \bibnamefont {Goddard}}, \bibinfo
		{author} {\bibfnamefont {S.~L.}\ \bibnamefont {French}}, \bibinfo {author}
		{\bibfnamefont {A.}~\bibnamefont {Helwak}},\ and\ \bibinfo {author}
		{\bibfnamefont {D.}~\bibnamefont {Tollervey}},\ }\bibfield  {title} {\bibinfo
		{title} {Nascent transcript folding plays a major role in determining {RNA}
			polymerase elongation rates},\ }\href
	{https://doi.org/https://doi.org/10.1016/j.molcel.2020.06.002} {\bibfield
		{journal} {\bibinfo  {journal} {Molecular cell}\ }\textbf {\bibinfo {volume}
			{79}},\ \bibinfo {pages} {488} (\bibinfo {year} {2020})}\BibitemShut
	{NoStop}%
	\bibitem [{\citenamefont {Sch{\"u}tz}\ and\ \citenamefont
		{Domany}(1993)}]{Schutz1993}%
	\BibitemOpen
	\bibfield  {author} {\bibinfo {author} {\bibfnamefont {G.}~\bibnamefont
			{Sch{\"u}tz}}\ and\ \bibinfo {author} {\bibfnamefont {E.}~\bibnamefont
			{Domany}},\ }\bibfield  {title} {\bibinfo {title} {Phase transitions in an
			exactly soluble one-dimensional exclusion process},\ }\href
	{https://doi.org/10.1007/BF01048050} {\bibfield  {journal} {\bibinfo
			{journal} {Journal of Statistical Physics}\ }\textbf {\bibinfo {volume}
			{72}},\ \bibinfo {pages} {277} (\bibinfo {year} {1993})}\BibitemShut
	{NoStop}%
	\bibitem [{\citenamefont {Derrida}\ \emph {et~al.}(1993)\citenamefont
		{Derrida}, \citenamefont {Evans}, \citenamefont {Hakim},\ and\ \citenamefont
		{Pasquier}}]{Derrida1993}%
	\BibitemOpen
	\bibfield  {author} {\bibinfo {author} {\bibfnamefont {B.}~\bibnamefont
			{Derrida}}, \bibinfo {author} {\bibfnamefont {M.~R.}\ \bibnamefont {Evans}},
		\bibinfo {author} {\bibfnamefont {V.}~\bibnamefont {Hakim}},\ and\ \bibinfo
		{author} {\bibfnamefont {V.}~\bibnamefont {Pasquier}},\ }\bibfield  {title}
	{\bibinfo {title} {Exact solution of a 1d asymmetric exclusion model using a
			matrix formulation},\ }\href {https://doi.org/10.1088/0305-4470/26/7/011}
	{\bibfield  {journal} {\bibinfo  {journal} {Journal of Physics A:
				Mathematical and General}\ }\textbf {\bibinfo {volume} {26}},\ \bibinfo
		{pages} {1493} (\bibinfo {year} {1993})}\BibitemShut {NoStop}%
	\bibitem [{\citenamefont {Tongaonkar}\ \emph {et~al.}(2005)\citenamefont
		{Tongaonkar}, \citenamefont {French}, \citenamefont {Oakes}, \citenamefont
		{Vu}, \citenamefont {Schneider}, \citenamefont {Beyer},\ and\ \citenamefont
		{Nomura}}]{tongaonkar2005histones}%
	\BibitemOpen
	\bibfield  {author} {\bibinfo {author} {\bibfnamefont {P.}~\bibnamefont
			{Tongaonkar}}, \bibinfo {author} {\bibfnamefont {S.~L.}\ \bibnamefont
			{French}}, \bibinfo {author} {\bibfnamefont {M.~L.}\ \bibnamefont {Oakes}},
		\bibinfo {author} {\bibfnamefont {L.}~\bibnamefont {Vu}}, \bibinfo {author}
		{\bibfnamefont {D.~A.}\ \bibnamefont {Schneider}}, \bibinfo {author}
		{\bibfnamefont {A.~L.}\ \bibnamefont {Beyer}},\ and\ \bibinfo {author}
		{\bibfnamefont {M.}~\bibnamefont {Nomura}},\ }\bibfield  {title} {\bibinfo
		{title} {Histones are required for transcription of yeast rrna genes by rna
			polymerase i},\ }\href {https://doi.org/10.1073/pnas.0504563102} {\bibfield
		{journal} {\bibinfo  {journal} {Proceedings of the National Academy of
				Sciences}\ }\textbf {\bibinfo {volume} {102}},\ \bibinfo {pages} {10129}
		(\bibinfo {year} {2005})}\BibitemShut {NoStop}%
	\bibitem [{\citenamefont {Schneider}\ \emph {et~al.}(2006)\citenamefont
		{Schneider}, \citenamefont {French}, \citenamefont {Osheim}, \citenamefont
		{Bailey}, \citenamefont {Vu}, \citenamefont {Dodd}, \citenamefont {Yates},
		\citenamefont {Beyer},\ and\ \citenamefont {Nomura}}]{schneider2006rna}%
	\BibitemOpen
	\bibfield  {author} {\bibinfo {author} {\bibfnamefont {D.}~\bibnamefont
			{Schneider}}, \bibinfo {author} {\bibfnamefont {S.}~\bibnamefont {French}},
		\bibinfo {author} {\bibfnamefont {Y.}~\bibnamefont {Osheim}}, \bibinfo
		{author} {\bibfnamefont {A.}~\bibnamefont {Bailey}}, \bibinfo {author}
		{\bibfnamefont {L.}~\bibnamefont {Vu}}, \bibinfo {author} {\bibfnamefont
			{J.}~\bibnamefont {Dodd}}, \bibinfo {author} {\bibfnamefont {J.}~\bibnamefont
			{Yates}}, \bibinfo {author} {\bibfnamefont {A.}~\bibnamefont {Beyer}},\ and\
		\bibinfo {author} {\bibfnamefont {M.}~\bibnamefont {Nomura}},\ }\bibfield
	{title} {\bibinfo {title} {{RNA} polymerase ii elongation factors spt4p and
			spt5p play roles in transcription elongation by {RNA} polymerase i and rrna
			processing},\ }\href {https://doi.org/10.1073/pnas.0605686103} {\bibfield
		{journal} {\bibinfo  {journal} {Proceedings of the National Academy of
				Sciences}\ }\textbf {\bibinfo {volume} {103}},\ \bibinfo {pages} {12707}
		(\bibinfo {year} {2006})}\BibitemShut {NoStop}%
	\bibitem [{\citenamefont {Claypool}\ \emph {et~al.}(2004)\citenamefont
		{Claypool}, \citenamefont {French}, \citenamefont {Johzuka}, \citenamefont
		{Eliason}, \citenamefont {Vu}, \citenamefont {Dodd}, \citenamefont {Beyer},\
		and\ \citenamefont {Nomura}}]{claypool2004tor}%
	\BibitemOpen
	\bibfield  {author} {\bibinfo {author} {\bibfnamefont {J.~A.}\ \bibnamefont
			{Claypool}}, \bibinfo {author} {\bibfnamefont {S.~L.}\ \bibnamefont
			{French}}, \bibinfo {author} {\bibfnamefont {K.}~\bibnamefont {Johzuka}},
		\bibinfo {author} {\bibfnamefont {K.}~\bibnamefont {Eliason}}, \bibinfo
		{author} {\bibfnamefont {L.}~\bibnamefont {Vu}}, \bibinfo {author}
		{\bibfnamefont {J.~A.}\ \bibnamefont {Dodd}}, \bibinfo {author}
		{\bibfnamefont {A.~L.}\ \bibnamefont {Beyer}},\ and\ \bibinfo {author}
		{\bibfnamefont {M.}~\bibnamefont {Nomura}},\ }\bibfield  {title} {\bibinfo
		{title} {Tor pathway regulates rrn3p-dependent recruitment of yeast {RNA}
			polymerase i to the promoter but does not participate in alteration of the
			number of active genes},\ }\href {https://doi.org/10.1091/mbc.e03-08-0594}
	{\bibfield  {journal} {\bibinfo  {journal} {Molecular biology of the cell}\
		}\textbf {\bibinfo {volume} {15}},\ \bibinfo {pages} {946} (\bibinfo {year}
		{2004})}\BibitemShut {NoStop}%
	\bibitem [{\citenamefont {MacDonald}\ \emph {et~al.}(1968)\citenamefont
		{MacDonald}, \citenamefont {Gibbs},\ and\ \citenamefont
		{Pipkin}}]{MacDonald1968}%
	\BibitemOpen
	\bibfield  {author} {\bibinfo {author} {\bibfnamefont {C.~T.}\ \bibnamefont
			{MacDonald}}, \bibinfo {author} {\bibfnamefont {J.~H.}\ \bibnamefont
			{Gibbs}},\ and\ \bibinfo {author} {\bibfnamefont {A.~C.}\ \bibnamefont
			{Pipkin}},\ }\bibfield  {title} {\bibinfo {title} {Kinetics of
			biopolymerization on nucleic acid templates},\ }\href
	{https://doi.org/10.1002/bip.1968.360060102} {\bibfield  {journal} {\bibinfo
			{journal} {Biopolymers}\ }\textbf {\bibinfo {volume} {6}},\ \bibinfo {pages}
		{1} (\bibinfo {year} {1968})}\BibitemShut {NoStop}%
	\bibitem [{\citenamefont {Little}(1961)}]{Little1961}%
	\BibitemOpen
	\bibfield  {author} {\bibinfo {author} {\bibfnamefont {J.~D.~C.}\
			\bibnamefont {Little}},\ }\bibfield  {title} {\bibinfo {title} {A proof for
			the queuing formula},\ }\href {https://doi.org/10.1287/opre.9.3.383}
	{\bibfield  {journal} {\bibinfo  {journal} {Operations Research}\ }\textbf
		{\bibinfo {volume} {9}},\ \bibinfo {pages} {383} (\bibinfo {year}
		{1961})}\BibitemShut {NoStop}%
	\bibitem [{\citenamefont {Elgart}\ \emph {et~al.}(2010)\citenamefont {Elgart},
		\citenamefont {Jia},\ and\ \citenamefont
		{Kulkarni}}]{Elgart2010-applications}%
	\BibitemOpen
	\bibfield  {author} {\bibinfo {author} {\bibfnamefont {V.}~\bibnamefont
			{Elgart}}, \bibinfo {author} {\bibfnamefont {T.}~\bibnamefont {Jia}},\ and\
		\bibinfo {author} {\bibfnamefont {R.~V.}\ \bibnamefont {Kulkarni}},\
	}\bibfield  {title} {\bibinfo {title} {Applications of little's law to
			stochastic models of gene expression},\ }\href
	{https://doi.org/10.1103/PhysRevE.82.021901} {\bibfield  {journal} {\bibinfo
			{journal} {Physical Review E}\ }\textbf {\bibinfo {volume} {82}},\ \bibinfo
		{pages} {021901} (\bibinfo {year} {2010})}\BibitemShut {NoStop}%
	\bibitem [{\citenamefont {{B. Derrida}}\ and\ \citenamefont {{M.R.
				Evans}}(1993)}]{Derrida93-correlations}%
	\BibitemOpen
	\bibfield  {author} {\bibinfo {author} {\bibnamefont {{B. Derrida}}}\ and\
		\bibinfo {author} {\bibnamefont {{M.R. Evans}}},\ }\bibfield  {title}
	{\bibinfo {title} {Exact correlation functions in an asymmetric exclusion
			model with open boundaries},\ }\href {https://doi.org/10.1051/jp1:1993132}
	{\bibfield  {journal} {\bibinfo  {journal} {J. Phys. I France}\ }\textbf
		{\bibinfo {volume} {3}},\ \bibinfo {pages} {311} (\bibinfo {year}
		{1993})}\BibitemShut {NoStop}%
	\bibitem [{\citenamefont {Derrida}\ \emph {et~al.}(1992)\citenamefont
		{Derrida}, \citenamefont {Domany},\ and\ \citenamefont
		{Mukamel}}]{Derrida1992}%
	\BibitemOpen
	\bibfield  {author} {\bibinfo {author} {\bibfnamefont {B.}~\bibnamefont
			{Derrida}}, \bibinfo {author} {\bibfnamefont {E.}~\bibnamefont {Domany}},\
		and\ \bibinfo {author} {\bibfnamefont {D.}~\bibnamefont {Mukamel}},\
	}\bibfield  {title} {\bibinfo {title} {An exact solution of a one-dimensional
			asymmetric exclusion model with open boundaries},\ }\href
	{https://doi.org/10.1007/BF01050430} {\bibfield  {journal} {\bibinfo
			{journal} {Journal of Statistical Physics}\ }\textbf {\bibinfo {volume}
			{69}},\ \bibinfo {pages} {667} (\bibinfo {year} {1992})}\BibitemShut
	{NoStop}%
	\bibitem [{\citenamefont {Cianci}\ \emph {et~al.}(2016)\citenamefont {Cianci},
		\citenamefont {Smith},\ and\ \citenamefont {Grima}}]{Cianci2016}%
	\BibitemOpen
	\bibfield  {author} {\bibinfo {author} {\bibfnamefont {C.}~\bibnamefont
			{Cianci}}, \bibinfo {author} {\bibfnamefont {S.}~\bibnamefont {Smith}},\ and\
		\bibinfo {author} {\bibfnamefont {R.}~\bibnamefont {Grima}},\ }\bibfield
	{title} {\bibinfo {title} {Molecular finite-size effects in stochastic models
			of equilibrium chemical systems},\ }\href {https://doi.org/10.1063/1.4941583}
	{\bibfield  {journal} {\bibinfo  {journal} {The Journal of Chemical Physics}\
		}\textbf {\bibinfo {volume} {144}},\ \bibinfo {pages} {084101} (\bibinfo
		{year} {2016})}\BibitemShut {NoStop}%
	\bibitem [{\citenamefont {Gillespie}(1977)}]{gillespie1977exact}%
	\BibitemOpen
	\bibfield  {author} {\bibinfo {author} {\bibfnamefont {D.~T.}\ \bibnamefont
			{Gillespie}},\ }\bibfield  {title} {\bibinfo {title} {Exact stochastic
			simulation of coupled chemical reactions},\ }\href
	{https://doi.org/10.1021/j100540a008} {\bibfield  {journal} {\bibinfo
			{journal} {The journal of physical chemistry}\ }\textbf {\bibinfo {volume}
			{81}},\ \bibinfo {pages} {2340} (\bibinfo {year} {1977})}\BibitemShut
	{NoStop}%
	\bibitem [{\citenamefont {Kendall}(1953)}]{Kendall1953}%
	\BibitemOpen
	\bibfield  {author} {\bibinfo {author} {\bibfnamefont {D.~G.}\ \bibnamefont
			{Kendall}},\ }\bibfield  {title} {\bibinfo {title} {{Stochastic Processes
				Occurring in the Theory of Queues and their Analysis by the Method of the
				Imbedded Markov Chain}},\ }\href {https://doi.org/10.1214/aoms/1177728975}
	{\bibfield  {journal} {\bibinfo  {journal} {The Annals of Mathematical
				Statistics}\ }\textbf {\bibinfo {volume} {24}},\ \bibinfo {pages} {338 }
		(\bibinfo {year} {1953})}\BibitemShut {NoStop}%
	\bibitem [{\citenamefont {Tak{\'a}cs}(1958)}]{Takacs1958}%
	\BibitemOpen
	\bibfield  {author} {\bibinfo {author} {\bibfnamefont {L.}~\bibnamefont
			{Tak{\'a}cs}},\ }\bibfield  {title} {\bibinfo {title} {On a coincidence
			problem concerning telephone traffic},\ }\href
	{https://doi.org/10.1007/BF02023865} {\bibfield  {journal} {\bibinfo
			{journal} {Acta Mathematica Academiae Scientiarum Hungarica}\ }\textbf
		{\bibinfo {volume} {9}},\ \bibinfo {pages} {45} (\bibinfo {year}
		{1958})}\BibitemShut {NoStop}%
	\bibitem [{\citenamefont {Krbálek}\ and\ \citenamefont
		{Hrabák}(2011)}]{Krbalek2011}%
	\BibitemOpen
	\bibfield  {author} {\bibinfo {author} {\bibfnamefont {M.}~\bibnamefont
			{Krbálek}}\ and\ \bibinfo {author} {\bibfnamefont {P.}~\bibnamefont
			{Hrabák}},\ }\bibfield  {title} {\bibinfo {title} {Inter-particle gap
			distribution and spectral rigidity of the totally asymmetric simple exclusion
			process with open boundaries},\ }\href
	{https://doi.org/10.1088/1751-8113/44/17/175203} {\bibfield  {journal}
		{\bibinfo  {journal} {Journal of Physics A: Mathematical and Theoretical}\
		}\textbf {\bibinfo {volume} {44}},\ \bibinfo {pages} {175203} (\bibinfo
		{year} {2011})}\BibitemShut {NoStop}%
	\bibitem [{\citenamefont {Ghosh}\ \emph {et~al.}(1998)\citenamefont {Ghosh},
		\citenamefont {Majumdar},\ and\ \citenamefont {Chowdhury}}]{Ghosh1998}%
	\BibitemOpen
	\bibfield  {author} {\bibinfo {author} {\bibfnamefont {K.}~\bibnamefont
			{Ghosh}}, \bibinfo {author} {\bibfnamefont {A.}~\bibnamefont {Majumdar}},\
		and\ \bibinfo {author} {\bibfnamefont {D.}~\bibnamefont {Chowdhury}},\
	}\bibfield  {title} {\bibinfo {title} {Distribution of time-headways in a
			particle-hopping model of vehicular traffic},\ }\href
	{https://doi.org/10.1103/PhysRevE.58.4012} {\bibfield  {journal} {\bibinfo
			{journal} {Phys. Rev. E}\ }\textbf {\bibinfo {volume} {58}},\ \bibinfo
		{pages} {4012} (\bibinfo {year} {1998})}\BibitemShut {NoStop}%
	\bibitem [{\citenamefont {Hrabák}(2020)}]{Hrabak2020}%
	\BibitemOpen
	\bibfield  {author} {\bibinfo {author} {\bibfnamefont {P.}~\bibnamefont
			{Hrabák}},\ }\bibfield  {title} {\bibinfo {title} {Time-headway distribution
			for random-sequential-update tasep with periodic and open boundaries},\
	}\href {https://doi.org/https://doi.org/10.1016/j.jtte.2019.03.006}
	{\bibfield  {journal} {\bibinfo  {journal} {Journal of Traffic and
				Transportation Engineering (English Edition)}\ }\textbf {\bibinfo {volume}
			{7}},\ \bibinfo {pages} {30} (\bibinfo {year} {2020})}\BibitemShut {NoStop}%
	\bibitem [{\citenamefont {Piovesan}\ \emph {et~al.}(2016)\citenamefont
		{Piovesan}, \citenamefont {Caracausi}, \citenamefont {Antonaros},
		\citenamefont {Pelleri},\ and\ \citenamefont {Vitale}}]{Piovesan2016}%
	\BibitemOpen
	\bibfield  {author} {\bibinfo {author} {\bibfnamefont {A.}~\bibnamefont
			{Piovesan}}, \bibinfo {author} {\bibfnamefont {M.}~\bibnamefont {Caracausi}},
		\bibinfo {author} {\bibfnamefont {F.}~\bibnamefont {Antonaros}}, \bibinfo
		{author} {\bibfnamefont {M.~C.}\ \bibnamefont {Pelleri}},\ and\ \bibinfo
		{author} {\bibfnamefont {L.}~\bibnamefont {Vitale}},\ }\bibfield  {title}
	{\bibinfo {title} {{GeneBase 1.1: a tool to summarize data from NCBI Gene
				datasets and its application to an update of human gene statistics}},\
	}\bibfield  {journal} {\bibinfo  {journal} {Database}\ }\textbf {\bibinfo
		{volume} {2016}},\ \href {https://doi.org/10.1093/database/baw153}
	{10.1093/database/baw153} (\bibinfo {year} {2016})\BibitemShut {NoStop}%
	\bibitem [{\citenamefont {Cox}(1967)}]{Cox1967}%
	\BibitemOpen
	\bibfield  {author} {\bibinfo {author} {\bibfnamefont {D.}~\bibnamefont
			{Cox}},\ }\href@noop {} {\emph {\bibinfo {title} {Renewal theory}}}\
	(\bibinfo  {publisher} {Methuen},\ \bibinfo {address} {London},\ \bibinfo
	{year} {1967})\BibitemShut {NoStop}%
	\bibitem [{\citenamefont {Filatova}\ \emph {et~al.}(2021)\citenamefont
		{Filatova}, \citenamefont {Popovic},\ and\ \citenamefont
		{Grima}}]{filatova2021statistics}%
	\BibitemOpen
	\bibfield  {author} {\bibinfo {author} {\bibfnamefont {T.}~\bibnamefont
			{Filatova}}, \bibinfo {author} {\bibfnamefont {N.}~\bibnamefont {Popovic}},\
		and\ \bibinfo {author} {\bibfnamefont {R.}~\bibnamefont {Grima}},\ }\bibfield
	{title} {\bibinfo {title} {Statistics of nascent and mature {RNA}
			fluctuations in a stochastic model of transcriptional initiation, elongation,
			pausing, and termination},\ }\href
	{https://doi.org/10.1007/s11538-020-00827-7} {\bibfield  {journal} {\bibinfo
			{journal} {Bulletin of Mathematical Biology}\ }\textbf {\bibinfo {volume}
			{83}},\ \bibinfo {pages} {1} (\bibinfo {year} {2021})}\BibitemShut {NoStop}%
	\bibitem [{\citenamefont {Thomas}\ \emph {et~al.}(2014)\citenamefont {Thomas},
		\citenamefont {Popovi{\'c}},\ and\ \citenamefont
		{Grima}}]{thomas2014phenotypic}%
	\BibitemOpen
	\bibfield  {author} {\bibinfo {author} {\bibfnamefont {P.}~\bibnamefont
			{Thomas}}, \bibinfo {author} {\bibfnamefont {N.}~\bibnamefont
			{Popovi{\'c}}},\ and\ \bibinfo {author} {\bibfnamefont {R.}~\bibnamefont
			{Grima}},\ }\bibfield  {title} {\bibinfo {title} {Phenotypic switching in
			gene regulatory networks},\ }\href {https://doi.org/10.1073/pnas.1400049111}
	{\bibfield  {journal} {\bibinfo  {journal} {Proceedings of the National
				Academy of Sciences}\ }\textbf {\bibinfo {volume} {111}},\ \bibinfo {pages}
		{6994} (\bibinfo {year} {2014})}\BibitemShut {NoStop}%
	\bibitem [{\citenamefont {Jia}\ and\ \citenamefont
		{Grima}(2020)}]{jia2020small}%
	\BibitemOpen
	\bibfield  {author} {\bibinfo {author} {\bibfnamefont {C.}~\bibnamefont
			{Jia}}\ and\ \bibinfo {author} {\bibfnamefont {R.}~\bibnamefont {Grima}},\
	}\bibfield  {title} {\bibinfo {title} {Small protein number effects in
			stochastic models of autoregulated bursty gene expression},\ }\href
	{https://doi.org/10.1063/1.5144578} {\bibfield  {journal} {\bibinfo
			{journal} {The Journal of chemical physics}\ }\textbf {\bibinfo {volume}
			{152}},\ \bibinfo {pages} {084115} (\bibinfo {year} {2020})}\BibitemShut
	{NoStop}%
	\bibitem [{\citenamefont {Jonkers}\ \emph {et~al.}(2014)\citenamefont
		{Jonkers}, \citenamefont {Kwak},\ and\ \citenamefont {Lis}}]{Jonkers2014}%
	\BibitemOpen
	\bibfield  {author} {\bibinfo {author} {\bibfnamefont {I.}~\bibnamefont
			{Jonkers}}, \bibinfo {author} {\bibfnamefont {H.}~\bibnamefont {Kwak}},\ and\
		\bibinfo {author} {\bibfnamefont {J.~T.}\ \bibnamefont {Lis}},\ }\bibfield
	{title} {\bibinfo {title} {Genome-wide dynamics of pol ii elongation and its
			interplay with promoter proximal pausing, chromatin, and exons},\ }\href
	{https://doi.org/10.7554/eLife.02407} {\bibfield  {journal} {\bibinfo
			{journal} {eLife}\ }\textbf {\bibinfo {volume} {3}},\ \bibinfo {pages}
		{e02407} (\bibinfo {year} {2014})}\BibitemShut {NoStop}%
	\bibitem [{\citenamefont {Szavits-Nossan}\ and\ \citenamefont
		{Grima}(2022)}]{Szavits2022}%
	\BibitemOpen
	\bibfield  {author} {\bibinfo {author} {\bibfnamefont {J.}~\bibnamefont
			{Szavits-Nossan}}\ and\ \bibinfo {author} {\bibfnamefont {R.}~\bibnamefont
			{Grima}},\ }\bibfield  {title} {\bibinfo {title} {Mean-field theory
			accurately captures the variation of copy number distributions across the
			{mRNA} life cycle},\ }\href {https://doi.org/10.1103/PhysRevE.105.014410}
	{\bibfield  {journal} {\bibinfo  {journal} {Phys. Rev. E}\ }\textbf {\bibinfo
			{volume} {105}},\ \bibinfo {pages} {014410} (\bibinfo {year}
		{2022})}\BibitemShut {NoStop}%
	\bibitem [{\citenamefont {Nagar}\ \emph {et~al.}(2011)\citenamefont {Nagar},
		\citenamefont {Valleriani},\ and\ \citenamefont {Lipowsky}}]{Nagar2011}%
	\BibitemOpen
	\bibfield  {author} {\bibinfo {author} {\bibfnamefont {A.}~\bibnamefont
			{Nagar}}, \bibinfo {author} {\bibfnamefont {A.}~\bibnamefont {Valleriani}},\
		and\ \bibinfo {author} {\bibfnamefont {R.}~\bibnamefont {Lipowsky}},\
	}\bibfield  {title} {\bibinfo {title} {Translation by ribosomes with {mRNA}
			degradation: Exclusion processes on aging tracks},\ }\href
	{https://doi.org/10.1007/s10955-011-0347-z} {\bibfield  {journal} {\bibinfo
			{journal} {Journal of Statistical Physics}\ }\textbf {\bibinfo {volume}
			{145}},\ \bibinfo {pages} {1385} (\bibinfo {year} {2011})}\BibitemShut
	{NoStop}%
	\bibitem [{\citenamefont {Karthika}\ and\ \citenamefont
		{Nagar}(2020)}]{Karthika2020}%
	\BibitemOpen
	\bibfield  {author} {\bibinfo {author} {\bibfnamefont {S.}~\bibnamefont
			{Karthika}}\ and\ \bibinfo {author} {\bibfnamefont {A.}~\bibnamefont
			{Nagar}},\ }\bibfield  {title} {\bibinfo {title} {Totally asymmetric simple
			exclusion process with resetting},\ }\href
	{https://doi.org/10.1088/1751-8121/ab6aef} {\bibfield  {journal} {\bibinfo
			{journal} {Journal of Physics A: Mathematical and Theoretical}\ }\textbf
		{\bibinfo {volume} {53}},\ \bibinfo {pages} {115003} (\bibinfo {year}
		{2020})}\BibitemShut {NoStop}%
	\bibitem [{\citenamefont {Lloyd-Price}\ \emph {et~al.}(2016)\citenamefont
		{Lloyd-Price}, \citenamefont {Startceva}, \citenamefont {Kandavalli},
		\citenamefont {Chandraseelan}, \citenamefont {Goncalves}, \citenamefont
		{Oliveira}, \citenamefont {H{\"a}kkinen},\ and\ \citenamefont
		{Ribeiro}}]{lloyd2016dissecting}%
	\BibitemOpen
	\bibfield  {author} {\bibinfo {author} {\bibfnamefont {J.}~\bibnamefont
			{Lloyd-Price}}, \bibinfo {author} {\bibfnamefont {S.}~\bibnamefont
			{Startceva}}, \bibinfo {author} {\bibfnamefont {V.}~\bibnamefont
			{Kandavalli}}, \bibinfo {author} {\bibfnamefont {J.~G.}\ \bibnamefont
			{Chandraseelan}}, \bibinfo {author} {\bibfnamefont {N.}~\bibnamefont
			{Goncalves}}, \bibinfo {author} {\bibfnamefont {S.~M.}\ \bibnamefont
			{Oliveira}}, \bibinfo {author} {\bibfnamefont {A.}~\bibnamefont
			{H{\"a}kkinen}},\ and\ \bibinfo {author} {\bibfnamefont {A.~S.}\ \bibnamefont
			{Ribeiro}},\ }\bibfield  {title} {\bibinfo {title} {Dissecting the stochastic
			transcription initiation process in live escherichia coli},\ }\href
	{https://doi.org/10.1093/dnares/dsw009} {\bibfield  {journal} {\bibinfo
			{journal} {DNA Research}\ }\textbf {\bibinfo {volume} {23}},\ \bibinfo
		{pages} {203} (\bibinfo {year} {2016})}\BibitemShut {NoStop}%
	\bibitem [{\citenamefont {Weidemann}\ \emph {et~al.}(2023)\citenamefont
		{Weidemann}, \citenamefont {Holehouse}, \citenamefont {Singh}, \citenamefont
		{Grima},\ and\ \citenamefont {Hauf}}]{weidemann2023minimal}%
	\BibitemOpen
	\bibfield  {author} {\bibinfo {author} {\bibfnamefont {D.~E.}\ \bibnamefont
			{Weidemann}}, \bibinfo {author} {\bibfnamefont {J.}~\bibnamefont
			{Holehouse}}, \bibinfo {author} {\bibfnamefont {A.}~\bibnamefont {Singh}},
		\bibinfo {author} {\bibfnamefont {R.}~\bibnamefont {Grima}},\ and\ \bibinfo
		{author} {\bibfnamefont {S.}~\bibnamefont {Hauf}},\ }\bibfield  {title}
	{\bibinfo {title} {The minimal intrinsic stochasticity of constitutively
			expressed eukaryotic genes is sub-poissonian},\ }\href
	{https://doi.org/10.1126/sciadv.adh5138} {\bibfield  {journal} {\bibinfo
			{journal} {Science Advances}\ }\textbf {\bibinfo {volume} {9}},\ \bibinfo
		{pages} {eadh5138} (\bibinfo {year} {2023})}\BibitemShut {NoStop}%
	\bibitem [{\citenamefont {Senecal}\ \emph {et~al.}(2014)\citenamefont
		{Senecal}, \citenamefont {Munsky}, \citenamefont {Proux}, \citenamefont {Ly},
		\citenamefont {Braye}, \citenamefont {Zimmer}, \citenamefont {Mueller},\ and\
		\citenamefont {Darzacq}}]{senecal2014transcription}%
	\BibitemOpen
	\bibfield  {author} {\bibinfo {author} {\bibfnamefont {A.}~\bibnamefont
			{Senecal}}, \bibinfo {author} {\bibfnamefont {B.}~\bibnamefont {Munsky}},
		\bibinfo {author} {\bibfnamefont {F.}~\bibnamefont {Proux}}, \bibinfo
		{author} {\bibfnamefont {N.}~\bibnamefont {Ly}}, \bibinfo {author}
		{\bibfnamefont {F.~E.}\ \bibnamefont {Braye}}, \bibinfo {author}
		{\bibfnamefont {C.}~\bibnamefont {Zimmer}}, \bibinfo {author} {\bibfnamefont
			{F.}~\bibnamefont {Mueller}},\ and\ \bibinfo {author} {\bibfnamefont
			{X.}~\bibnamefont {Darzacq}},\ }\bibfield  {title} {\bibinfo {title}
		{Transcription factors modulate c-fos transcriptional bursts},\ }\href
	{https://doi.org/https://doi.org/10.1016/j.celrep.2014.05.053} {\bibfield
		{journal} {\bibinfo  {journal} {Cell reports}\ }\textbf {\bibinfo {volume}
			{8}},\ \bibinfo {pages} {75} (\bibinfo {year} {2014})}\BibitemShut {NoStop}%
	\bibitem [{\citenamefont {Fritzsch}\ \emph {et~al.}(2018)\citenamefont
		{Fritzsch}, \citenamefont {Baumg{\"a}rtner}, \citenamefont {Kuban},
		\citenamefont {Steinshorn}, \citenamefont {Reid},\ and\ \citenamefont
		{Legewie}}]{fritzsch2018estrogen}%
	\BibitemOpen
	\bibfield  {author} {\bibinfo {author} {\bibfnamefont {C.}~\bibnamefont
			{Fritzsch}}, \bibinfo {author} {\bibfnamefont {S.}~\bibnamefont
			{Baumg{\"a}rtner}}, \bibinfo {author} {\bibfnamefont {M.}~\bibnamefont
			{Kuban}}, \bibinfo {author} {\bibfnamefont {D.}~\bibnamefont {Steinshorn}},
		\bibinfo {author} {\bibfnamefont {G.}~\bibnamefont {Reid}},\ and\ \bibinfo
		{author} {\bibfnamefont {S.}~\bibnamefont {Legewie}},\ }\bibfield  {title}
	{\bibinfo {title} {Estrogen-dependent control and cell-to-cell variability of
			transcriptional bursting},\ }\href
	{https://doi.org/https://doi.org/10.15252/msb.20177678} {\bibfield  {journal}
		{\bibinfo  {journal} {Molecular systems biology}\ }\textbf {\bibinfo {volume}
			{14}},\ \bibinfo {pages} {e7678} (\bibinfo {year} {2018})}\BibitemShut
	{NoStop}%
	\bibitem [{\citenamefont {French}\ and\ \citenamefont
		{Miller}(1989)}]{French1989}%
	\BibitemOpen
	\bibfield  {author} {\bibinfo {author} {\bibfnamefont {S.~L.}\ \bibnamefont
			{French}}\ and\ \bibinfo {author} {\bibfnamefont {O.~L.}\ \bibnamefont
			{Miller}},\ }\bibfield  {title} {\bibinfo {title} {Transcription mapping of
			the escherichia coli chromosome by electron microscopy},\ }\href
	{https://doi.org/10.1128/jb.171.8.4207-4216.1989} {\bibfield  {journal}
		{\bibinfo  {journal} {Journal of Bacteriology}\ }\textbf {\bibinfo {volume}
			{171}},\ \bibinfo {pages} {4207} (\bibinfo {year} {1989})}\BibitemShut
	{NoStop}%
	\bibitem [{\citenamefont {Padovan-Merhar}\ \emph {et~al.}(2015)\citenamefont
		{Padovan-Merhar}, \citenamefont {Nair}, \citenamefont {Biaesch},
		\citenamefont {Mayer}, \citenamefont {Scarfone}, \citenamefont {Foley},
		\citenamefont {Wu}, \citenamefont {Churchman}, \citenamefont {Singh},\ and\
		\citenamefont {Raj}}]{Padovan2015}%
	\BibitemOpen
	\bibfield  {author} {\bibinfo {author} {\bibfnamefont {O.}~\bibnamefont
			{Padovan-Merhar}}, \bibinfo {author} {\bibfnamefont {G.}~\bibnamefont
			{Nair}}, \bibinfo {author} {\bibfnamefont {A.}~\bibnamefont {Biaesch}},
		\bibinfo {author} {\bibfnamefont {A.}~\bibnamefont {Mayer}}, \bibinfo
		{author} {\bibfnamefont {S.}~\bibnamefont {Scarfone}}, \bibinfo {author}
		{\bibfnamefont {S.}~\bibnamefont {Foley}}, \bibinfo {author} {\bibfnamefont
			{A.}~\bibnamefont {Wu}}, \bibinfo {author} {\bibfnamefont {L.}~\bibnamefont
			{Churchman}}, \bibinfo {author} {\bibfnamefont {A.}~\bibnamefont {Singh}},\
		and\ \bibinfo {author} {\bibfnamefont {A.}~\bibnamefont {Raj}},\ }\bibfield
	{title} {\bibinfo {title} {Single mammalian cells compensate for differences
			in cellular volume and dna copy number through independent global
			transcriptional mechanisms},\ }\href
	{https://doi.org/https://doi.org/10.1016/j.molcel.2015.03.005} {\bibfield
		{journal} {\bibinfo  {journal} {Molecular Cell}\ }\textbf {\bibinfo {volume}
			{58}},\ \bibinfo {pages} {339} (\bibinfo {year} {2015})}\BibitemShut
	{NoStop}%
	\bibitem [{\citenamefont {Zhou}\ and\ \citenamefont {Zhang}(2012)}]{Zhou2012}%
	\BibitemOpen
	\bibfield  {author} {\bibinfo {author} {\bibfnamefont {T.}~\bibnamefont
			{Zhou}}\ and\ \bibinfo {author} {\bibfnamefont {J.}~\bibnamefont {Zhang}},\
	}\bibfield  {title} {\bibinfo {title} {Analytical results for a multistate
			gene model},\ }\href {https://doi.org/10.1137/110852887} {\bibfield
		{journal} {\bibinfo  {journal} {SIAM Journal on Applied Mathematics}\
		}\textbf {\bibinfo {volume} {72}},\ \bibinfo {pages} {789} (\bibinfo {year}
		{2012})}\BibitemShut {NoStop}%
	\bibitem [{\citenamefont {Shao}\ and\ \citenamefont
		{Zeitlinger}(2017)}]{Shao2017}%
	\BibitemOpen
	\bibfield  {author} {\bibinfo {author} {\bibfnamefont {W.}~\bibnamefont
			{Shao}}\ and\ \bibinfo {author} {\bibfnamefont {J.}~\bibnamefont
			{Zeitlinger}},\ }\bibfield  {title} {\bibinfo {title} {Paused {RNA}
			polymerase ii inhibits new transcriptional initiation},\ }\href
	{https://doi.org/10.1038/ng.3867} {\bibfield  {journal} {\bibinfo  {journal}
			{Nature Genetics}\ }\textbf {\bibinfo {volume} {49}},\ \bibinfo {pages}
		{1045} (\bibinfo {year} {2017})}\BibitemShut {NoStop}%
	\bibitem [{\citenamefont {Epshtein}\ and\ \citenamefont
		{Nudler}(2003)}]{Epshtein2003}%
	\BibitemOpen
	\bibfield  {author} {\bibinfo {author} {\bibfnamefont {V.}~\bibnamefont
			{Epshtein}}\ and\ \bibinfo {author} {\bibfnamefont {E.}~\bibnamefont
			{Nudler}},\ }\bibfield  {title} {\bibinfo {title} {Cooperation between rna
			polymerase molecules in transcription elongation},\ }\href
	{https://doi.org/10.1126/science.1083219} {\bibfield  {journal} {\bibinfo
			{journal} {Science}\ }\textbf {\bibinfo {volume} {300}},\ \bibinfo {pages}
		{801} (\bibinfo {year} {2003})}\BibitemShut {NoStop}%
	\bibitem [{\citenamefont {Kim}\ \emph {et~al.}(2019)\citenamefont {Kim},
		\citenamefont {Beltran}, \citenamefont {Irnov},\ and\ \citenamefont
		{Jacobs-Wagner}}]{Kim2019}%
	\BibitemOpen
	\bibfield  {author} {\bibinfo {author} {\bibfnamefont {S.}~\bibnamefont
			{Kim}}, \bibinfo {author} {\bibfnamefont {B.}~\bibnamefont {Beltran}},
		\bibinfo {author} {\bibfnamefont {I.}~\bibnamefont {Irnov}},\ and\ \bibinfo
		{author} {\bibfnamefont {C.}~\bibnamefont {Jacobs-Wagner}},\ }\bibfield
	{title} {\bibinfo {title} {Long-distance cooperative and antagonistic rna
			polymerase dynamics via dna supercoiling},\ }\href
	{https://doi.org/https://doi.org/10.1016/j.cell.2019.08.033} {\bibfield
		{journal} {\bibinfo  {journal} {Cell}\ }\textbf {\bibinfo {volume} {179}},\
		\bibinfo {pages} {106} (\bibinfo {year} {2019})}\BibitemShut {NoStop}%
	\bibitem [{\citenamefont {Chatterjee}\ \emph {et~al.}(2021)\citenamefont
		{Chatterjee}, \citenamefont {Goldenfeld},\ and\ \citenamefont
		{Kim}}]{Chatterjee2021}%
	\BibitemOpen
	\bibfield  {author} {\bibinfo {author} {\bibfnamefont {P.}~\bibnamefont
			{Chatterjee}}, \bibinfo {author} {\bibfnamefont {N.}~\bibnamefont
			{Goldenfeld}},\ and\ \bibinfo {author} {\bibfnamefont {S.}~\bibnamefont
			{Kim}},\ }\bibfield  {title} {\bibinfo {title} {Dna supercoiling drives a
			transition between collective modes of gene synthesis},\ }\href
	{https://doi.org/10.1103/PhysRevLett.127.218101} {\bibfield  {journal}
		{\bibinfo  {journal} {Phys. Rev. Lett.}\ }\textbf {\bibinfo {volume} {127}},\
		\bibinfo {pages} {218101} (\bibinfo {year} {2021})}\BibitemShut {NoStop}%
	\bibitem [{\citenamefont {Tripathi}\ \emph {et~al.}(2021)\citenamefont
		{Tripathi}, \citenamefont {Brahmachari}, \citenamefont {Onuchic},\ and\
		\citenamefont {Levine}}]{Tripathi2021}%
	\BibitemOpen
	\bibfield  {author} {\bibinfo {author} {\bibfnamefont {S.}~\bibnamefont
			{Tripathi}}, \bibinfo {author} {\bibfnamefont {S.}~\bibnamefont
			{Brahmachari}}, \bibinfo {author} {\bibfnamefont {J.~N.}\ \bibnamefont
			{Onuchic}},\ and\ \bibinfo {author} {\bibfnamefont {H.}~\bibnamefont
			{Levine}},\ }\bibfield  {title} {\bibinfo {title} {{DNA supercoiling-mediated
				collective behavior of co-transcribing RNA polymerases}},\ }\href
	{https://doi.org/10.1093/nar/gkab1252} {\bibfield  {journal} {\bibinfo
			{journal} {Nucleic Acids Research}\ }\textbf {\bibinfo {volume} {50}},\
		\bibinfo {pages} {1269} (\bibinfo {year} {2021})}\BibitemShut {NoStop}%
	\bibitem [{\citenamefont {Klindziuk}\ and\ \citenamefont
		{Kolomeisky}(2021)}]{Klindziuk2021}%
	\BibitemOpen
	\bibfield  {author} {\bibinfo {author} {\bibfnamefont {A.}~\bibnamefont
			{Klindziuk}}\ and\ \bibinfo {author} {\bibfnamefont {A.~B.}\ \bibnamefont
			{Kolomeisky}},\ }\bibfield  {title} {\bibinfo {title} {Long-range
			supercoiling-mediated rna polymerase cooperation in transcription},\ }\href
	{https://doi.org/10.1021/acs.jpcb.1c01859} {\bibfield  {journal} {\bibinfo
			{journal} {The Journal of Physical Chemistry B}\ }\textbf {\bibinfo {volume}
			{125}},\ \bibinfo {pages} {4692} (\bibinfo {year} {2021})}\BibitemShut
	{NoStop}%
	\bibitem [{\citenamefont {Szavits-Nossan}\ and\ \citenamefont
		{Evans}(2020)}]{Szavits2020-dynamics}%
	\BibitemOpen
	\bibfield  {author} {\bibinfo {author} {\bibfnamefont {J.}~\bibnamefont
			{Szavits-Nossan}}\ and\ \bibinfo {author} {\bibfnamefont {M.~R.}\
			\bibnamefont {Evans}},\ }\bibfield  {title} {\bibinfo {title} {Dynamics of
			ribosomes in {mRNA} translation under steady- and nonsteady-state
			conditions},\ }\href {https://doi.org/10.1103/PhysRevE.101.062404} {\bibfield
		{journal} {\bibinfo  {journal} {Phys. Rev. E}\ }\textbf {\bibinfo {volume}
			{101}},\ \bibinfo {pages} {062404} (\bibinfo {year} {2020})}\BibitemShut
	{NoStop}%
	\bibitem [{\citenamefont {de~Gier}\ and\ \citenamefont
		{Essler}(2005)}]{deGier2005}%
	\BibitemOpen
	\bibfield  {author} {\bibinfo {author} {\bibfnamefont {J.}~\bibnamefont
			{de~Gier}}\ and\ \bibinfo {author} {\bibfnamefont {F.~H.~L.}\ \bibnamefont
			{Essler}},\ }\bibfield  {title} {\bibinfo {title} {Bethe ansatz solution of
			the asymmetric exclusion process with open boundaries},\ }\href
	{https://doi.org/10.1103/PhysRevLett.95.240601} {\bibfield  {journal}
		{\bibinfo  {journal} {Phys. Rev. Lett.}\ }\textbf {\bibinfo {volume} {95}},\
		\bibinfo {pages} {240601} (\bibinfo {year} {2005})}\BibitemShut {NoStop}%
\end{thebibliography}
\end{document}